\documentstyle[epsfig,12pt]{article}

\def\be{\begin{equation}}
\def\ee{\end{equation}}
\def\bea{\begin{eqnarray}}
\def\eea{\end{eqnarray}}

\def\'#1{\if#1i{\accent 19\i}\else{\accent 19 #1}\fi}

\def\e{\-men\-te }
\def\o{\-men\-to }

\def\b{ }

\def\6{\partial} \def\al{\alpha} \def\b{\beta}
\def\g{\gamma} \def\d{\delta} \def\ve{\varepsilon}
\def\e{\epsilon} \def\S{\Sigma}
\def\z{\zeta} \def\h{\eta} \def\th{\theta}
 \def\k{\kappa} \def\l{\lambda}
\def\m{\mu}   \def\p{\pi}
\def\r{\rho} \def\s{\sigma} \def\t{\tau}
\def\p{\phi} \def\ph{\varphi} \def\ps{\psi}
\def\o{\omega}  
 \def\L{\Lambda} \def\S{\Sigma}
  \def\O{\Omega}
 \def\la{\large} 
  
\def\ti{\tilde}  \def\a{\dag}
\def\non{\nonumber\\}
\def\ve{\vec}

\def\ii {\'{\i}}

\def\ti{\tilde}
\def\v{\upsilon}
  
\def\le{\left}
\def\ri{\right}
\def\fr{\frac}
\def\bo{\boldmath}
\def\n{\nabla}
\def\la{\langle}
\def\ra{\rangle}

\def\w{\widetilde}
\def\g{\gamma}
\newcommand{\fourmat}[4]{\left(\begin{array}{cc}  
{#1} & {#2} \\ {#3} & {#4} \end{array} \right)}
\renewcommand{\baselinestretch}{1.2}
\begin{document}
\pagestyle{plain}


\def\baselinestretch{1.2}
\hoffset=-1.0 true cm
\voffset=-2 true cm
\topmargin=1.0cm
\thispagestyle{empty}
\def\thefootnote{\fnsymbol{footnote}}

\thicklines
\begin{picture}(370,60)(0,0)
\setlength{\unitlength}{1pt}
\put(40,53){\line(2,3){15}}
\put(40,53){\line(5,6){19}}
\put(40,53){\line(1,1){27}}
\put(40,53){\line(6,5){33}}
\put(40,53){\line(3,2){25}}
\put(40,53){\line(2,1){19}}
\put(40,53){\line(5,-6){17}}
\put(40,53){\line(1,-1){22}}
\put(40,53){\line(6,-5){30}}
\put(40,53){\line(3,-2){22}}
\put(40,53){\line(-2,1){15}}
\put(40,53){\line(-3,1){23}}
\put(40,53){\line(-4,1){26}}
\put(40,53){\line(-6,1){36}}
\put(40,53){\line(-1,0){40}}
\put(40,53){\line(-6,-1){32}}
\put(40,53){\line(-3,-1){20}}
\put(40,53){\line(-2,-1){10}}
\put(75,45){\Huge \bf IFT}
\put(180,56){\small \bf Instituto de F\'\i sica Te\'orica}
\put(165,42){\small \bf Universidade Estadual Paulista} 
\put(-25,2){\line(1,0){433}}
\put(-25,-2){\line(1,0){433}}
\end{picture}  


\vskip .3cm
\noindent
{DISSERTA\c C\~AO DE MESTRADO}
\hfill    IFT--D.000/98\\


\vspace{3cm}
\begin{center}

{\large {\bf {Corda Bos\^onica \`a Temperatura Finita}} }


\vspace{1.2cm}
Wagner Paniago de Souza
\end{center}

\vskip 3cm
\hfill Orientador
\vskip 0.4cm

\hfill {\em }
Prof. Dra. Maria Cristina Batoni Abdalla
\vskip 4cm
\vfill
\begin{center}
Julho de 2002
\end{center}

\newpage

\pagenumbering{roman}

\begin{center}
{\Large \bf Agradecimentos}
\end{center}
\vskip 2.0cm


\newpage

\begin{center}
{\Large \bf Resumo}
\end{center}
\vskip 2.0cm

Inicialmente, apresentamos a constru\c{c}\~ao de Umezawa e Takahashi para uma din\^amica de campos t\'ermicos (DCT) e obtemos algumas grandezas estat\'{\i}sticas, tais como o n\'umero m\'edio de part\'{\i}culas, entropia e energia livre de Helmholtz para sistemas em equil\'{\i}brio t\'ermico. Com o objetivo de aplicarmos estes conceitos em uma teoria de cordas, quantizamos a corda bos\^onica cl\'assica (aberta e fechada) em um calibre manifestamente covariante e no calibre de cone de luz. Derivamos ent\~ao uma descri\c{c}\~ao para a corda \`a temperatura finita no contexto da DCT. A entropia dos estados associados \`as equa\c{c}\~oes de movimento da corda aberta com diferentes combina\c{c}\~oes das condi\c{c}\~oes de contorno de Neumann e Dirichlet s\~ao dadas. Discutimos as trasforma\c{c}\~oes, do espa\c{c}o do Fock e dos operadores, geradas por transforma\c{c}\~oes unit\'arias de Bogoliubov mais gerais, cujos operadores formam uma \'algebra $SU(1,1)$. Neste contexto, obtivemos tamb\'em a entropia da corda fechada.   


\vskip 1.0cm
\noindent
{\bf Palavras Chaves}: Corda Bos\^onica e Temperatura Finita
\vskip 0.5cm

\noindent
{\bf \'Areas do Conhecimento}:
Teoria de Campos
\newpage

\begin{center}
{\Large \bf Abstract}
\end{center}
\vskip 2.0cm

Inicially, we present the Umezawa e Takahashi's construction for a Thermal Fields Dynamics (TFD) and we obtain some statistical quantities as the mean number of particles, entropy and Helmholtz's free energy for systems in thermal equilibrium. With the objectiv of apply these concepts in a strings theory, we quantize the classical bosonic string (open and close) in manifestly covariant gauge and in the light-cone gauge. So, we derive a description for the strings at finite tempe- \\ rature based on the TFD background. The entropy of the states associated with the moviment equations of the opened strings with different Neumann and Dirichlet boundary condictions are given, we discuss the Fock space transformations (and of the operators) generated by more general unitary Bogoliubov transformations, end even the entropy of the closed strings.


\vfill \eject

\newpage

\tableofcontents

\newpage

\section{Introdu\c{c}\~ao}
\pagenumbering{arabic}
${}$

       Nas \'ultimas d\'ecadas, um dos maiores problemas da f\'{\i}sica das altas energias foi encontrar uma descri\c{c}\~ao consistente para o mundo na escala subat\^omica. Para atingirmos este objetivo, devemos tamb\'em ser capazes de entender a gravidade tanto no n\'ivel cl\'assico quanto no n\'ivel qu\^antico. Pode-se dizer que neste s\'eculo as duas teorias que mais tiveram sucesso neste \^ambito foram a teoria qu\^antica dos campos e a teoria da relatividade geral. No entanto, estas duas teorias n\~ao convivem muito bem quando aplicadas ao mesmo dom\'{\i}nio. Neste contexto, a teoria que cont\'em o maior n\'umero de caracter\'{\i}sticas desej\'aveis \'e a teoria de cordas. Esta \'e a teoria qu\^antica na qual seus constituintes fundamentais s\~ao objetos matem\'aticos extensos, unidimensionais, descritos em um espa\c{c}o-tempo de dimens\~oes superiores. Esta teoria tem uma variedade de aplica\c{c}\~oes fundamentais em F\'{\i}sica Te\'orica. Atualmente, a mais importante aplica\c{c}\~ao da teoria de cordas \'e a unifica\c{c}\~ao de todas as for\c{c}as fundamentais  e part\'{\i}culas elementares em uma \'unica teoria. A id\'eia b\'asica desta teoria \'e que toda a mat\'eria \'e formada por cordas min\'usculas, por exemplo, se um el\'etron \'e visto como um ponto de $10^{-13} cm$, se magnificarmos a resolu\c{c}\~ao da escala para $10^{-33} cm$, ent\~ao iriamos verificar que ele \'e um objeto extenso como uma corda. A unifica\c{c}\~ao \'e feita de modo que todas as part\'{\i}culas s\~ao formadas por um \'unico tipo de corda, e diferentes tipos de part\'{\i}culas s\~ao somente excita\c{c}\~oes de um mesmo tipo de corda, ou seja, se excitarmos a corda, um modo de excita\c{c}\~ao ser\'a o el\'etron, o outro um f\'oton, e assim por diante. Devemos ainda ressaltar que o modo correspondente a um estado de massa nula e spin 2 pode ser identificado com o gr\'aviton, e desta maneira garantimos que a teoria necessariamente cont\'em a gravidade qu\^antica.

        Para chegarmos nestes resultados, apresentaremos uma breve introdu\c{c}\~ao  \`a teoria relativ\'{\i}stica da part\'{\i}cula cl\'assica e, em seguida, obteremos a a\c{c}\~ao de Nambu-Goto para uma corda cl\'assica, cuja qual possui claramente uma interpreta\c{c}\~ao geom\'etrica, e, a partir desta, deduziremos uma a\c{c}\~ao mais geral, a a\c{c}\~ao de Polyakov, e mostraremos que estas duas s\~ao classicamente equivalentes. Analisaremos tamb\'em as simetrias e invari\^ancias da a\c{c}\~ao e obteremos o tensor energia-momento. Al\'em disso, obteremos e resolveremos a equa\c{c}\~ao de movimento em um dado calibre covari\^ante e sujeita a diferentes condi\c{c}\~oes de contorno, cujas solu\c{c}\~oes nos fornecem as amplitudes de vibra\c{c}\~ao da corda em termos dos coeficientes do oscilador $\al_{n}^{\mu}$. Escreveremos tamb\'em os colchetes de Poisson para as vari\'aveis din\^amicas da corda e do centro de massa e para os  $\al_{n}^{\mu}$. A \'algebra cl\'assica de Virassoro e os estados de massa para a corda ser\~ao tamb\'em obtidos. Em seguida, quantizaremos a corda bos\^onica (aberta e fechada) e veremos como o conte\'udo de part\'{\i}culas aparece nesta teoria. A quantiza\c{c}\~ao (can\^onica) ser\'a realizada tanto em um calibre manifestamente covariante quanto em um n\~ao manifestamente covariante (cone de luz) sendo que este possui a vantagem de estar livre de estados de norma negativa. Verificaremos tamb\'em que para a teoria ter certos aspectos desej\'aveis, a dimens\~ao do espa\c{c}o-tempo no qual a corda est\'a imersa deve ser $D=26$. 

       Visando estudar alguns aspectos termodin\^amicos da teoria de cordas, iremos a-\\presentar no primeiro cap\'{\i}tulo o formalismo de Umezawa e Takahashi~{\cite{ut}}, que consiste basicamente na constru\c{c}\~ao de uma din\^amica para campos t\'ermicos, na qual o v\'acuo depende da temperatura de tal forma que o valor esperado neste v\'acuo de uma vari\'avel din\^amica de um sistema seja igual \`a sua m\'edia estat\'{\i}stica no ensemble grand-can\^onico. Neste formalismo, os estados \`a temperatura finita s\~ao mapeados no estado de v\'acuo \`a temperatura nula atrav\'es de uma transforma\c{c}\~ao de Bogoliubov. O formalismo de Umezawa \'e essencialmente baseado em conceitos quanto-mec\^anicos e permite ser escrito no formalismo Lagrangeano, podendo ser aplicado em teorias de campos de forma mais abrangente. Em seguida, mostraremos como a din\^amica de campos t\'ermicos pode ser implementada em uma teoria de campos qualquer partindo de certos axiomas. Iremos aplicar este formalismo para uma teoria de cordas bos\^onicas. Como iremos verificar, a constru\c{c}\~ao de uma teoria de cordas \`a temperatura finita via transforma\c{c}\~oes de Bogoliubov conservam muitas das propriedades da teoria de cordas \`a temperatura nula. A temperatura ser\'a introduzida atrav\'es do mapeamento do v\'acuo e das solu\c{c}\~oes das equa\c{c}\~oes de movimento, as quais s\~ao expandidas em termos dos  coeficientes dos osciladores, \`a temperatura nula, nos respectivos v\'acuo e solu\c{c}\~oes das equa\c{c}\~oes de movimento \`a temperatura finita via tais transforma\c{c}\~oes. Deste modo, podemos interpretar a corda t\'ermica como um modelo de excita\c{c}\~oes de um v\'acuo t\'ermico bos\^onico. Calcularemos a entropia do sistema que \'e dada pelo valor esperado de um  operador definido como operador de entropia. Para cada um dos setores da corda, a entropia do v\'acuo t\'ermico deve ser calculada  para cada uma das dire\c{c}\~oes espa\c{c}o-temporais, sendo que, devido \`a sua extensividade, a entropia total \'e dada pela soma das entropias dos campos escalares em cada uma das dire\c{c}\~oes. 

         Analizaremos para a corda aberta, se as diferentes condi\c{c}\~oes de contorno sob as quais as equa\c{c}\~oes de movimento podem estar sujeitas interferem nas propriedades termodin\^amicas do sistema. Em geral, se um dado estado depende das condi\c{c}\~oes de contorno, a entropia, que \'e calculada como a m\'edia do operador entropia neste estado, tamb\'em depende. Uma vez que o estado de v\'acuo da corda bos\^onica depende de certa forma das condi\c{c}\~oes de contorno impostas sobre as equa\c{c}\~oes de movimento, desejamos olhar a entropia dos estados associados \`as solu\c{c}\~oes gerais das equa\c{c}\~oes de movimento com diferentes combina\c{c}\~oes  das condi\c{c}\~oes de contorno  em suas extremidades. Em resumo, construiremos uma din\^amica de campos t\'ermicos para a corda bos\^onica aberta e calcularemos a entropia dos estados no calibre do cone de luz com depend\^encia expl\'{\i}cita das condi\c{c}\~oes de contorno. Isto ser\'a importante para a obtens\~ao do v\'acuo t\'ermico em teoria de cordas e D-branas t\'ermicas no contexto da din\^amica de campos t\'ermicos e no entendimento da entropia das D-branas (que \'e dada pelo valor m\'edio do operador entropia para a corda bos\^onica no estado de D-brana t\'ermica)  no limite perturbativo da teoria de cordas.

         Em seguida, introduziremos temperatura em uma teoria de cordas bos\^onica fechada de um modo um pouco diferente do usado para a corda aberta. Os estados de v\'acuo e os operadores ser\~ao constru\'{\i}dos usando o mesmo formalismo da  din\^amica de campos t\'ermicos, mas diferindo da constru\c{c}\~ao anterior, pelo fato de que aqui os operadores que mapeiam objetos em $T=0$ em objetos \`a temperatura finita \'e uma combina\c{c}\~ao linear de operadores que satisfazem a \'algebra $SU(1,1)$.  Segundo \cite{chu,hu}, podemos definir operadores de Bogoliubov para qualquer teoria, cujos quais formam uma representa\c{c}\~ao de osciladores do grupo $SU(1,1)$ para b\'osons, e se todos estes operadores forem considerados, as transforma\c{c}\~oes t\'ermicas podem ser geradas por uma combina\c{c}\~ao linear destes geradores do $SU(1,1)$. Os coeficientes destes ge-\\radores determinam se as transforma\c{c}\~oes s\~ao ou n\~ao unit\'arias  e se as caracter\'{\i}sticas b\'asicas da din\^amica de campos t\'ermicos s\~ao satisfeitas pelos geradores. Uma vez constru\'{\i}do o sistema \`a temperatura finita, obteremos a entropia do sistema neste formalismo  do mesmo modo j\'a citado acima.    


\newpage

\section{Corda Bos\^onica}
\subsection{Part\'{\i}cula Cl\'assica Relativ\'{\i}stica}
${}$

        Inicialmente, vamos descrever a a\c{c}\~ao de uma part\'{\i}cula puntual cl\'assica relativ\'{\i}stica. Este estudo servir\'a para nos guiar na obten\c{c}\~ao da a\c{c}\~ao para a corda bos\^onica cl\'assica. A part\'{\i}cula est\'a se movendo em um espa\c{c}o-tempo de coordenadas $(X^{0},X^{1},\cdots,X^{D-1})$, e sua trajet\'oria neste \'e chamada de linha-mundo. A m\'etrica do espa\c{c}o-tempo usada \'e a de Minkowski com $D-1$ autovalores negativos e um positivo. 

        Sendo $x^{\mu}(\tau)$ a trajet\'oria cl\'assica da part\'{\i}cula, na qual $\tau$  \'e um par\^ametro real que rotula os pontos ao longo da linha mundo no espa\c{c}o-tempo, ent\~ao, $\dot{x}^{\mu}=\partial_{\tau}x^{\mu}$ representa a velocidade da part\'{\i}cula (vetor tangente). Sendo a linha-mundo parametrizada pelo tempo pr\'oprio da part\'{\i}cula, ou seja, $\dot{x}^{\mu}\dot{x}_{\mu}=1$, ent\~ao, o momento da part\'{\i}cula  \'e dado por $p^{\mu}=m\dot {x}^{\mu}$, onde $m$ \'e a massa de repouso da part\'{\i}cula. Deste modo, $p^{\mu}p_{\mu}=m^{2}\dot{x}^{\mu}\dot{x}_{\mu}=m^{2}=p^{2}$, portanto
\be
p^{2}-m^{2}=0. \label{empc}
\ee
Esta \'e a equa\c{c}\~ao de movimento de uma part\'{\i}cula relativ\'{\i}stica que nos fornece um v\ii nculo entre a massa de repouso e seu momento, e \'e conhecida com condi\c{c}\~ao de concha de massa. Agora, se $\tau$ \'e algum outro par\^ ametro que n\~ao o tempo pr\'oprio, o momento da part\'{\i}cula ser\'a
\be
p^{\mu}=\frac{m\dot {x}^{\mu}}{\sqrt{(\dot {x}^{\nu})^{2}}}. \label{mpc}
\ee
Pode-se verificar que $p^{\mu}$ dado pela eq.(\ref{mpc}) satifaz a condi\c{c}\~ao de concha de massa, eq.(\ref{empc}).

        Dada a a\c{c}\~ao
\be
S=\int^{\tau_2}_{\tau_1} d\tau L(\dot {x}^{\mu},x^{\mu},\tau ), \label{apc}
\ee
o momento $p^\mu$ da part\ii cula \'e dado pela equa\c{c}\~ao $p^{\mu}=-\frac {\6L}{\6\dot {x}_{\mu}}$, a qual, quando integrada, resulta em 
\be
L=-p_{\mu}\dot {x}^{\mu}, \label{lpc}
\ee
onde $p^{\mu}$ \'e o momento conjugado de $x^{\mu}$. Substituindo a eq.(\ref{mpc}) na eq.(\ref{lpc}) obtemos
\be
L=-m\sqrt{\dot {x}_{\mu}\dot {x}^{\mu}}. \label{lpc1}
\ee

        Desta forma a a\c{c}\~ao para uma part\'{\i}cula relativ\ii stica de massa $m$ \'e
\be
S=-m \int^{\tau_2}_{\tau_1} d\tau \sqrt{\dot {x}_{\mu}\dot {x}^{\mu}},\label{apc1}
\ee  
onde $\tau$  \'e um par\^ ametro arbitr\'ario da linha mundo.

        Qualquer parametriza\c{c}\~ao da linha mundo \'e permitida, desde que, como veremos, a a\c{c}\~ao seja invariante por reparametriza\c{c}\~ao (porque esta representa um objeto geom\'etrico, ou seja, \'e proporcional ao comprimento invariante da linha mundo). Reparametrizando a eq.(\ref{apc1}) de $\tau$ para $\tau^{'}$ temos que: $\tau \longrightarrow \tau^{'}$ e $x^\mu(\tau) \longrightarrow x^\mu (\tau^{'})$. Portanto, diferenciando as coordenadas antigas $(x^\mu(\tau))$ em rela\c{c}\~ao aos novos par\^ametros $(\t^{'})$ obtemos
\be
\dot {x}^{\mu^{'}}=\frac{\6}{\6\t^{'}}x^{\mu}(\t)={\frac{\6}{\6\t}x^{\mu}(\t)}{\frac{\6\t}{\6\t^{'}}=\dot x^{\mu}\6_{\t^{'}}\t}, \label{rep}
\ee
e a diferencial de $\tau^\prime$ \'e 
\be
d\tau^{'}=\frac{\6\tau^{'}}{\6\tau}d\tau. \label{dr}
\ee
Ent\~ao, uma nova a\c{c}\~ao $(S^{'})$ escrita em termos dos novos par\^ametros ser\'a
\be
S^{'}=-m\int d\tau^{'} \sqrt{\dot {x}^{\mu^{'}}\dot {x}^{'}_{\mu}},     
\ee
que, por substitui\c{c}\~ao das equa\c{c}\~oes (\ref{rep}) e (\ref{dr}) obt\'em-se
\be
S^{'}=-m\int\frac{ \6{\tau}^{'}}{\6{\tau}} d\tau  \sqrt{\dot {x}^{\mu}\frac{\6{\tau}}{\6{\tau}^{'}}\dot {x}_{\mu}\frac{\6{\tau}}{\6{\tau^\prime}}}      
\ee
ou seja
$$
S^{'}=-m \int^{\tau_2}_{\tau_1}d\tau \sqrt{\dot {x}_{\mu}\dot {x}^{\mu}}=S.
$$
Logo, como vemos, a a\c{c}\~ao \'e invariante por reparametriza\c{c}\~ao da linha mundo.

        Em termos da m\'etrica do espa\c{c}o-tempo o quadrado da dist\^ancia entre dois pontos vizinhos \'e
\be
ds^2=g_{\mu\nu}dx^{\mu}dx^{\nu}.
\ee
Ent\~ao, o integrando da a\c{c}\~ao (\ref{apc1}) torna-se
\be
d\tau \sqrt{\dot {x}_{\mu}\dot {x}^{\mu}}= d\tau \sqrt{g_{\mu\nu}d\dot {x}^{\mu}d\dot {x}^{\nu}}=\sqrt{g_{\mu\nu}dx^{\mu}dx^{\nu}}=\sqrt{ds^2}= ds. \label{clm}
\ee

        Logo, ultilizando-se a eq.(\ref{clm}), a a\c{c}\~ao (\ref{apc1}) pode ser escrita como:
\be
S=-m \int ds
\ee
ou seja, \'e o produto da massa de repouso pelo comprimento da linha mundo (como j\'a hav\'{\i}amos citado).

        A a\c{c}\~ao escrita na forma (\ref{apc1}) n\~ao \'e muito \'util, pois n\~ao pode ser usada para uma part\'{\i}cula com massa nula e apresenta problemas quando tentamos quantizar a teoria, devido a presen\c{c}a da raiz quadrada no integrando. Uma Lagrangeana alternativa que podemos usar para obter uma a\c{c}\~ao classicamente equivalente a a\c{c}\~ao (\ref{apc1}) \'e a seguinte:
\be
L=-p_\mu\dot x^\mu +\frac{1}{2}e(\tau)(p^2 -m^2)   \label{lapc}
\ee
onde temos a Lagrangeana dada pela eq.(\ref{lpc}), mais um v\'{\i}nculo dado pela eq.(\ref{empc})  vezes o multiplicador de Lagrange $\frac {1}{2}e(\tau)$, onde $e(\t)$ \'e um campo fict\'{\i}cio.

        Devemos eliminar $p^\mu$ da eq.(\ref{lapc}) escrevendo-o em termos de $\dot x^\mu$ e $e$. Para isto, usaremos a equa\c{c}\~ao de movimento para $p_\mu$:
$$
\frac{\delta L}{\delta p_\mu}=-\dot x^\mu + ep^\mu =0,
$$
ou seja,
$$
p^\mu=\frac{\dot x^\mu}{e}
$$
que, substituindo na eq.(14) obt\^em-se
$$
L= -\frac{1}{2}(e^{-1}\dot x^2 + em^2).
$$
Portanto a nova a\c{c}\~ao pode ser escrita como
\be
S= -\frac{1}{2}\int d\tau (e^{-1}\dot x^2 + em^2). \label{apc2}
\ee

        Agora, mostraremos que esta \'utima a\c{c}\~ao \'e classicamente equivalente \`a primeira. Das equa\c{c}\~oes de Euler-Lagrange temos:
$$
\frac{\d S}{\d e}=\frac{\6L}{\6e}
$$
$$
\frac{\6L}{\6e}=-\frac{1}{2}(-e^{-2}\dot {x}^2 + m^2)=0
$$
$$
\dot x^2-m^2e^2=0;
$$
$$
e=\frac{\sqrt{\dot x^2}}{m}.
$$
Substituindo esta \'ultima equa\c{c}\~ao na eq.(\ref{apc2}) obtemos
\begin{eqnarray}
S&=& -\frac{1}{2}\int d\tau (\frac{m}{\sqrt{\dot x^2}}\dot x^2 +\frac{\sqrt{\dot x^2}}{m} m^2) \nonumber \\
&=&-\frac{1}{2}\int d\tau (m\sqrt{\dot x^2}+m\sqrt{\dot x^2}) \nonumber \\
&=&-m\int^{\tau_2}_{\tau_1} d\tau \sqrt{\dot {x}_{\mu}\dot {x}^{\mu}} \nonumber
\end{eqnarray}
que \'e id\^entica a primeira a\c{c}\~ao obtida, eq.(\ref{apc1}), usando a lagrangeana (\ref{lpc1}).

        As a\c{c}\~oes (\ref{lpc1}) e (\ref{apc1}) s\~ao chamadas respectivamente de a\c{c}\~ao de Nambu-Goto (NG) e Polyakov (PL), sendo que esta \'ultima n\~ao apresenta os problemas da primeira e \'e mais geral. Qu\^anticamente n\~ao sabemos se as duas a\c{c}\~oes s\~ao equivalentes.

\subsection{Corda Bos\^onica Cl\'assica}
${}$

        Para constru\'{\i}rmos a a\c{c}\~ao da corda cl\'assica, estenderemos a id\'eia de que a a\c{c}\~ao da part\'{\i}cula cl\'assica \'e proporcional ao comprimento da linha mundo. Ou seja, a a\c{c}\~ao para uma corda cl\'assica deve ser proporcional a \'area da superf\'{\i}cie descrita pela corda quando a mesma evolui no espa\c{c}o-tempo, a saber,
\be
S=-T\int dA,   \label{acao}
\ee
onde a constante de proporcionalidade \'e menos a tens\~ao da corda. Esta generaliza\c{c}\~ao da linha mundo \'e chamada folha mundo e esta ilustrada na fig. 2.1 abaixo.
\begin{figure}
\center{\epsfig{figure=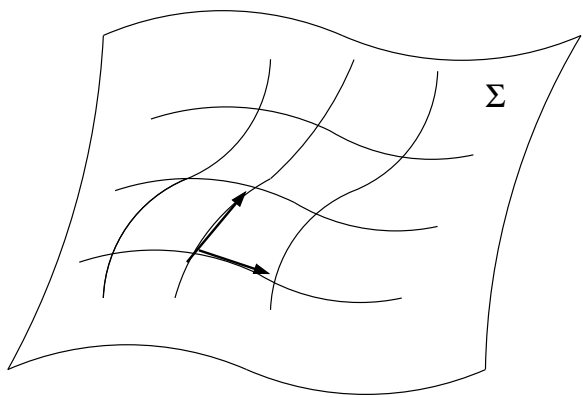}}
\caption{Superf\'{\i}cie de coordenadas $\s$ e $\t$ e descrita pelo campo $X^{\mu}(\t,\s)$}
\end{figure}

        Se $\s$ \'e uma coordenada tipo espa\c{c}o que denota a posi\c{c}\~ao ao longo da corda e $\t$ uma coordenada tipo tempo que parametriza sua evolu\c{c}\~ao temporal, ambos definidos respectivamente nos intervalos $0 \leq \s \leq \pi$ e $\tau_1 \leq \t \leq \t_2$, ent\~ao $X^\mu(\t,\s) \equiv X^\mu(\s^0,\s^1) \equiv X^\mu(\s^a)$ descreve matematicamente a folha mundo e nos fornece a posi\c{c}\~ao da corda para $\s$ e $\t$ espec\'{\i}ficos.

        Para maior clareza das express\~oes, estamos convencionando, neste cap\'{\i}tulo, que os \'{\i}ndices gregos referem-se ao espa\c{c}o-tempo e variam de $0$ a $D-1$, onde $D$ \'e a dimens\~ao do espa\c{c}o-tempo, e os \'{\i}ndices latinos referem-se \`a folha mundo e assumem os valores $0$ e $1$.  Uma outra conven\c{c}\~ao usada \'e a das unidades nas quais $ \hbar= c=1 $. Nestas unidades, a tens\~ao da corda  possui unidade de $[comprimento]^{-2}$.

        Precisamos obter uma forma expl\'{\i}cita para a a\c{c}\~ao dada pela eq.~(\ref{acao}). Para isto, derivaremos uma express\~ao integral para a \'area de uma  superf\'{\i}cie curva (folha mundo), dotada de uma m\'etrica gen\'erica $g_{ab}(\s^{a})$ e imersa em um espa\c{c}o-tempo $D$-dimensional com  m\'etrica $G_{\mu \nu}(X)$.

        Dividindo a folha mundo em diversos paralelogramos infinitesimais  (ver fig. 2.2) de lados $\vec{d\t}$ e $\vec{d\s}$, onde  $\vec{d\t}$ e  $\vec{d\s}$ s\~ao vetores tangentes \`as linhas de coordenadas constantes $\t$ e $\s$ respectivamente. A \'area do paralelogramo \'e dada por
\be
dA= \|\ve{d\t} \times \ve{d\s} \|=\|\ve{d\s^0} \times \ve{d\s^1} \|=\|\ve{d\s^0}\|\|\ve{d\s^1}\|\sin\th \label{dA}
\ee
onde $\th$ \'e o \^angulo entre $\ve{d\s^0}$ e $\ve{d\s^1}$. Usando a  defini\c{c}\~ao $ \|\ve{B}\|^2=g_{ab}B^aB^b $,
onde $\ve{B}$ \'e um vetor arbitr\'ario, e o fato da m\'etrica ser um tensor sim\'etrico, podemos escrever a eq.(\ref{dA}) na forma
\bea
dA&=&\sqrt{\|\ve{d\s^0}\|^2\|\ve{d\s^1}\|^2(1-\cos^2\theta)} \nonumber \\
&=&\sqrt{\|\ve{d\s^0}\|^2\|\ve{d\s^1}\|^2 -(\|\ve{d\s^0}\|\|\ve{d\s^1}\|\cos^2\theta)^2} \nonumber \\
&=&\sqrt{\|\ve{d\s^0}\|^2\|\ve{d\s^1}\|^2 -(\|\ve{d\s^0}\|.\|\ve{d\s^1}\|)^2}  \nonumber \\
&=&\sqrt{\mid g_{00} g_{11}-(g_{01})^2 \mid}d\s^0d\s^1.
\eea
Mas $\mid g_{00} g_{11}-(g_{01})^2 \mid = \mid det \, g_{ab} \mid=\mid g \mid$, logo
\be
dA=d^{2}\s\sqrt{\mid g \mid}=d^{2}\s\sqrt{- g}=d^{2}\s\sqrt{-det \; g_{ab}}. \label{dA1}
\ee
O sinal negativo na eq. (\ref{dA1}) aparece porque assumimos que $\t$\ \'e uma coordenada do tipo tempo e $\s$ \'e do tipo espa\c{c}o. Isto nos leva \`a uma m\'etrica $(g_{ab})$ indeterminada  na superf\'{\i}cie, cujo  determinante \'e menor do que zero. Ent\~ao $-g>0$.
\begin{figure}
\center{\epsfig{figure=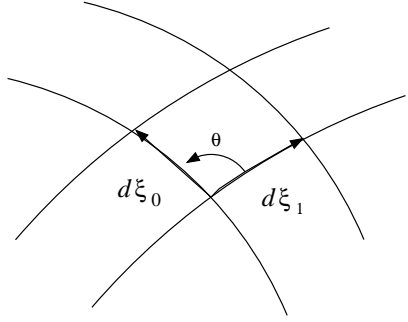}}
\caption{Em cada ponto da superf\'{\i}cie existem dois vetores de deslocamento infinitesimais  $d\t$ e $d\s$ tangentes as linhas constantes $\t$ e $\s$.}
\end{figure}
%
%

        Usando o fato de  que a folha mundo est\'a ``imersa'' no espa\c{c}o-tempo,  podemos usar a m\'etrica do espa\c{c}o-tempo $ \left(G_{\mu\nu}(X)\right)$ para medirmos dist\^ancias na superf\'{\i}cie da folha mundo, ou seja, obtermos uma express\~ao para a m\'etrica da folha mundo em termos da m\'etrica e coordenadas do espa\c{c}o-tempo. Neste caso, chamamos esta m\'etrica obtida para a folha mundo de m\'etrica induzida e a representamos por $h_{ab}$ para a diferenciarmos da m\'etrica gen\'erica $g_{ab}$.

        Consideremos uma dist\^ancia $ds$ no espa\c{c}o-tempo. Em termos da m\'etrica do espa\c{c}o-tempo, o quadrado da dist\^ancia entre dois pontos vizinhos $X^\mu$ e $X^\mu+dX^\mu$ \'e dado por
\be
ds^2=G_{\mu\nu} dX^\mu dX^\nu. \label{ds2}
\ee
Se os pontos $X^\mu$ e $X^\mu+dX^\mu$ est\~ao sobre a superf\'{\i}cie da folha mundo, ent\~ao $X^\mu=X^\mu(\s^a)$, e sua diferencial \'e
\be
dX^\mu=\6_aX^\mu d\s^a. \label{dX}
\ee
Por outro lado, como os pontos est\~ao sobre a superf\'{\i}cie, ent\~ao  $ds^2$ pode ser escrito em termos da m\'etrica induzida na superf\'{\i}cie $(h_{ab})$
\be
ds^2= h_{ab}d\s^ad\s^b. \label{ds3}
\ee
Substituindo a eq.(\ref{dX}) na eq.(\ref{ds2}) e igualando com a eq.(\ref{ds3}) obtemos:

\be
h_{ab}=G_{\mu\nu}\6_aX^\mu \6_bX^\nu. \label{mi}
\ee
\'E importante notar que fizemos uma mudan\c{c}a na m\'etrica de $g_{ab}(\s^c)$ para $h_{ab}\left (X^\mu(\s^c)\right)$. Ent\~ao, calculando as componentes da m\'etrica~(\ref{mi}), obtemos as seguintes rela\c{c}\~oes:
\begin{eqnarray}
h_{00}&=&G_{\mu\nu}\6_0X^\mu \6_0X^\nu=\6_0X^\mu \6_0X_\mu=\dot X^\mu\dot X_\mu=(\dot X)^2 \nonumber \\
h_{11}&=&G_{\mu\nu}\6_1X^\mu \6_1X^\nu=\6_1X^\mu \6_1X_\mu=X^{\mu^{'}} X^{'}_\mu=(X^{'})^2  \nonumber\\
h_{01}&=&h_{10}=G_{\mu\nu}\6_0X^\mu \6_1X^\nu=\6_0X^\mu \6_1X_\mu=\dot X^{\mu}X^{'\mu}.   \label{cm}
\eea

        Trocando na eq.(\ref{dA1}) a m\'etrica gen\'erica $g_{ab}$ pela m\'etrica induzida $h_{ab}$ e usando as componentes de $h_{ab}$ deduzidas em (\ref{cm}), podemos escrever
\be
dA=d^{2}\s \sqrt{-\det h_{ab}}=d^{2}\s \sqrt{-h_{00}h_{11}+(h_{01})^2}=d^{2}\s \sqrt{(\dot X \cdot X^{'})^{2}-(\dot X)^2(X^{'})^2}. \label{dA2}
\ee
Substituindo a eq.(\ref{dA2}) na eq.(\ref{acao}), obtemos
\be
S_{NG}=-T\int d\t d\s \sqrt{(\dot X \cdot X^{'})^{2}-(\dot X)^2(X^{'})^2}=-T \int d\t d\s \sqrt{-h} \label{acao ng}
\ee
onde $h=\det{h_{ab}}$.

        A a\c{c}\~ao dada pela eq.(\ref{acao ng}) \'e conhecida como  a\c{c}\~ao de Nambu-Goto (NG). Sendo um objeto geom\'etrico, uma \'area, a a\c{c}\~ao de Nambu-Goto \'e invariante por reparametri-za\c{c}\~ao, ou seja, invariante sob transforma\c{c}\~oes de coordenadas da folha mundo, 
 \be
(\t,\s) \longrightarrow \left(\t^{'}(\t,\s),\s^{'}(\t,\s)\right ).
\ee

        A presen\c{c}a de uma raiz quadrada no integrando expressa a n\~ao linearidade da  a\c{c}\~ao de Nambu-Goto. Devido a dificuldade de trabalharmos com esta a\c{c}\~ao n\~ao linear, tentaremos eliminar a raiz quadrada da eq.(\ref {acao ng}) escrevendo uma outra  a\c{c}\~ao classicamente equivalente.
        
        Para isto, faremos a varia\c{c}\~ao da a\c{c}\~ao de Nambu-Goto, eq.(\ref{acao ng}), onde usaremos a  varia\c{c}\~ao de $h$ dada por 
\be
\d h=\d \det h_{ab}=\d e^{tr \; \ln(h_{ab})}=h\; tr\; \d \ln h_{ab}=h\; tr \; \frac{1}{h_{ab}}\d h_{ab}=hh^{ab}\d h_{ab},
\ee
que \'e obtida usando-se a condi\c{c}\~ao de ortogonalidade do tensor m\'etrico $h^{ab}h_{bc}=\d^a_{ c}$: 
\be
\delta S_{NG}=-\frac{T}{2} \int d\t d\s \sqrt{-h}h^{ab}\d h_{ab}. \label{pvang}
\ee
A varia\c{c}\~ao em $h_{ab}$ \'e devido a varia\c{c}\~ao de $X^\mu$, porque $h_{ab}$ \'e a m\'etrica induzida por $G_{\mu\nu}$. Temos como resultado da varia\c{c}\~ao da eq.(\ref{mi}) e da simetria do tensor m\'etrico $\d h_{ab}=\d h_{ba}$ a equa\c{c}\~ao  
\be
\d h_{ab}=G_{\mu\nu}\d(\6_aX^\mu \6_bX^\nu)=\6_a\d X^\mu\6_bX_\mu+\6_aX^\mu\6_b\d X_\mu=2\6_aX^\mu \6_b\d X_\mu. \label{vmi}
\ee
Desta forma, substituindo a eq.(\ref{vmi}) na eq.(\ref{pvang}) e usando $ \6_b({\sqrt{-h}}h^{ab}\6_aX^\mu \d X_\mu)=\6_{b}({\sqrt{-h}}h^{ab}\6_aX^\mu ) \d X_\mu +{\sqrt{-h}}h^{ab}\6_aX^\mu \6_{b}(\d X_\mu)$, a eq.(\ref{pvang}) pode ser colocada na forma
$$
\delta S_{NG}=-T\int d^{2}\s \6_b({\sqrt{-h}}h^{ab}\6_aX^\mu \d X_\mu)+T\int d^{2}\s \6_b({\sqrt{-h}}h^{ab}\6_aX^\mu ) \d X_\mu.
$$
Pelo Teorema de Gauss, a primeira integral da equa\c{c}\~ao acima torna-se
\be
\delta S_{NG}=-T \int_ {\6 \S}  d\s_b {\sqrt{-h}}h^{ab}\6_aX^\mu \d X_\mu + T\int d^{2}\s \6_b({\sqrt{-h}}h^{ab}\6_aX^\mu) \d X_\mu, \label{vang}
\ee
onde $\6 \S$ \'e o contorno da superf\'{\i}cie que representa a folha mundo. Com o uso da condi\c{c}\~ao de contorno na folha mundo $\left. \d X^\mu \right |_{\6 M}=0$, ou seja, a varia\c{c}\~ao dos campos \'e nula no contorno que delimita a superf\'{\i}cie, a primeira integral da eq.(\ref{vang}) se anula.

        Da eq.(\ref{vang}), obtemos a equa\c{c}\~ao de movimento para o campo $X^\mu$, que \'e dada por 
\be
\frac{\d S}{\d X^\mu}=T \6_b({\sqrt{-h}}h^{ab}\6_aX^\mu), \label{variacao}
\ee
onde assumimos aqui que a m\'etrica induzida $h_{ab}$ n\~ao depende do campo $X^\mu$, ou seja, $h_{ab}=g_{ab}$ e tamb\'em omitimos o r\'otulo NG da a\c{c}\~ao. Agora, se os campos n\~ao interagem, ou seja, a Lagrangeana n\~ao apresenta termos de auto-intera\c{c}\~ao do tipo $(X^{\mu} X_{\mu})^{n}$ ent\~ao, comparando a eq.(\ref{variacao}) com a equa\c{c}\~ao de Euler-Lagrange 
$$
\frac{\d S}{\d X^\mu}=\frac {\6 \cal L}{\6X^\mu}-\6_b \left (\frac{\6 \cal L}{\6(\6_bX^\mu)}\right ),
$$
obtemos
$$
\6_b \left (\frac{\6 \cal L}{\6(\6_bX^\mu)} \right )=-T \6_b({\sqrt-g}g^{ab}\6_aX^\mu).
$$
Esta \'ultima igualdade ser\'a satisfeita somente se $\frac{\6 {\cal L}}{\6(\6_b X^{\mu})} =-T({\sqrt{-g}}g^{ab}\6_aX^\mu)$, que por integra\c{c}\~ao resulta em
\be
{\cal L}=-\frac{T}{2}{\sqrt{-g}}g^{ab}\6_aX^\mu \6_b X_\mu. \label{Lagran}
\ee

        Como o tensor $g^{ab}$ \'e sim\'etrico, a densidade de Lagrangeana foi dividida por dois para evitarmos contagem dupla dos \ii ndices. A a\c{c}\~ao respectiva \'e chamada de a\c{c}\~ao de Polyakov $(S_P)$ e \'e escrita como
\be
S_P=-\frac{T}{2}\int d^{2}\s {\sqrt{-g}}g^{ab}\6_aX^\mu \6_b X_\mu.  \label{acaopol}
\ee

        Esta a\c{c}\~ao tamb\'em possui invari\^ancia por reparametriza\c{c}\~ao. As reparametriza\c{c}\~oes locais sob as quais esta a\c{c}\~ao \'e invariante s\~ao
\bea
\d g^{ab}&=& \xi^c\6_cg^{ab}-\6_c\xi^ag^{cb}-\6_c\xi^bg^{ac}, \non
\d X^\mu&=&\xi^a\6_aX^\mu, \non
\d(\sqrt{-g})&=&\6_a(\xi^a\sqrt{-g}),
\eea
onde $\xi^a$ \'e um deslocamento infinitesimal nas coordenadas $ (\t,\s)$. 

       Uma outra invari\^ancia local, pelo menos no n\'{\i}vel cl\'assico, \'e a de Weyl ou reescalonamento conforme da m\'etrica da folha mundo: $\d X^\mu=0$ e $\d g_{ab}=\L g_{ab}$, onde $\L=\L(\t,\s)$ \'e uma fun\c{c}\~ao infinitesimal arbitr\'aria de $\s^{a}$.
Existe ainda uma simetria global que reflete a simetria do espa\c{c}o-tempo no qual a corda est\'a se propagando. Para o espa\c{c}o plano, esta \'e justamente a invari\^ancia de Lorentz ou Poincar\'e descrita por $\d X^\mu= a^{\mu}_{\nu}X^\nu + b^\nu$ onde $b^\nu$ \'e um vetor constante e $a_{\mu\nu}=\h_{\mu\r}a^{\r}_{\nu}$ \'e um tensor anti-sim\'etrico ($\h_{\mu\r}$ \'e a m\'etrica de Minkowski).
Note que a exig\^encia fundamental para se obter a a\c{c}\~ao de Polyakov \'e  que $h^{ab}$  n\~ao dependa de $X^\mu$. Podemos  mostrar que as a\c{c}\~oes de Nambu-Goto e Polyakov s\~ao classicamente equivalentes mostrando que a solu\c{c}\~ao cl\'assica da equa\c{c}\~ao de movimento para $g^{ab}$ \'e a eq.(\ref{mi}), isto \'e, $\left . \frac{\d S}{\d g_{ab}} \right|_{g_{ab}=h_{ab}=0}$. Isto \'e feito considerando que a a\c{c}\~ao de Polyakov \'e um invariante de Weyl (que ser\'a discutido adiante) e com a restri\c{c}\~ao que $h_{ab}$ n\~ao dependa do campo $X^{\mu}$, ou seja, quando  $h_{ab}=g_{ab}$.

\subsection{Tensor Energia-Momento}
${}$

        O tensor energia-momento bidimensional, que \'e representado por $T_{ab}$, \'e proporcional \`a derivada variacional da a\c{c}\~ao com respeito a m\'etrica $g^{ab}$ da folha mundo. Uma vez que as a\c{c}\~oes de Nambu-Goto e de Polyakov n\~ao dependem de derivadas da m\'etrica, o tensor energia-momento \'e dado por
\be
T_{ab}=-\frac {2}{T} \frac {1}{\sqrt {-g}} \frac {\d S}{\d g^{ab}}=-\frac {2}{T} \frac {1}{\sqrt {-g}} \frac {\6 \cal {L}}{\6 g^{ab}}.
\ee
A densidade de lagrangeana $\cal{L}$ \'e dada pela eq.(\ref{Lagran}). Ent\~ao, calculado $T_{ab}$ obtemos
\bea
T_{ab}&=&-\frac {2}{T} \frac {1}{\sqrt {-g}} {\frac {\6}{\6 g^{ab}}} \le (-\frac{T}{2}{\sqrt{-g}}g^{cd}\6_cX^\mu \6_d X_\mu \ri ) \non
&=&\frac {1}{2} \left ( g^{cd} \frac{\6 g_{cd}}{\6g^{ab}} \right )g^{cd} \6_cX^\mu \6_d X_\mu  + \6_a X^\mu \6_b X_\mu.
\eea
Usando que $g_{ab}=-g^{mn}\frac{\d g_{mn}}{\d g^{ab}}$, a equa\c{c}\~ao acima torna-se
\be
T_{ab}=-\frac {1}{2}  g_{ab}g^{cd} \6_cX^\mu \6_d X_\mu  + \6_a X^\mu \6_b X_\mu.
\ee

        O tensor energia-momento possui duas propriedades importantes. 

{\bf 1 )} A primeira delas \'e que, devido a invari\^ancia de Weyl da a\c{c}\~ao, seu tra\c{c}o \'e nulo. De fato,
\be
tr T_{ab}=g^{ab}T_{ab}= g^{ab} \6_a X^\mu \6_b X_\mu -\frac {1}{2} tr g_{ab}\;g^{cd} \6_cX^\mu \6_d X_\mu. \label{ttem}
\ee
Devemos ent\~ao calcular $tr g_{ab}$ que aparece nesta equa\c{c}\~ao, para isto, usaremos o fato de que a a\c{c}\~ao da corda \'e um invariante de Weyl. Uma transforma\c{c}\~ao finita de Weyl pode ser escrita como $g_{ab}\longrightarrow g_{ab}^{'} = e^{\L(\t,\s)} g_{ab}$ e sua inversa \'e $g^{ab}\longrightarrow g^{'ab} = e^{-\L(\t,\s )} g^{ab}$. Desta forma, podemos transformar a m\'etrica gen\'erica $g_{ab}$ na m\'etrica plana de Minkowski $\eta_{ab}$, ou seja, $g_{ab}^{'} =\eta _{ab}=e^{\L(\s )} g_{ab}$, que substituindo na express\~ao para o tra\c{c}o de $g_{ab}$ obtemos $tr\; g_{ab}=g^{ab}g_{ab}=\eta ^{ab}e^{-\L(\s )}\eta _{ab}e^{\L(\s )}=tr\; \eta_{ab}=2$. Portanto, substituindo este resultado na equa\c{c}\~ao (\ref{ttem}), obtemos $tr T_{ab}=0$. \\

{\bf 2)} A outra propriedade \'e que $T_{ab}=0$ pois, como a equa\c{c}\~ao de movimento para o campo $g^{ab}$ \'e igual a zero, e $T_{ab}$ \'e proporcional a esta equa\c{c}\~ao, ent\~ao $T_{ab}=0$. 

        Estes s\~ao os dois v\'{\i}nculos cl\'assicos da teoria. No processo de quantiza\c{c}\~ao, tentaremos implementar estes v\'{\i}nculos quanticamente e veremos que surgir\~ao resultados interessantes.

\subsection{Din\^amica da Corda Cl\'assica}
${}$

        Agora, iremos obter as equa\c{c}\~oes de movimento cl\'assicas para uma corda bos\^onica aberta. Partindo de uma forma geral para a a\c{c}\~ao (Nambu-Goto ou Polyakov)
\be
S= \int d^2 \s {\cal L}= \int d\s d\t {\cal L}, \label{acao geral}
\ee
consideraremos o efeito de uma evolu\c{c}\~ao  da corda entre duas configura\c{c}\~oes fixas $X^\mu(\t_{1},\s)$ e $X^\mu(\t_{2},\s)$ onde $\t \in [\t_{1},\t_{2}]$ e $\s \in[0,\pi]$.

        Variando a a\c{c}\~ao~(\ref{acao geral}) e integrando por partes, obtemos
\be
\d S=\int d^2 \s \frac{\d {\cal L}}{\d X^\mu} \d X^\mu + \int d\s \left. \frac {\6 {\cal L}}{\6 (\6_{\t}X^{\mu})}\d  X^\mu \right |^{\t_2}_{\t_1} + \int d\t \left. \frac {\6 {\cal L}}{\6(\6_{\s}X^{\mu})} \d X^\mu \right |^{\s = \pi}_{\s=0}, \label{vag}
\ee
onde a segunda integral da eq.(\ref{vag}) \'e nula porque a configura\c{c}\~ao  do sistema \'e fixa em $\t_1$ e $\t_2$, ou seja, $\d X^\mu (\t_{1})=\d X^\mu (\t_{2})=0$.  Usando a equa\c{c}\~ao de Euler-Lagrange na primeira integral da eq.(\ref{vag}) (observando que $ \frac {\6 \cal L}{\6 X^\mu}=0$), obtemos:
\be
\d S=-\int d^2 \s \6_{a} \left ( \frac {\6 {\cal L}}{\6 (\6_{a}X^{\mu})} \right ) \d X^\mu + \int d\t \left.\frac {\6 {\cal L}}{\6 (\6_{\s} X^\mu)} \d X^\mu \right |^{\s = \pi} _{\s = 0}. \label{vag2}
\ee

        Pelo princ\'{\i}pio variacional, a primeira integral da eq.(\ref{vag2}) nos fornece a equa\c{c}\~ao de movimento
\be
\6_{a} \left( \frac {\6 {\cal L}}{\6 (\6_{a}X^\mu)} \right)=0,
\ee
enquanto que  a segunda fornece as seguintes condi\c{c}\~oes de contorno
\be
\left. \frac {\6 {\cal L}}{\6 X^{'\mu}} \right |^{\s = \pi} _{\s = 0}=0, \ \ \ \ \mbox{ou} \ \ \ \ \ \left. \d {X^\mu } \right |^{\s = \pi} _{\s = 0}=0.
\ee
Os momentos conjugados a $\dot {X}^\mu=\6_{\t}X^{\mu}$ e $X^{'\mu}=\6_{\s}X^{\mu}$ s\~ao
\be
P^{\mu}_{\t}=-\frac {\6 {\cal L}}{\6 \dot {X}_\mu} \ \ \ \ \ \mbox{e} \ \ \ \ \ P^{\mu}_{\s}=-\frac {\6 {\cal L}}{\6 {X^{'}}_\mu}
\ee
respectivamente, onde $P^{\mu}_{\s}$ \'e o fluxo de momento de uma extremidade a outra da corda (na dire\c{c}\~ao $\s$) e $P^{\mu}_{\t}$ \'e o fluxo de momento transverso a $P^{\mu}_{\s}$. Em termos destas correntes de momentos, a equa\c{c}\~ao de movimento pode ser escrita como
\be
\frac {\6P^{\mu}_{\t}}{\6 \t} + \frac {\6 P^{\mu}_{\s}}{\6 \s}=0,\label{eq. de mov}
\ee
e a primeira condi\c{c}\~ao de contorno pode ser expressa na forma
\be
P^{\mu}_{\s}(\s=0) - P^{\mu}_{\s}(\s=\pi)=0. \label{ccm}
\ee
Esta equa\c{c}\~ao mostra que o fluxo do momento da corda \'e conservado, isto quer dizer que o fluxo de momento que entra por um lado da corda \'e igual em m\'odulo ao fluxo que sai pelo outro lado. Para uma corda aberta isto fica claro, porque \'e razo\'avel dizer que o momento pode fluir para dentro e para fora da corda por ambas as extremidades. 

        O momento total da corda \'e a integral do momento ao longo de uma curva arbitr\'aria $\g$ que liga duas extremidades opostas da folha mundo, onde uma destas extremidades \'e limitada pela condi\c{c}\~ao $\s=0$ e a outra pela condi\c{c}\~ao $\s=\pi$, ent\~ao:
\be
P^{\mu}=\int d\s P^{\mu}_{\t} + d\t P^{\mu}_{\s}=\int^{\pi}_{0} d\s P^{\mu}_{\t}.
\ee

        Derivando o momento em rela\c{c}\~ao a $\t$ e usando a equa\c{c}\~ao de movimento, eq.(\ref{eq. de mov}), resulta
\be
\frac {dP^{\mu}}{d\t}=-\int^{\pi}_{0} d\s \frac { \6 P^{\mu}_{\s}}{\6 \s}=P^{\mu}_{\s}(\s=0) -P^{\mu}_{\s}(\s=\pi) =0
\ee
ou seja, o momento total \'e conservado.
\subsection{Equa\c{c}\~ao de Movimento e suas Solu\c{c}\~oes}
${}$

        Para uma an\'alise subseq\"uente da din\^amica e quantiza\c{c}\~ao da corda, faremos uma escolha de um calibre conveniente. Esta escolha servir\'a para diminuirmos os graus de liberdade das vari\'aveis din\^amicas que aparecem explicitamente na a\c{c}\~ao. Em duas dimens\~oes, uma transforma\c{c}\~ao geral de coordenadas ($\s,\t \longrightarrow \s^{'},\t^{'}$) depende de duas fun\c{c}\~oes livres, a saber, $\s^{'}$ e $\t^{'}$. Por esta transforma\c{c}\~ao, quaisquer duas das tr\^es componentes de $h$ podem ser eliminadas. Uma forma conveniente de fazermos isto \'e escolhermos uma parametriza\c{c}\~ao da folha mundo, tal que $g_{ab}=e^\L \h_{ab}$, onde $\h_{ab}$ \'e a m\'etrica plana da folha mundo ($\h _{ab}=diag \; (-1,1)$), ou seja, m\'etrica bi-dimensional de Minkovski, e $e^\L$ \'e conhecido como fator conforme. \'E sempre poss\'{\i}vel fazer esta escolha, pelo menos localmente. Esta escolha \'e chamada de calibre conforme.  Substituindo este calibre conforme na a\c{c}\~ao de Polyakov~(\ref{acaopol}) , esta se reduz a
\be
S=-\frac{T}{2}\int d^{2}\s \h^{ab}\6_aX^\mu \6_b X_\mu,  \label{acao gauge}
\ee
e o tensor  energia-momento bidimencional
\be
T_{ab}=\6_aX^\mu \6_b X_\mu - \h_{ab}\6_cX^\mu \6^c X_\mu.  
\ee

        Como $T_{ab}=0$, os graus de liberdade do campo $X^\mu$ est\~ao sujeitos aos seguintes v\ii nculos:
\bea
T_{00}&=&T_{11}= \frac{1}{2} (\dot X^2 + X^{'2})=0, \non 
T_{10}&=&T_{01}=\dot X \cdot X^{'}=0. \label{vinc}
\eea

        Se fizermos a varia\c{c}\~ao da a\c{c}\~ao~(\ref{acao gauge}), esta ir\'a conter um termo de volume e um de superf\'{\i}cie. Para assegurarmos que a a\c{c}\~ao seja estacion\'aria, pelo princ\'{\i}pio da m\'{\i}nima a\c{c}\~ao, ambos os termos devem ser iguais a zero e isto nos conduz \`a equa\c{c}\~ao de movimento 
\be 
{\Box} X^{\mu}(\t,\s) =\6^a \6_a  X^{\mu}=\left (\frac{\6^{2}}{\6 \s^{2}}-\frac {\6^{2}}{\6 \t^{2}} \right) X^{\mu}(\t,\s) =0 \label{em}
\ee
e a condi\c{c}\~ao de contorno peri\'odica 
\be
X^{\mu}(\t,0)=X^{\mu}(\t,\pi) \label{ccp}
\ee
 para a corda fechada, ou \`a condi\c{c}\~ao de contorno para a corda aberta,
\be
\left. \6_{\s} X^{\mu} \d X_{\mu} \right|_{0}^{\pi}=0 
\ee
onde esta \'ultima  pode ser satisfeita de dois modos
\be 
\cases{\left. \6_\s X^\mu \right |_{0}^{\pi}=0 \rightarrow \mbox{condi\c{c}\~oes de contorno de Neumann (N)}, \cr
\left. \d X^\mu \right |_{\6M}=0 \rightarrow \mbox{condi\c{c}\~oes de contorno de Dirichlet (D).} \cr } \label{cc}
\ee
A eq.(\ref{em}) \'e uma equa\c{c}\~ao de onda sem massa em duas dimens\~oes, cuja solu\c{c}\~ao pode ser escrita de forma gen\'erica como uma combina\c{c}\~ao dos movimentos para direita $X^{\mu}_{R}$ e para esquerda $X^{\mu}_ {L}$, $X^{\mu}(\s^{a})=X^{\mu}_{R}(\s^{-})+X^{\mu}_{L}(\s^{+})$, onde $\s^{-}=\t - \s$ e $\s^{+}=\t + \s$.  Desta forma, \'e conveniente mudar as coordenadas da folha mundo para coordenadas de cone de luz $(\s^{+},\s^{-})$. Nestas novas coordenadas, o elemento de comprimento sobre a folha mundo torna-se $ds^{2}=-d\s^{+}ds^{-}$, o que nos leva a dizer que as novas componentes da m\'etrica plana de Minkowski s\~ao $\eta_{-+}=\eta_{+-}=-\frac{1}{2}$ e sua inversa $\eta^{-+}=\eta^{+-}=2$ e $\eta_{++}=\eta_{--}=\eta^{++}=\eta^{--}=0$. As derivadas conjugadas a $\s^{\pm}$ s\~ao $\6_{\t}= \6_{+}+\6_{-}$ e $\6_{\s}= \6_{+}-\6_{-}$. As condi\c{c}\~oes de contorno (\ref{cc}) implicam que os movimentos para a esquerda e para direita da corda n\~ao s\~ao independentes. As solu\c{c}\~oes da equa\c{c}\~ao de onda livre podem estar sujeitas tanto \`as condi\c{c}\~oes de contorno Neumann quanto \`as de Dirichelet, embora esta \'ultima quebre a invari\^ancia de Poincar\'e. Desta forma, a \'unica solu\c{c}\~ao que \'e invariante de Poincar\'e \'e a solu\c{c}\~ao na qual ambas as extremidades da corda est\~ao sujeitas \`as condi\c{c}\~oes de contorno Neumann, e \'e dada por
\be
X^\mu(\t,\s)=x^{\mu} +2\al^{'}p^\mu \t +i\sqrt{2\al^{'}}\sum _{n\neq 0}\frac{1}{n}\al^{\mu}_{n}e^{-in\t}\cos{n\s}, \label{sol}
\ee
e a solu\c{c}\~ao geral compat\'{\i}vel \`a condi\c{c}\~ao peri\'odica (\ref{ccp}) \'e 
\be
X^\mu(\t,\s)=x^{\mu} +2\al^{'}p^\mu \t+i\sqrt{2\al^{'}}\sum _{n\neq 0}\frac{1}{2n}\le( \al^{\mu}_{n}e^{-2in(\t - \s)}+{\b}^{\mu}_{n}e^{-2in(\t + \s)}\ri ) \label{scf},
\ee
sendo as componentes do movimento para a direita e para a esquerda dadas respectivamente por
\be
X^{\mu}_{R}(\t,\s)=\frac{1}{2}x^{\mu} +\frac{1}{2}2\al^{'} p^\mu (\t- \s) +\frac{i}{2}\sqrt{2\al^{'}}\sum _{n\neq 0}\frac{1}{n}\al^{\mu}_{n}e^{-2in(\t - \s)},
\ee
\be
X^{\mu}_{L}(\t,\s)=\frac{1}{2}x^{\mu} +\frac{1}{2}l^2 p^{\mu} (\t + \s) +\frac{i}{2}l\sum _{n\neq 0}\frac{1}{n} {\b}^{\mu}_{n}e^{-2in(\t + \s)},
\ee
onde $n$ \'e um inteiro, $\al^{'}$ \'e o par\^ametro de Regge que est\'a relacionado com a tens\~ao da corda atrav\'es da rela\c{c}\~ao $T=(2\pi \al^{'})^{-1}$. $\al^{\mu}_{n} \  \mbox{e} \ \b^{\mu}_{n} $ s\~ao os coeficientes de Fourier, que ser\~ao interpretados como coeficientes do oscilador. Deste modo, vemos que a solu\c{c}\~ao para a corda fechada envolve uma superposi\c{c}\~ao linear dos movimentos da corda para a esquerda e para direita, com coeficientes de Fourier  $\al^{\mu}_{n} \  \mbox{e} \ \b^{\mu}_{n} $, respectivamente. \'E importante observarmos que no caso da corda fechada estes coeficientes d\~ao origem aos modos de vibra\c{c}\~ao para a direita e esquerda, respectivamente. As constantes $x^\mu$ e $p^\mu$ s\~ao interpretadas como sendo, respectivamente, a coordenada e o momento do centro de massa da corda. O fato de $X^\mu$ ser real exige que $x^\mu$ e $p^\mu$ tamb\'em o sejam. O mesmo fato nos conduz \`a condi\c{c}\~ao de $\al^{\mu}_{-n}$ $(\b^{\mu}_{-n})$ ser o adjunto de $\al^{\mu\dag }_{n}$ $(\b^{\mu\dag }_{n})$, isto \'e, $\al^{\mu\dag }_{n}=\al^{\mu}_{-n}$ $(\b^{\mu\dag }_{n}=\b^{\mu}_{-n})$. Esta condi\c{c}\~ao ser\'a importante para determinarmos os colchetes de Poisson para os coeficientes $\al^{\mu}_{n}$ $(\b^{\mu}_{n})$.

      Quando fixamos $\t$, os colchetes de Poisson para as vari\'aveis din\^amicas do sistema  s\~ao: 
\be
\left \{X^{\mu} (\t,\s ),X^{\nu}(\t,\s^{'}) \right \}=0, \ \ \ \left \{\dot X^{\mu} (\t,\s ),\dot X^{\nu}(\t,\s^{'}) \right \}=0,
\ee
\be
\left \{P^{\mu} (\t, \s ),X^{\nu}(\t,\s^{'}) \right \}=T\left \{\dot {X}^{\mu} (\t, \s ),X^{\nu}(\t,\s^{'}) \right \} = \d (\s-\s^{'}) \h_{\mu\nu}. \label{cp}
\ee
\'E essencial escrevermos os colchetes de Poisson para os coeficientes do oscilador $\al^{\mu}_{n}$ e $\b^{\mu}_{n}$, pois, quando quantizarmos a teoria, estes coeficientes ser\~ao os operadores de cria\c{c}\~ao e destrui\c{c}\~ao dos modos normais de vibra\c{c}\~ao da corda. Para isto, inserimos a solu\c{c}\~ao (\ref{scf}) nos colchetes de Poisson~(\ref{cp}) e obtemos
\be
\left \{\al^{\mu}_{m},\al^{\mu}_{n} \right \}=  \left \{\b^{\mu}_{m},\b^{\mu}_{n} \right \}=im\d_{m+n} \h^{\mu\nu}, \label{rco}
\ee 
e tamb\'em
\be
\left \{\al^{\mu}_{m},{\b}^{\mu}_{n} \right \}=0.
\ee
O colchete de Poisson correspondente \`as vari\'aveis din\^amicas do centro de massa fica sendo
\be
\{p^\mu,x^\nu\}=\h^{\mu\nu}.
\ee

  A Hamiltoniana \'e contru\'{\i}da por integra\c{c}\~ao sobre todo o comprimento da corda da densidade de Hamiltoniana dada por ${\cal H}=P^\mu_{\t} (\t,\s) \dot {X}_{\mu}-{\cal L}$. Neste calibre conforme, $g_{ab}=\h^{ab}$, a Hamiltoniana pode ser escrita em termos dos coeficientes do oscilador. Para a corda aberta, obtemos   
\be
H=\int d\s {\cal H}=\frac{T}{2}\int^{\pi}_{0} d^2\s (\dot {X}^2 +{X}^{'2})=\frac {1}{2} \sum _{-\infty}^{\infty} \al_{-n}^{\mu}\al_{\mu n}=\sum _{n \neq 0} \al_{-n}^{\mu}\al_{\mu n} + \frac{1}{2}\sqrt{2\al^{'}}p^{\mu}p_{\mu}, \label{hca}
\ee
onde definimos o coeficiente $\al_{0}^{\mu}=\sqrt{2\al^{'}}p^{\mu}$. Para a corda fechada, temos
\be
H=\frac {1}{2}\sum _{-\infty}^{\infty} (\al_{-n}^{\mu}\al_{\mu  n} + {\b}^{\mu}_{-n} {\b}_{\mu n})=\sum _{n \neq 0} (\al_{-n}^{\mu}\al_{\mu n} + {\b}_{-n}^{\mu} {\b}_{\mu n})+\frac{1}{2}\sqrt{2\al^{'}}p^{\mu}p_{\mu}, \label{hcf}
\ee
onde para a corda fechada usamos  $\al_{0}^{\mu}={\b}_{0}^{\mu}=\frac{1}{2}lp^{\mu}$.   
     
        Podemos ainda escrever para o caso da corda aberta solu\c{c}\~oes da equa\c{c}\~ao de onda~(\ref{em}) sujeitas \`as outras combina\c{c}\~oes das condi\c{c}\~oes de contorno em suas extremidades que n\~ao s\~ao invariantes de Poincar\'e. Por exemplo, para uma corda aberta sujeita \`as condi\c{c}\~oes de contorno de Dirichlet nas duas extremidades, temos como solu\c{c}\~ao
\be
X^{\mu}(\t,\s)=\frac{c^{\mu}(\pi - \s)+d^{\mu}\s}{\pi} - \sqrt{2 \al^{'}} \sum_{n \neq 0} \le( \frac{\al^{\mu}_{n}}{n}e^{-in\t}\sin{n\s} \ri),
\ee
e para condi\c{c}\~oes de contorno mistas, temos
\be
X^{\mu}(\t,\s)=c^{\mu}- \sqrt{2 \al^{'}} \sum_{r \in {\it Z^{'}}} \le( \frac{\al^{\mu}_{n}}{n}e^{-in\t}\sin{n\s} \ri), \ \ (\mbox{caso Neumann-Dirichlet}),
\ee
e 
\be
X^{\mu}(\t,\s)=d^{\mu}+ i\sqrt{2 \al^{'}} \sum_{r \in {\it Z^{'}}} \le( \frac{\al^{\mu}_{n}}{n}e^{-in\t}\cos{n\s} \ri),\ \ (\mbox{caso Dirichlet-Neumann}).
\ee 
Em ambos os casos, ${\it Z^{'}}={\it Z}+\frac{1}{2}$, $c^{\mu}$ e  $d^{\mu}$ s\~ao vetores constantes das duas extremidades da corda imersa no espa\c{c}o-tempo. Em todo o texto, a menos que seja mencionado, estaremos trabalhando somente com a solu\c{c}\~ao $X^{\mu}(\t,\s)$ invariante por Poincar\'e.

\subsection{Algebra de Virasoro e Estados de Massa para a Corda}
${}$
      
        Os graus de liberdade no calibre covariante devem satisfazer n\~ao somente a equa\c{c}\~ao de onda (\ref{em}) que nos conduz aos modos de expans\~ao, como tamb\'em os v\'{\i}nculos que s\~ao originados da equa\c{c}\~ao de movimento para $g_{ab}$ e s\~ao conseq\"u\^encia da invari\^ancia por reparametriza\c{c}\~ao da corda. Para analisarmos estes v\ii nculos, \'e conveniente expressarmos as componentes do tensor energia-momento nas coordenadas de cone de luz, tal que  as combina\c{c}\~oes $T_{++}$ e $T_{--}$ do tensor  energia-momento bidimencional $T_{ab}$ sejam dadas por
\bea
T_{++} &\equiv& \frac {1}{2} \left (T_{00}+T_{01} \right)=\6_{+}X \cdot \6_{+}X,\non 
T_{--} &\equiv& \frac {1}{2} \left (T_{00}-T_{01} \right)=\6_{-}X \cdot \6_{-}X . \label{T}
\eea
Escrevendo $T_{++}$ e $T_{--}$ desta forma, as equa\c{c}\~oes de v\'{\i}nculo~(\ref{vinc}) podem ser escritas como  $T_{++}$=$T_{--}=0$. Estas combina\c{c}\~oes das componentes do tensor energia-momento s\~ao v\'alidas tanto para a corda aberta quanto para a fechada. Vamos agora considerar os modos de expans\~ao do v\'{\i}nculo $T_{ab}=0$.

         Para a corda fechada, podemos verificar que $\dot{X}^{\mu}_{L}={X}^{' \mu}_{L} \  \mbox{e} \ \dot{X}^{\mu}_{R}=-{X}^{' \mu}_{R}$. Ent\~ao, $T_{++}$ pode ser escrito somente em termos de $X^{\mu}_{L}$ e a componente $T_{--}$ em termos de ${X}^{\mu}_{R}$, ou seja, $T_{++}=\dot{X}^{\mu}_{L}\dot{X}_{\mu L} \  \mbox{e} \  T_{--}=\dot{X}^{\mu}_{R}\dot{X}_{\mu R}$. Classicamente, estamos livres para implementarmos estes v\'{\i}nculos, mas como veremos, quanticamente qualquer express\~ao contendo operadores que n\~ao comutam n\~ao \'e bem definida sem antes especificarmos uma prescri\c{c}\~ao normal para os mesmos. Sendo assim, \'e \'util neste caso trabalharmos com as componentes de Fourier de $T_{--}$  e $T_{++}$  para os movimentos para direita e para a esquerda da corda fechada definidas em $\t = 0$, respectivamente como
\bea
L_m &=& \frac{T}{2} \int ^{\pi}_{0} d\s \left (e^{-2im\s} T_{--} \right) \non
&=&\frac{T}{2} \int ^{\pi}_{0} d\s \left (e^{-2im\s} \dot{X}^{2}_{ R} \right), \ \ \ \ \ \ \ \ \  m \neq 0, 
\eea
e
\bea
{\overline {L}}_m &=& \frac{T}{2} \int ^{\pi}_{0} d\s \left (e^{2im\s} T_{++} \right) \non
&=&\frac{T}{2} \int ^{\pi}_{0} d\s \left (e^{2im\s} \dot{X}^{2}_{L} \right), \ \ \ \ \ \ \ \ \  m \neq 0 .
\eea
Em termos dos modos de Fourier, estas componentes podem ser escritas, respectivamente, como
\be
L_{m}=\frac {1}{2} \sum^{\infty}_{n=-\infty} \al_{m-n}^{\mu} \al_{\mu n}, \ \ \ \ \ \ \ \ \ \ \ \ \ \ \ \ \ \ \ m \neq 0,
\ee
\be
{\overline {L}}_{m}=\frac {1}{2} \sum^{\infty}_{n=-\infty} {\b}_{m-n}^{\mu} {\b}_{\mu n}, \ \ \ \ \ \ \ \ \ \ \ \ \ \ \ \ \ \ m \neq 0,
\ee
onde  $\al_{0}=\b_{0}=\frac{1}{2}\sqrt{2\al^{'}}p^{\mu}$. $L_m$ e ${\overline {L}}_m$ s\~ao conhecidos como operadores de Virasoro para a corda fechada. Estes dois operadores formar\~ao duas \'algebras de Virasoro independentes, uma para o movimento para a esquerda e uma para a direita. Como j\'a citamos anteriormente, as condi\c{c}\~oes de contorno para uma corda aberta significam que a separa\c{c}\~ao da solu\c{c}\~ao da equa\c{c}\~ao de onda em movimento para a esquerda e para direita n\~ao \'e poss\'{\i}vel porque os mesmos n\~ao s\~ao independentes. Portanto, a expans\~ao da corda aberta envolve somente um conjunto de osciladores $\al^{\mu}_{n}$ e, conseq\"uentemente, iremos definir uma \'unica \'algebra de Virasoro em vez de \'algebras independentes para os movimentos para a direita e para a esquerda como no caso da corda fechada. Podemos ent\~ao definir os geradores da \'algebra de Virassoro $(L_m)$, tomando a combina\c{c}\~ao das componentes de Fourier de  $T_{++}$ e $T_{--}$ em $\t=0$ da seguinte maneira
\bea
L_m &=& T\int ^{\pi}_{0} d\s \left (e^{im\s} T_{++} +e^{-im\s} T_{--} \ri ) \non
&=&\frac{T}{2} \int ^{\pi}_{0} d\s \le \{e^{im\s}  \le ( \dot {X}+{X}^{'} \ri)^2+e^{-im\s}  \le ( \dot {X}+{X}^{'} \ri)^2\ri \} \non 
&=& \frac {1}{2} \sum ^{\infty}_{-\infty} \al_{m-n} \cdot \al_{n}, \label{ov}
\eea
onde definimos $\al^{\mu}_{0}=\sqrt{2\al^{'}}p^{\mu}$. Notamos que, em particular, a Hamiltoniana para as cordas aberta e fechada podem ser escritas, respectivamente, como $H=L_0$ e $H=L_0 + {\overline {L}}_{0}$.

        O quadrado da massa $(M)$ de  uma corda em um dado estado de oscila\c{c}\~ao \'e dado por $M^2= -p_\mu p^\mu$. A equa\c{c}\~ao de v\ii nculo $L_0={\overline {L}}_{0}=0$ nos fornece uma importante equa\c{c}\~ao de movimento que determina $M^2$ em termos dos modos internos de oscila\c{c}\~ao da corda, isto \'e, 
\be
M^2=\frac {1}{\al^{'}} \sum^{\infty}_{n=1} \al_{-n} \cdot \al_{n}
\ee
para a corda aberta, e
\be
M^2=\frac {2}{\al^{'}} \sum^{\infty}_{n=1} (\al_{-n} \cdot \al_{n}+\b_{-n} \cdot \b_{n}) \label{cmcf}
\ee 
para a corda fechada. Estas equa\c{c}\~oes s\~ao conhecidas como condi\c{c}\~ao de concha de massa para as cordas aberta e fechada, respectivamente. Quanticamente, estas condi\c{c}\~oes ser\~ao modificadas devido ao efeito do ordenamento normal. Devido ao fato de que $L_0={\overline {L}}_{0}$ para a corda fechada, temos que os dois termos nas equa\c{c}\~oes ({\ref{cmcf}}) ou ({\ref{hcf}}) d\~ao contribui\c{c}\~oes iguais. 

        Os modos de Fourier do tensor energia-momento $L_m \ \mbox{e} \  {\overline {L}}_{m}$ s\~ao  chamados de operadores de Virasoro. Os par\^enteses de Poisson dos operadores de Virassoro podem ser calculados a partir  da defini\c{c}\~ao de $L_m$ e dos colchetes de Poisson j\'a conhecidos dos osciladores, eq.(\ref{rco}). Da defini\c{c}\~ao de $L_n$, equa\c{c}\~ao (\ref{ov}), temos
\be
\left \{L_m,L_n \right \}=\frac{1}{4} \sum _{k,l} \left \{\al_{m-k}.\al_{k},\al_{n-l}.\al_{l} \right \}.
\ee
A identidade $ \left \{AB,CD \right \}=A  \left \{B,C \right \} D + AC\left \{B,C \right \} +  \left \{A,C \right \}DB + C \left \{A,D \right \}B$, mais as rela\c{c}\~oes de comuta\c{c}\~ao dos osciladores (\ref{rco}), nos levam a
\bea
\left \{L_{m},L_{n} \right \}=\frac{i}{4} \sum _{k,l} \left ( k\al_{m-k} \cdot \al_{l} \d_{k+n-l} +   k\al_{m-k} \cdot \al_{n-l} \d_{k+l}+ (m-k)\al_{l} \cdot \al_{k} \d_{m-k+n-l} \ri . \non
\left. + (m-k)\al_{n-l} \cdot \al_{k} \d_{m-k+l} \right), \nonumber
\eea
onde $\d_n$ \'e 1 se $n=0$, e $0$ para outros valores. Efetuando-se a soma em $l$ a equa\c{c}\~ao acima se reduz a
\be
\left \{L_m,L_n \right \}=\frac{i}{2} \sum _{k}  k\al_{m-k} \cdot \al_{k+n}+\frac{i}{2} \sum _{k}(m-k)\al_{m-k+n}\cdot \al_{k}. \label{pro}
\ee
Fazendo uma mudan\c{c}a de vari\'aveis de $k \rightarrow k^{'}=k+n$ na primeira soma, a eq.(\ref{pro}) se reduz \`a \'algebra de Virassoro 
\be
\left \{L_m,L_n \right \}= i(m-n)L_{m+n}. \label{acv}
\ee
Analogamente, partindo da defini\c{c}\~ao de ${\overline {L}}_{m}$, obtemos
\be
\left \{{\overline {L}}_m,{\overline {L}}_n \right \}= i(m-n){\overline {L}}_{m+n}. \label{acvf}
\ee        
Uma vez que os coeficientes $\al^{\mu}_{m}$ e $\b^{\nu}_{n}$ comutam, \'e natural escrevermos que $\left \{L_m,{\overline L}_n \right \}=0$.

        Ap\'os o processo de quantiza\c{c}\~ao a forma do operadores de Virasoro ser\'a ligeiramente modificada e veremos que um termo extra surgir\'a na \'algebra de Virasoro. O termo aparece devido \`as rela\c{c}\~oes de comuta\c{c}\~ao dos coeficientes de Fourier $\al^{\mu}_{n}$ e $\b^{\nu}_{n}$.

\subsection{Quantiza\c{c}\~ao Can\^onica}
${}$

        Quando quantizamos a corda livre, o conte\'udo de part\'{\i}culas aparece, pois o espa\c{c}o de Fock da teoria de cordas cont\'em v\'arios estados de part\'{\i}culas. Como veremos, anomalias que surgir\~ao na teoria devido ao ordenamento normal dos ope- \\radores de cria\c{c}\~ao e destrui\c{c}\~ao desaparecer\~ao se impormos restri\c{c}\~oes na dimens\~ao do espa\c{c}o-tempo, no qual a folha mundo est\'a imersa. Existem muitos procedimentos diferentes para se quantizar um sistema cl\'assico~{\cite{gsw,bl,bh,jp}}. Quando usados corretamente, todos s\~ao equivalentes, embora estas equival\^encias n\~ao sejam triviais. 

        Nesta se\c{c}\~ao, apresentaremos a quantiza\c{c}\~ao can\^onica, que \'e baseada em termos do campo $X^{\mu}$ com restri\c{c}\~oes f\'{\i}sicas somente no espa\c{c}o de Fock. Tais restri\c{c}\~oes prov\^em de v\'{\i}nculos sobre o tensor energia-momento e d\~ao origem a graus de liberdade n\~ao f\'{\i}sicos. A quantiza\c{c}\~ao destes graus de liberdade n\~ao f\'{\i}sicos s\~ao respons\'aveis pelo  surgimento de estados com norma negativa (fantasmas) na teoria. A situa\c{c}\~ao \'e an\'aloga \`as condi\c{c}\~oes de Gupta-Bleuler na eletrodin\^amica, onde o v\'{\i}nculo cl\'assico $\6_{\mu}A^{\mu}=0$ \'e trocado pela exig\^encia de que componentes de freq\"u\^encia positiva dos correspondentes operadores qu\^anticos aniquilam estados f\'{\i}sicos do f\'oton. 

       Para fazermos a quantiza\c{c}\~ao can\^onica da teoria, usaremos o calibre conforme onde fixamos $h_{ab}=\h_{ab}$ (como mencionado na se\c{c}.~2.2.3). A escolha deve ser cuidadosa, pois uma vez fixado o calibre o tra\c{c}o do tensor energia-momento \'e nulo. 

        No processo de quantiza\c{c}\~ao can\^onica, ou primeira quantiza\c{c}\~ao, da corda bos\^onica cl\'assica devemos considerar os campos $X^{\mu}$ como operadores quanto-mec\^anicos (isto \'e equivalente a fazermos a transi\c{c}\~ao da mec\^anica cl\'assica para mec\^anica qu\^antica em primeira quantiza\c{c}\~ao via rela\c{c}\~oes can\^onicas de comuta\c{c}\~ao para as coordenadas e seus momentos canonicamente conjugados). Para isto, trocaremos os par\^enteses de Poisson por comutadores de acordo com a rela\c{c}\~ao $\{ \ \ , \ \ \} \longrightarrow \frac{1}{i}[ \ \  ,\ \  ]$, ou seja, usaremos o Princ\'{\i}pio da Correspond\^encia. Deste modo,  obtemos as rela\c{c}\~oes de comuta\c{c}\~ao a tempos iguais  
\bea
\left [X^{\mu} (\t,\s ),X^{\nu}(\t,\s^{'}) \right ]=\left [\dot X^{\mu} (\t,\s ),\dot X^{\nu}(\t,\s^{'}) \right ]=0, \non
\left [P^{\mu}_{\t} (\t, \s ),X^{\nu}(\t,\s^{'}) \right ]=T\left [\dot {X}^{\mu} (\t, \s ),X^{\nu}(\t,\s^{'}) \right ] =-i \d (\s-\s^{'}) \h^{\mu\nu}. \label{rc}
\eea

        As componentes espaciais do campo $X^{\mu} (\t,\s )$ comportam-se como campos escalares independentes sujeitos \`as rela\c{c}\~oes de comuta\c{c}\~ao usuais para os campos de Klein-Gordon. No entanto, a componente temporal $X^{0} (\t,\s )$ satisfaz estas rela\c{c}\~oes de comuta\c{c}\~ao com o sinal trocado (negativo). Isto \'e inevit\'avel se escrevermos as rela\c{c}\~oes de comuta\c{c}\~ao a tempos iguais na forma covariante, uma vez que o tensor m\'etrico que aparece do lado direito possui ambos os sinais. Este sinal negativo implicar\'a em uma mudan\c{c}a na estrutura do espa\c{c}o de Hilbert do campo $X^{\mu} (\t,\s )$, pois aparecem estados de norma negativa.

       Podemos observar que o v\'{\i}nculo cl\'assico $T_{ab}=0$, os quais em termos dos campos $X^{\mu} (\t,\s )$ \'e representado pela eq.(\ref{T}), n\~ao pode ser implementado quanticamente como uma equa\c{c}\~ao operatorial, pois isto estar\'a em conflito com as rela\c{c}\~oes de comuta\c{c}\~ao~(\ref{rc}). Isto significa que devemos aplicar estes v\'{\i}nculos aos estados f\'{\i}sicos. No entanto, n\~ao podemos definir um estado f\'{\i}sico $|\psi \ra$ aceitav\'el via $T_{ab}|\psi \ra=0$, pois isto tamb\'em est\'a em conflito com as rela\c{c}\~oes de comuta\c{c}\~ao. Para mostrarmos isto, realizamos o c\'alculo de $[T_{ab},X^{\mu}]$ e obtivemos os seguintes resultados: $[T_{00},X^{\mu}]=[T_{11},X^{\mu}]=-iT^{-1}\dot{X}^{\mu}$ e $[T_{01},X^{\mu}]=[T_{10},X^{\mu}]=-iT^{-1}{X}^{'\mu}$. Isto implica que $T_{ab}$ \'e um operador que certamente n\~ao \'e zero. Desta forma, temos uma incompatibilidade entre a quantiza\c{c}\~ao can\^onica e a fixa\c{c}\~ao do calibre $h_{ab}=\eta_{ab}$. 
Somente ser\~ao admitidos vetores de estado para os quais o valor esperado de $T_{ab}$ seja nulo
\be
\la \psi |T_{ab}|\psi \ra=0. \label{vetem}
\ee
Podemos ainda separar o operador tensor energia-momento em duas partes, uma de freq\"u\^encia positiva $T_{ab}^{(+)}$ e outra de freq\"u\^encia negativa $T_{ab}^{(-)}$ que consistem de operadores de aniquila\c{c}\~ao e de cria\c{c}\~ao, respectivamente $\le ( T_{ab}=T_{ab}^{(+)}+T_{ab}^{(-)} \ri)$, onde $ \le(T_{ab}^{(+) \a}=T_{ab}^{(-)}\ri)$. Desta forma, a eq.(\ref{vetem}) pode ser escrita como uma equa\c{c}\~ao de auto-valor, cuja parte de freq\"u\^encias positivas aniquila os estados f\'{\i}sicos $T_{ab}^{(+)}|\psi \ra=0$. A correspondente condi\c{c}\~ao adjunta \'e $\la \psi |T_{ab}^{(-)}=0$. Estas duas condi\c{c}\~oes s\~ao suficientes para garantirmos (\ref{vetem}), pois
\be
\la \psi |T_{ab}|\psi \ra=\la \psi |T_{ab}^{(+)}|\psi \ra+\la \psi |T_{ab}^{(-)}|\psi \ra= 0.
\ee 
Como veremos adiante, sob certas restri\c{c}\~oes, os estados obtidos desta forma ser\~ao livres de fantasmas. 

        Uma vez feita a quantiza\c{c}\~ao can\^onica para os campos $X^{\mu}$, devemos tamb\'em escrever as rela\c{c}\~oes de comuta\c{c}\~ao a tempos iguais para as vari\'aveis canonicamente conjugadas do centro de massa, $x^{\mu}$ e $p^{\mu}$, e para os coeficientes de Fourier $\al^{\mu}_{n}$ e $\b^{\mu}_{n}$, que s\~ao, respectivamente,
\be
[x^\mu,p^\nu]=i\h^{\mu\nu}, 
\ee
$$
 \left [\al^{\mu}_{m},\al^{\nu}_{n} \right ]=\left [\b^{\mu}_{m},\b^{\nu}_{n} \right ]=m\d_{m+n} \h^{\mu \nu}, 
$$
\be
\left [\al^{\mu}_{m},\b^{\nu}_{n} \right ]=0.
\ee
Os coeficientes $\al^{\mu}_{n}$ e $\b^{\mu}_{n}$ s\~ao interpretados como operadores de cria\c{c}\~ao para $n$ negativo e operadores de destrui\c{c}\~ao para $n$ positivo. Todas as rela\c{c}\~oes escritas com $\al^{\mu}_{n}$ s\~ao v\'alidas tanto para a corda aberta quanto para corda fechada, e as escritas com $\b^{\mu}_{n}$ s\~ao v\'alidas somente para a corda fechada, quando isto n\~ao for verdadeiro, deixaremos claro no texto. A imposi\c{c}\~ao de que $X^\mu$ seja Hermitiano exige que $x^\mu$ e $p^\mu$ tamb\'em o sejam. O mesmo fato nos conduz  a impor a condi\c{c}\~ao de Hermiticidade conjugada dos operadores para todo e qualquer valor inteiro de $n$, ou seja, 
\be
(\al^{\mu}_{n})^{\dag}=\al^{\mu}_{-n} \ \ \ \mbox{e} \ \ \ (\b^{\mu}_{n})^{\dag}=\b^{\mu}_{-n}.
\ee 
 
        Os operadores $\al^{\mu}_{n}$ e $\b^{\mu}_{n}$ est\~ao relacionados com os operadores de cria\c{c}\~ao e aniquila\c{c}\~ao (convenientemente normalizados) do oscilador harm\^onico da forma 
\be
a^{\mu}_{n}=\frac{1}{\sqrt{n}}\al^{\mu}_{n} \ \ \ \mbox{e} \ \ \ b^{\mu}_{n}=\frac{1}{\sqrt{n}}\al^{\mu}_{n},
\ee
e seus Hermitianos conjugados 
\be
a^{\mu \a}_{n}=\frac{1}{\sqrt{n}}\al^{\mu}_{-n} \ \ \ \mbox{e} \ \ \ b^{\mu \a}_{n}=\frac{1}{\sqrt{n}}\b^{\mu}_{-n}. 
\ee
Estas rela\c{c}\~es s\~ao v\'alidas somente para $n>0$. Os $a^{\mu}_{n}$'s e os $b^{\mu}_{n}$'s obedecem as rela\c{c}\~oes de comuta\c{c}\~ao padr\~ao do oscilador harm\^onico somente para $\mu \neq 0$ e $\nu \neq 0$, devido \`a presen\c{c}a do tensor m\'etrico  $\eta^{\mu \nu}$, cuja componente temporal $(\eta^{00})$ \'e negativa, ou seja, $[a^{\mu}_{m},a^{\mu}_{n}]=\d_{m+n} \h^{\mu\nu}$.

          O estado fundamental do oscilador \'e definido como sendo aquele que \'e aniquilado por $\al^{\mu}_{n}$ com $n>0$, $\al^{\mu}_{n}|0\ra _{\al}=0$, para a corda aberta, e aniquilado por $\al^{\mu}_{n}$ e $\b^{\mu}_{n}$ para a corda fechada $\al^{\mu}_{n}|0\ra _{\al}|0 \ra _{\b}=\b^{\mu}_{n}|0\ra _{\al}|0 \ra _{\b}=0$. Especificando que os osciladores est\~ao no estado fundamental, n\~ao determinamos completamente o estado da corda. Um outro grau de liberdade \'e seu momento do centro de massa $p^{\mu}$. Quando desejamos ent\~ao especificar completamente o estado aniquilado por $\al^{\mu}_{n}$ $(\b^{\mu}_{n})$, para $n > 0$, e com momento do centro de massa $p^{\mu}$ escrevemos $|0,p^{\mu}\ra_{\al} $ $(|0,p^{\mu}\ra_{\b})$.

        Um ponto de fundamental import\^ancia nesta teoria \'e que o espa\c{c}o de Fock cons-\\tru\'{\i}do por aplica\c{c}\~oes sucessivas do operador de cria\c{c}\~ao $a_{m}^{\mu \a}$ no estado fundamental $|0,p^{\mu}\ra_{\al} $ n\~ao \'e positivo definido, pois as componentes temporais possuem um sinal negativo devido a m\'etrica do espa\c{c}o-tempo ser de Minkowski. As rela\c{c}\~oes de comuta\c{c}\~ao destas componentes temporais s\~ao $\left [a^{0}_{m},a^{0 \a}_{m} \right ]=-1$ e, desta forma, o vetor de estado $a^{0 \a}_{m}|0,p^{\mu}\ra _{\al}$ possui norma negativa, pois $_{\al}\langle p^{\mu}, 0|a^{0}_{m},a^{0 \a}_{m}|0, p^{\mu}\rangle_{\al} = \ _{\al}\langle p^{\mu}, 0|\left [a^{0}_{m},a^{0 \a}_{m} \right ]|0,p^{\mu}\ra_{\al} =-1$, e o v\'acuo \'e normalizado.  Este mesmo argumento \'e v\'alido para os operadores de cria\c{c}\~ao, $a_{m}^{\mu \a}$ e $b_{m}^{\mu \a}$, que geram o espa\c{c}o de Fock a partir do v\'acuo $|0\ra _{\al}|0 \ra _{\b}$  para a corda fechada.  Deste modo, o espa\c{c}o f\'{\i}sico permitido para os estados da corda  \'e um subespa\c{c}o do espa\c{c}o completo de Fock, e \'e especificado por certas condi\c{c}\~oes subsidi\'arias, que ser\~ao apresentadas adiante. 

        Para termos uma teoria causal \'e nescess\'ario que o subespa\c{c}o f\'{\i}sico seja livre de estados de norma negativa, que s\~ao usualmente chamados ``fantasmas''. Estados de norma negativa s\~ao estados n\~ao aceit\'aveis fisicamente, pois est\~ao em conflito com a interpreta\c{c}\~ao probabil\'{\i}stica da mec\^anica qu\^antica. Esperamos ent\~ao que os estados fantasmas desacoplem do espa\c{c}o de Hilbert. Verificaremos que estes fantasmas de fato desacoplam do espa\c{c}o de Hilbert se a dimens\~ao do espa\c{c}o-tempo for  menor ou igual a 26, e  a constante $a$ que surge devido ao ordenamento normal for menor ou igual a 1. 

        As condi\c{c}\~oes subsidi\'arias usadas para determinarmos o espa\c{c}o dos estados f\'{\i}sicos, que correspondem \`as equa\c{c}\~oes cl\'assicas de v\'{\i}nculo $T_{ab}=0$, s\~ao equivalente a impor que $L_{m}|\psi \ra=\overline{L}_{m}|\psi \ra=0$ para $m>0$, uma vez que $L_{m}$ para a corda aberta \'e definida como sendo a componente de Fourier da combina\c{c}\~ao das componentes $T_{++}$ e $T_{--}$ definidas em (\ref{T}), e para a corda fechada $L_{m}$ e $\overline{L}_{m}$ s\~ao definidos em termos das componentes de Fourier de $T_{--}$ e   $T_{++}$, respectivamente. O caso $m=0$ requer um pouco mais de cuidado, uma vez que quanticamente o operador de Virasoro (\ref{ov}) apresenta problemas, pois, sendo operadores, os coeficientes $\al^{\mu}_{m+n}$ $(\b^{\mu}_{m+n})$ e $\al^{\mu}_{n}$ $(\b^{\mu}_{n})$ n\~ao comutam para $m=0$, surgindo deste modo o problema de ordenamento normal nas express\~oes para $L_{0}$ e $\overline{L}_{0}$, que podem ser escritos como
\be
L_{0}=\frac{1}{2}\al_{0}^{2}+\sum_{n=1}^{\infty}\al_{-n} \cdot \al_{n}+ a \  \ \ \mbox{e} \ \ \  \overline{L}_{0}=\frac{1}{2}\b_{0}^{2}+\sum_{n=1}^{\infty}\b_{-n} \cdot \al_{n}+ a,
\ee
onde $a=\frac{D}{2}\sum_{n=1}^{\infty}n$ \'e uma constante que aparece devido as rela\c{c}\~oes de comuta\c{c}\~ao. Podemos resolver este problema definindo a express\~ao normalmente ordenada para $L_{0}$ e $\overline{L}_{0}$ da forma 
\be
:L_{0}:=\sum_{n=-\infty}^{\infty}:\al_{-n} \cdot \al_{n}:=\frac{1}{2}\al_{0}^{2}+\sum_{n=1}^{\infty}\al_{-n} \cdot \al_{n},
\ee
\be
:\overline{L}_{0}:=\sum_{n=-\infty}^{\infty}:\b_{-n} \cdot \b_{n}:=\frac{1}{2}\b_{0}^{2}+\sum_{n=1}^{\infty}\b_{-n} \cdot \b_{n}.
\ee
Devido a esta ambig\"uidade do ordenamento normal, o v\'{\i}nculo cl\'assico $H=L_{0}=\overline{L}_{0}=0$ deve ser implementado quanticamente como 
\be
(L_{0}-a)|\psi \ra=(\overline{L}_{0}-a)|\psi \ra=0. \label{cvs}
\ee
Portanto, os estados f\'{\i}sicos devem satisfazer a condi\c{c}\~ao de Virasoro\be
 (L_{m}-a\d_{m})|\psi \ra=0, \ \ \ \ \ \ \ \ \ \ \  m \geq 0 ,
\ee
\be
 (\overline{L}_{m}-a\d_{m})|\psi \ra=0, \ \ \ \ \ \ \ \ \ \ \  m \geq 0 .
\ee
Estas condi\c{c}\~oes, mais as propriedades de Hermiticidade $L_{-m}=L_{m}^{\a}$ e $\overline{L}_{-m}=\overline{L}_{m}^{\a}$, mostram que, se $|\psi \ra$ e $|\phi \ra$ s\~ao dois estados f\'{\i}sicos, ent\~ao $\la \phi |(L_{m}-a\d_{m})|\psi \ra=\la \phi |(\overline{L}_{m}-a\d_{m})|\psi \ra=0, \ \ \forall m$.

\subsection{\'Algebra Qu\^antica de Virasoro e Estados F\'{\i}sicos}
${}$

        Nesta se\c{c}\~ao, veremos qual \'e a estrutura da anomalia que aparece na \'algebra de Virasoro, que classicamente \'e dada pela eq.~(\ref{acv}), devido ao ordenamento normal dos operadores. Todos os passos usados para obtermos a express\~ao (\ref{acv}) s\~ao v\'alidos tanto no n\'{\i}vel cl\'assico quanto no n\'{\i}vel qu\^antico, o problema surge no segundo termo da  express\~ao (\ref{pro}) para $m+n=0$, deste modo, \'e razo\'avel dizer que a \'algebra qu\^antica de Virasoro tem a forma
\be
\left [L_m,L_n \right ]= (m-n)L_{m+n}+A(m)\d_{m+n}. \label{aqv}
\ee   
Devemos ent\~ao calcular explicitamente a forma do termo $A(m)$. Usando a identidade de Jacobi $[L_{k},[L_{m},L_{n}]]+[L_{m},[L_{n},L_{k}]]+[L_{n},[L_{k},L_{m}]]=0$, e o fato de que $A(m)=-A(-m)$, que pode ser mostrado trocando-se $m$ por $n$ e vice-versa no comutador (\ref{aqv}), obtemos 
\be 
(n-m)A(n+m)+(2n+m)A(m)-(n+2m)A(n)=0. \label{eq.1}
\ee
Se escolhermos $n=2m$ e considerarmos que $n\neq 0$, pois se $m=0$ temos que $A(0)=0$, a eq.~(\ref{eq.1}) torna-se
\be
A(3m)+5A(m)-4A(2m)=0. \label{eq.2}
\ee
Assumindo que $A(m)$ tem uma forma polinomial $A(m)\!=\!\sum_{p}\!a_{p}\!m^{p}$, ent\~ao da eq.(\ref{eq.2}), sempre que $a_{p} \neq 0$, podemos obter a seguinte equa\c{c}\~ao expon\^encial em $p$, $3^{p}+5-2^{p+2}=0$. Para $p \geq 4$, $3^{p}$ come\c{c}a a ficar muito maior que  $2^{p+2}$, ent\~ao as \'unicas solu\c{c}\~oes desta equa\c{c}\~ao s\~ao $p=1$ ou $p=3$. Portanto, $A(m)=a_{1}m+a_{3}m^{3}$. Os valores dos coeficientes $a_{1}$ e $a_{3}$ podem ser encontrados se calcularmos o valor esperado do comutador $[L_{m},L_{-m}]$ entre dois estados f\'{\i}sicos para dois valores diferentes de $m$. Por simplicidade, escolhemos tal estado f\'{\i}sico como sendo o estado fundamental de uma corda com momento do centro de massa nulo $(p^{\mu}=0)$  representado por $|0;0 \ra $. Para $m=1$, obtemos que $a_{1}=-a_{3}$ e para $m=2$, obtemos $2a_{1}+ 8a_{3}= \frac{1}{2}\eta^{\mu \nu}\eta_{\mu \nu}=\frac{1}{2}D$. Desta forma, dos resultados obtidos para $m=1$ e $m=2$, obtemos que os coeficientes $a_{1}$ e $a_{3}$ valem, respectivamente, $\frac{-D}{12}$ e $\frac{D}{12}$, portanto, $A(m)=\frac{D}{12}(m^{3}-m)$ e a \'algebra qu\^antica de Virasoro torna-se

\be
\left [L_m,L_n \right ]= (m-n)L_{m+n}+\frac{D}{12}(m^{3}-m)\d_{m+n}, \label{aqvc}
\ee 
onde $D$ \'e chamado de carga central e nos fornece a dimens\~ao do espa\c{c}o-tempo no qual a folha mundo est\'a imersa. Em outras palavras, nos fornece o n\'umero de campos escalares livres na folha mundo, uma vez que o campo bidimensional $X^{\mu}$ possui $D$ componentes independentes. Isto quer dizer que cada campo escalar livre  contribui com uma unidade para a carga central. 

         A estrutura da \'algebra qu\^antica de Virasoro \'e tal que $L_{-1}$,$L_{0}$ e $L_{1}$ geram uma subalgebra fechada, sem anomalia, isom\'orfica aos grupos SU(1,1) ou SL(2,R). 

        Verificaremos agora que, devido \`a presen\c{c}a da anomalia na \'algebra qu\^antica de Virasoro, o v\'{\i}nculo cl\'assico $L_{m}=0,  \forall m$ n\~ao pode ser implementado em estados quanto-mec\^anicos, uma vez que,
\be
\la \psi |[L_m,L_n ]|\psi \ra= \la \psi |L_{0}|\psi \ra +\frac{D}{12}(m^{3}-m)\la \psi |\psi \ra.
\ee 
Ent\~ao, se assegurarmos que $L_n |\psi \ra=0,  \forall m $, ent\~ao $\la \psi |[L_m,L_n ]|\psi \ra=0$, o que n\~ao \'e verdade. No caso da corda fechada, teremos tamb\'em os operadores $\overline{L}_{n}$, que comutam com $L_{n}$ e satisfaz a mesma \'algebra de Virasoro, como na equa\c{c}\~ao ({\ref{aqvc}}). Para a corda fechada, segue-se ainda a seguinte condi\c{c}\~ao adicional
\be 
( L_{0}-\overline{L}_{0})|\psi \ra=0. \label{cvs2}
\ee
   O quadrado da massa $(M)$ de uma corda em um dado estado de oscila\c{c}\~ao \'e dado por
\be
M^2= -p_\mu p^\mu. 
\ee
  
       Para a corda aberta, temos:
\be 
L_{0}= \sum ^{\infty}_{n=1} \al^{\mu}_{-n}\al_{\mu n} + \al^{'}p^{\mu} p_{\mu}.
\ee
Ent\~ao, a condi\c{c}\~ao~(\ref{cvs}) implica que 
\be
(\sum ^{\infty}_{n=1} \al^{\mu}_{-n}\al_{\mu n}-a)|\ph \rangle= \al^{'} M^{2}|\ph\ra, 
\ee
que \'e chamada condi\c{c}\~ao de concha de massa, e o operador de massa para a corda \'e 
\be
 (N-a) = \al ^{'}M^{2},
\ee
onde definimos o operador n\'umero de n\'{\i}veis como 
\be
 N=\sum _{n> 0}N_{n} =\sum _{n>0}\al_{-n}^{\mu}\al _{\mu n}.
\ee
Este operador conta o n\'umero de excita\c{c}\~oes em um dado estado, $N_{n}$, ponderado pelo n\'umero do osciladores. Os autovalores de $N$ s\~ao $\sum^{\infty}_{n=1}nN_{n}$. Portanto, a massa de cada estado da corda \'e determinada pelo n\'{\i}vel de excita\c{c}\~ao.
 
        A massa $M$ dos estados f\'{\i}sicos da corda fechada pode ser obtida  a partir do v\'{\i}nculo~(\ref{cvs2}). Usando $L_{0}=\frac {1}{4} \al^{'}p^{\mu} p_{\mu} + \sum^{\infty}_{n=1} \al_{-n}^{\mu} \al_{n \mu}$ e a  express\~ao  ${\overline{L}}_{0}=\frac {1}{4} \al^{'}p^{\mu} p_{\mu} + \sum^{\infty}_{n=1}{ \b}_{-n}^{\mu}{\b}_{n \mu}$, este v\'{\i}nculo pode ser escrito como 
\be
M^{2} =-p^{\mu} p_{\mu}= M^{2}_{L}+ M^{2}_{R},
\ee
onde 
\be
\al ^{'}M^{2}_{L}=2(\overline{N}-a) \ \ \ \ \  \mbox{e} \ \ \ \ \ \al^{'}M^{2}_{R}=2(N-a).
\ee 
Ainda, como conseq\"u\^encia do v\'{\i}nculo $(L_{0}-{\overline{L}}_{0})|\ph\rangle =0$ para todos os estados f\'{\i}sicos $|\ph\rangle $, obtemos 
$$
N|\ph \ra= \overline{N}|\ph \ra
$$
e, conseq\"uentemente
$$
 M^{2}_{L} = M^{2}_{R}.
$$
Este resultado nos diz que o quadrado das massas possui contribui\c{c}\~oes iguais para direita e para a esquerda. 


        As Hamiltonianas (\ref{hca}) e (\ref{hcf}) est\~ao relacionadas  com os operadores de Virasoro, respectivamente da seguinte forma:
\be
H=L_{0}-a,
\ee
para a corda aberta e
\be
H=2(L_{0}^{'}+\overline{L}_{0}^{'})=2(L_{0}+{\overline{L}}_{0}-2a),
\ee
para a corda fechada.

       Agora, faremos uma an\'alise das condi\c{c}\~oes para as quais n\~ao existem estados f\'{\i}sicos de norma negativa para o caso da corda aberta. Veremos que estes estados de norma negativa existem somente para certas regi\~oes dos valores que podem ser assumidos pelo par\^ametro $a$ e pela dimens\~ao $D$. Para isto, olharemos regi\~oes do espa\c{c}o de Hilbert para as quais os estados f\'{\i}sicos possuem norma zero. Tais estados f\'{\i}sicos estar\~ao sempre presentes nas fronteiras que dividem o espa\c{c}o  f\'{\i}sico de Hilbert em uma regi\~ao onde os estados possuem norma positiva e outra regi\~ao onde estes vetores possuem norma negativa. Estes vetores de norma nula nos fornecer\~ao os valores cr\'{\i}ticos de $a$ e $D$, para os quais o espa\c{c}o de Hilbert est\'a livre de fantasmas.

       Denotemos o estado fundamental de uma corda de momento $k^{\mu}$ como $|0;k^{\mu} \ra $. A condi\c{c}\~ao de concha de massa implica que $\al^{'}k^{2}=a$. Agora, consideremos o primeiro estado excitado $\xi \cdot \al_{-1}|0;k^{\mu} \ra $, onde $\xi^{\mu}(k)$  \'e o vetor de polariza\c{c}\~ao com $D$ componentes independentes. A condi\c{c}\~ao de concha de massa agora implica que $\al^{'}k^{2}=a-1$ e a condi\c{c}\~ao de estado f\'{\i}sico $L_{1}| \psi \ra=0$ nos fornece $ \xi \cdot k =0$. Esta condi\c{c}\~ao nos conduz \`a $D-1$ polariza\c{c}\~oes permitidas. A norma deste vetor de estado \'e $\xi \cdot \xi $. Se escolhermos que o vetor $k$ se encontra no plano $(0,1)$,  ent\~ao teremos $D-2$ estados (tipo espa\c{c}o) com polariza\c{c}\~ao normal ao plano e com norma positiva. Podemos ver que os autovalores do operador de massa para o estado fundamental e para o primeiro estado excitado s\~ao, respectivamente, 
\be 
M^{2}=-\frac{a}{\al^{'}} \ \ \ \ \mbox{e} \ \ \ \   M^{2}=\frac{a}{\al^{'}}(1-a). 
\ee
Analisaremos agora algumas condi\c{c}\~oes sobre o par\^ametro $a$ e suas implica\c{c}\~oes. Se $a<1$, ent\~ao $M^{2}>0$ e, conseq\"uentemente, $k^{\mu}$ ser\'a um vetor tipo tempo e podemos escolher um sistema de refer\^encia, no qual este s\'o possua componente temporal $(k_{0}, 0,0,\dots )$. Neste caso, $\xi^{\mu}$ \'e tipo espa\c{c}o e possui norma positiva. Se $a>1$, $M^{2}<0$ e $k^{\mu}$ \'e um vetor tipo espa\c{c}o com somente coordenadas espaciais em um referencial apropriado $(0,k_{1},k_{2},\cdots)$ e $\xi^{\mu}$ \'e um quadrivetor tipo tempo com norma negativa. O \'ultimo caso \'e o qual a=1, o que implica em $M^{2}=0$ e, conseq\"uentemente, $k^{\mu}$ \'e um vetor tipo luz, sendo assim,  $\xi^{\mu}$ \'e proporcional a $k^{\mu}$ e possui norma zero. Portanto, uma vez que a norma de  $\xi \cdot \al_{-1}|0;k^{\mu} \ra $ \'e $\xi \cdot \xi $, obtemos a primeira condi\c{c}\~ao para aus\^encia de fantasmas $a \leq 1$. Na fronteira onde $a=1$ (estados de norma zero), a part\'{\i}cula vetorial \'e um f\'oton e o estado fundamental \'e um t\'aquion, pois possui o quadrado da massa negativa. O t\'aquion  viaja com velocidade maior que a da luz e pode ser excitado para qualquer energia negativa, tornando desta forma a teoria inconsistente. A presen\c{c}a do t\'aquion mostra que o estado fundamental da teoria de cordas, se existe \'e inst\'avel.

        Agora, para esta fronteira na qual $a=1$, devemos examinar tamb\'em quais s\~ao os valores cr\'{\i}ticos de $D$. Para isto, consideremos o seguinte estado
\be
|\psi \ra= \le[c_{1}\al_{-1} \cdot \al_{-1}+c_{2}\al_{0} \cdot \al_{-2}+c_{3}(\al_{0} \cdot \al_{-2})^{2}\ri ]|0;p \ra.
\ee
Da condi\c{c}\~ao de concha de massa para $(L_{0}-a)|\psi \ra=0$, obtemos $p^{2}=2a-4$, logo $p^{2}=-2$, pois $a=1$. Tal estado tamb\'em deve satisfazer as condi\c{c}\~oes de estado f\'{\i}sico $L_{1}|\psi \ra=L_{2}|\psi \ra=0$. A condi\c{c}\~ao $L_{1}|\psi \ra=0$ nos conduz \`a $(2c_{1}+2c_{2}-4c_{3})|0;p \ra=0$ e a   condi\c{c}\~ao $L_{2}|\psi \ra=0$  \`a $(Dc_{1}-4c_{2}-2c_{3})|0;p \ra=0$. Desta forma, podemos dizer que $| \psi \ra$ somente ser\'a um estado f\'{\i}sico se os coeficientes $c_{1},c_{2} \ \mbox{e} \ c_{3}$ satisfizerem as seguintes rela\c{c}\~oes
\be 
c_{2}=c_{1}\frac{D-1}{5} \ \ \ \ \ \mbox{e} \ \ \ \ \ c_{3}=c_{1}\frac{D+1}{10}.
\ee
Neste caso, a norma \'e $\la \psi | \psi \ra = \frac{2c_{1}^{2}}{25}(D-1)(26-D)$, de onde podemos observar que o espectro \'e livre de fantasmas para $D \leq 26$. 

        Esta an\'alise realizada aqui para a corda aberta pode ser estendida para a corda fechada, desde que um estado de uma corda fechada pode ser escrito como um produto tensorial de dois estados de corda aberta, podemos tomar duas c\'opias id\^enticas dos nossos resultados para a corda aberta. O estado fundamental da corda fechada \'e um t\'aquion, cuja massa \'e dada po $M^{2}=-4/\al^{'}$ e com todas as conseq\"u\^encias que vimos para o caso da corda aberta. Os demais estados podem ser constru\'{\i}dos por aplica\c{c}\~oes de um n\'umero igual de operadores de cria\c{c}\~ao no v\'acuo dos setores esquerdo e direito.                                                                        
\subsection{Quantiza\c{c}\~ao no Calibre de Cone de Luz}
${}$
 
        Fixando o calibre conforme, n\~ao eliminamos totalmente as simetrias de calibre da a\c{c}\~ao da corda, o que nos possibilita impor condi\c{c}\~oes adicionais de calibre que reduzam o n\'umero de componentes n\~ao-triviais de $X^{\mu}$ e nos conduzam somente aos graus de liberdade din\^amicos. Podemos verificar que ainda resta uma simetria residual, pois a a\c{c}\~ao em um calibre conforme possui ainda uma invari\^ancia por reparametriza\c{c}\~ao das coordenadas da folha mundo
\be
(\s,\t) \longrightarrow (\ti{\s}(\s,\t),\ti{\t}(\s,\t)),
\ee
onde $(\s,\t)$ e $(\ti{\s},\ti{\t})$ s\~ao sistemas de coordenadas ortogonais sobre a folha mundo.  Usaremos esta simetria residual para impor uma condi\c{c}\~ao de calibre extra que, como veremos, n\~ao \'e covariante, mas muito conveniente, pois nos conduz a uma teoria livre de fantasmas. Esta invari\^ancia deve ser compat\'{\i}vel com os v\'{\i}nculos gerados pelo tensor energia-momento. Usando a regra da cadeia, estes dois sistemas de coordenadas podem ser relacionados na forma
\bea
\frac{\6 X^{\mu}}{\6 \t}&=&\frac{\6 X^{\mu}}{\6 \ti{\t}}\frac{\6 \ti{\t}}{\6 \t}+\frac{\6 X^{\mu}}{\6 \ti{\s}}\frac{\6 \ti{\s}}{\6 \t}, \non
\frac{\6 X^{\mu}}{\6 \s}&=&\frac{\6 X^{\mu}}{\6 \ti{\t}}\frac{\6 \ti{\t}}{\6 \t}+\frac{\6 X^{\mu}}{\6 \ti{\s}}\frac{\6 \ti{\s}}{\6 \s}.
\eea
Ent\~ao, impondo que o lado esquerdo destas equa\c{c}\~oes devam satisfazer os v\'{\i}nculos de Virasoro, equa\c{c}\~oes (\ref{T}), obtemos as seguintes equa\c{c}\~oes diferenciais para as novas coordenadas:
\be
\le (\frac{{\6}^{2}}{\6 {\t}^{2}}-\frac{{\6}^{2}}{\6 {\s}^{2}} \ri )\ti{\s}=0,
\ee
e similarmente,
\be
\le (\frac{{\6}^{2}}{\6 {\t}^{2}}-\frac{{\6}^{2}}{\6 {\s}^{2}} \ri )\ti{\t}=0.
\ee
Portanto, os v\'{\i}nculos implicam que $\ti{\s}$ e $\ti{\t}$ satisfazem as mesmas equa\c{c}\~oes de movimento, no calibre conforme, que as coordenadas espa\c{c}o-temporais $X^{\mu}(\s,\t)$. Isto significa que podemos identificar $\ti{\s}$ ou $\ti{\t}$ como sendo uma das componentes de $X^{\mu}$. Podemos fazer uma reparametriza\c{c}\~ao tal que uma das coordenadas na folha mundo seja id\^entica a uma das coordenadas espa\c{c}o-temporais. A escolha na qual $X^{\mu}$ \'e id\^entica a $\t$ nos conduz ao calibre de cone de luz como segue. 

         Se $X^0$ \'e a dimens\~ao temporal do espa\c{c}o-tempo, para um espa\c{c}o-tempo $D$-dimensional ent\~ao temos $D-1$ dimens\~oes espaciais. Definamos as coordenadas do cone de luz $X^{+}$ e $ X^{-}$ como:
\be
X^{\pm}=\frac{1}{\sqrt{2}}(X^{0} \pm X^{D-1}). \label{ccl}
\ee 

         Existe uma grande diferen\c{c}a entre estas e as coordenadas de cone de luz introduzidas anteriormente para folha mundo. No espa\c{c}o-tempo temos um total de $D$ coordenadas e ({\ref{ccl}}) envolve somente duas delas, a saber, $X^{0}\ \mbox{e}\  X^{D-1}$, de um modo arbitr\'ario e n\~ao covariante. Na folha mundo, existem somente duas coordenadas, desta forma a escolha de $\s^{\pm}$ n\~ao \'e arbitr\'aria.  Neste sistema de coordenadas, o produto escalar de dois vetores $V=(V^{+},V^{-},V^{i})$  e $U=(U^{+},U^{-},U^{i})$ \'e definido como
\be
U.V= U^{+}V^{-}+U^{-}V^{+}-U^{i}V^{i},
\ee
onde o \'{\i}ndice $i$ vai de $1$ a $D-2$, e os \'{\i}ndices s\~ao levantados e abaixados de acordo com a regra: $V^{+}=-V_{-}$, $V^{-}=-V_{+}$, $V^{i}=V_{i}$. Estas regras nos dizem que as componentes da m\'etrica s\~ao $\h_{ij}=1$ e $\h_{+-}=\h_{-+}=-1$. A id\'eia b\'asica do calibre de cone de luz \'e escolhermos a dire\c{c}\~ao de $X^{+}$ ao longo de $\ti{\t}$, desta forma $X^{+}$ deve satisfazer a mesma equa\c{c}\~ao de movimento para $\ti{\t}$, cuja solu\c{c}\~ao pode ser escolhida como
\be
X^{+}=x^{+}+l^{2} p^{+}\t   \label{X+},
\ee
onde $p^{+}$ e $x^{+}$ s\~ao, respectivamente, o momento e a coordenada do centro de massa da corda no calibre do cone de luz. Com esta escolha, usaremos os v\'{\i}nculos para encontrar $X^{-}$.

         A solu\c{c}\~ao da equa\c{c}\~ao de movimento (\ref{em}) para uma corda aberta com a condi\c{c}\~ao de contorno de Neumann \'e
\be
 X^{\mu}= x^{\mu}+ 2\al^{'} p^{\mu}\t+i\sqrt{2\al^{'}}\sum_{n \neq 0}\frac{1}{n} \al^{\mu}_{n} e^{in\t } \cos n \s , 
\ee 
e a componente temporal do momento canonicamente conjugado no calibre conforme \'e 
\be
P^{\mu}_{\t} = \frac{ \6{\cal{L}}}{\6 \dot{X}_{\mu}}=T\dot{X}^{\mu}.
\ee 
Notemos ent\~ao que as condi\c{c}\~oes de contorno (\ref{T}) podem ser escritas, respectivamente, como: 
\be
P^{\mu}_{\t}X^{'}_ {\mu}=0,        \label{pv}
\ee 
\be
P^{\mu}_{\t}P_{\t \mu}+ T^{2}X^{'\mu}X_{\mu}^{'}= 0.  \label{sv}
\ee 

Nas coordenadas do cone de luz, a componente temporal da corrente de momento $P^{\mu}_{\t}$ pode ser escrita como
\be
P^{\pm}_{\t}= T\dot{X}^{\pm},    \label{p+-}
\ee
e as derivadas temporais e espaciais da solu\c{c}\~ao ({\ref{X+}}) s\~ao, respectivamente,
\be
\dot{X}^{+}=l^{2}p^{+} \ \ \ \mbox{e} \ \ \ X^{'+}= 0.
\ee
Ent\~ao, a componente $+$ da equa\c{c}\~ao~(\ref{p+-}) torna-se
\be
P^{+}_{\t}= 2\al^{'}Tp^{+}.
\ee
Logo, para o primeiro v\'{\i}nculo, equa\c{c}\~ao~({\ref{pv}), obtemos
\be
 P^{\mu}_{\t}X^{'}_{\mu}=P_{ \t }^{+}X^{-'}+ P_{\t}^{-}X^{'+}-P^{i}_{\t}X^{'}_{i}=2\al^{'}T p^{+}X^{-'}-P^{i}_{\t}X^{'}_{i}=0,
\ee
que, em termos das componentes $i$ da corrente de momento $P^{\mu}_{\t}$ pode ser escrito como
\be
 X^{-'}= \frac{1}{2\al^{'}Tp^{+}}P^{i}_{\t}X^{'}_{i}=\frac{1}{2\al^{'}p^{+}}\dot{X}^{i}X^{'}_{i}. \label{X'}
\ee
J\'a o segundo v\'{\i}nculo \'e dado por:
$$
P^{\mu}_{\t}P_{\t \mu}+T^{2}X^{'\mu}X_{\mu}^{'}=0
$$
$$
2P^{+}_{\t}P^{-}_{\t}-P^{i}_{\t}P^{i}_{\t}+2T^{2}X^{+'}X^{-'}-T^{2}X^{'}_{i}X^{'}_{i}=0
$$
\bea
2P^{+}_{\t}P^{-}_{\t}&=&T^{2}{X}^{'}_{i}X^{'}_{i}+P^{i}_{\t}P^{i}_{\t}  \non
4\al^{'}Tp^{+}P^{-}_{\t}&=&T^{2}X^{'}_{i}X^{'}_{i}+T^{2}\dot{X}^{i}\dot{X}^{i} \non
P^{-}_{\t}&=&\frac{T}{4\al^{'}p^{+}} \le[(\dot{X}^{i})^{2}+(X^{'}_{i})^{2}\ri] \non
\dot{X}^{-}&=&\frac{1}{4\al^{'}p^{+}} \le[(\dot{X}^{i})^{2}+(X^{'}_{i})^{2}\ri]  \label{X-}.  \label{X.}
\eea
Logo, o primeiro v\'{\i}nculo determina $X^{-'}$ e o segundo $\dot{X}^{-}$. Deste modo, podemos combinar estes dois v\'{\i}nculos na forma abaixo para obtermos uma \'unica express\~ao que \'e equivalente aos v\'{\i}nculos de Virasoro (\ref{T})
\be
\dot{X}^{-}+X^{-'}=\frac{1}{4\al^{'}p^{+}}\le(\dot{X}^{i}+X^{'}_{i}\ri )^{2}.  \label{vv}
\ee

          Nas coordenadas do cone de luz, a componente $X^{-}$ da expans\~ao para $X^{\mu}$ \'e   
\be
X^{-}=x^{-}+2\al^{'}p^{-}\t+i\sqrt{2\al^{'}}\sum_{n \neq 0}\frac{1}{n}\al^{-}_{n}e^{-in \t}\cos{n \s}.
\ee
A combina\c{c}\~ao das suas derivadas espacial e temporal, e definindo $\al_{0}^{-}=lp^{-}$,  fornece
\bea
\dot{X}^{-}+X^{-'}&=&2\al^{'}p^{-}\t+\sqrt{2\al^{'}}\sum_{n \neq 0}\al^{-}_{n}e^{-in \t}e^{-in \s} \non
\dot{X}^{-}+X^{-'}&=&\sqrt{2\al^{'}}\sum_{n=-\infty}^{\infty}\al^{-}_{n}e^{-in \t}e^{-in \s}.   \label{-}
\eea
Da mesma forma, a componente $X^{i}$ da expans\~ao para $X^{\mu}$ \'e   
\be
X^{i}=x^{i}+2\al^{'}p^{i}\t+i\sqrt{2\al^{'}}\sum_{n \neq 0}\frac{1}{n}\al^{i}_{n}e^{-in \t}\cos{n \s}.
\ee
E definindo $\al_{0}^{i}=\sqrt{2\al^{'}}p^{i}$, podemos escrever 
\be
\le(\dot{X}^{i}+X^{'}_{i}\ri )^{2}=2\al^{'}\sum_{m=-\infty}^{\infty}\sum_{l=-\infty}^{\infty}\al^{i}_{m}\al^{i}_{l}e^{-i(m+l) \t}e^{-i(m+l) \s}. \label{i}
\ee
Portanto,  substituindo as equa\c{c}\~oes~(\ref{-}) e (\ref{i}) na equa\c{c}\~ao~(\ref{vv}) e fazendo a seguinte mudan\c{c}a dos \'{\i}ndices $n=m+l$, esta combina\c{c}\~ao de v\'{\i}nculos pode ser escrita como
\be
\al^{-}_{n}=\frac{1}{4\al^{'}p^{+}}\sum_{m=-\infty}^{\infty}\al^{i}_{m}\al^{i}_{n-m},
\ee
que \'e uma forma alternativa de escrevermos os v\'{\i}nculos de Virasoro.
       Em resumo, implementando o calibre do cone de luz e fixando $X^{+}$, pudemos resolver $X^{+}$ e $X^{-}$ em termos de $X^{i}$ e $\al^{-}_{n}$ em termos de $\al^{i}_{n}$, uma vez que escolhemos a solu\c{c}\~ao $X^{+}$ de modo que $\al^{-}_{n}$=0 para $n \neq 0$. Portanto, este calibre nos conduz somente a graus de liberdade transversais  $X^{i}$. 

        A densidade de Lagrangeana em um calibre conforme \'e
$$
{\cal{L}}= - \frac{T}{2}(\dot{X}^{\mu}\dot{X}_{\mu}-X^{'\mu}X^{'}_{\mu})
$$
e a Lagrangeana
$$
 L = \int d \s{\cal{L}}= \frac{-T}{2}\int d \s (\dot{X}^{\mu}\dot{X}_{\mu}-X^{'\mu}X^{'}_{\mu}).
$$
No calibre do cone de luz, $ X^{+}$ \'e especificado e $ X^{-}$  \'e eliminado. $X^{i}$ deve satisfazer a mesma equa\c{c}\~ao de onda que $ X^{\mu}$ em um calibre covariante. Ent\~ao, a Lagrangeana no calibre de cone de luz \'e
\be
 L_{gcl} = -\frac{T}{2}\int d\s ((\dot{X}^{i})^{2}-(X^{'}_{i})^{2}).
\ee
Integrando a express\~ao para a densidade de Hamiltoniana ${\cal{H}}=P^{\mu}_{\t}\cdot{X}_{\mu}-{\cal{L}}$, obtemos a Hamiltoniana correspondente no calibre de cone de luz
\bea
H_{gcl}&=& \frac{T}{2}\int d\s \le ( (\dot{X}^{i})^{2}+(X^{'}_{i})^{2}\ri) \non
&=& \frac{1}{2T}\int d\s \le((\dot{P}^{i}_{\t})^{2}+T^{2}(X^{'}_{i})^{2}\ri) \non
&=& 2\al^{'}p^{+} \int d\s P^{-}_{\t} \non
&=&2\al^{'}{p^{+} p^{-}},    \label{hlcg}
\eea
onde usamos a equa\c{c}\~ao~(\ref{vv}). Em termos dos modos de expans\~ao de Fourier, $p^{+}$ e $p^{-}$ podem ser escritos, respectivamente, como
\be
p^{+}=\frac{1}{4\al^{'} \al^{-}_{0}}\sum^{\infty}_{m= -\infty}\al^{i}_{m}\al^{i}_{-m}
 \ \ \ \mbox{e} \ \ \ p^{-}=\frac{\al^{-}_{0}}{\sqrt{2\al^{'}}}.
\ee
Logo a Hamiltoniana acima torna-se
\be
H_{gcl}=\frac{1}{2}\sum^{\infty}_{m= -\infty}\al^{i}_{m}\al^{i}_{-m}.
\ee

        Agora iremos quantizar a teoria, para isto $X^{i}(\t,\s)$ e $ P^{i}_{\t}$ ser\~ao tratados como operadores qu\^anticos e,  conseq\"uentemente, os modos de Fourier ser\~ao quantizados. Os comutadores destes operadores em tempos iguais s\~ao 
$$
[X^{i}(\t,\s ), P^{j}_{\t}(\t, \s^{'})]= i \d^{ij}\d(\s-\s^{'}),
$$
$$
[X^{i}(\t,\s ), X^{j}(\t, \s^{'})]=[ P^{i}_{\t}(\t, \s), P^{j}_{\s}(\t, \s^{'})]=0,
$$
\be
[x^{-}, p^{+}]=-i,                \label{rclcg}
\ee
$$
[x^{-}, X^{i}]=[x^{-}, P^{j}_{\t}]=[p^{+}, X^{i}]=[p^{+}, P^{i}_{\t}]=0.
$$
Das express\~oes para $ X^{i}$ e $P^{j}_{\t}$ nas rela\c{c}\~oes de comuta\c{c}\~ao acima, podemos mostrar que:
\be
[\al^{i}_{n },\al^{j}_{m }]= n\d^{ij}\d_{n+m}.   \label{rcc}
\ee

        Como na quantiza\c{c}\~ao can\^onica, n\'os podemos interpretar estes coeficientes de Fourier quantizados como um conjunto de osciladores harm\^onicos qu\^anticos se identificarmos
\be 
\al^{i}_{-n}= (\al^{i}_{n})^\a,\ \ \ \ \ \ \ \ \ \ \ n>0,
\ee 
 e tamb\'em temos problemas de diverg\^encia da Hamiltoniana devido a n\~ao comutatividade dos operadores $\al^{i}_{m}$ e $\al^{i}_{-m}$. Para resolvermos este problema, devemos ordenar normalmente os operadores que aparecem na Hamiltoniana, e explicitamente separarmos a diverg\^encia constante, que representa a energia de ponto zero dos osciladores, da seguinte forma
\be
H=\frac{1}{2}\sum^{\infty}_{m= -\infty}:\al^{i}_{m}\al^{i}_{-m}:-a,
\ee
onde das rela\c{c}\~oes de comuta\c{c}\~ao (\ref{rcc}) temos
\be
a=-\frac{D-2}{2}\sum^{\infty}_{n=1}n.
\ee
Podemos regularizar esta diverg\^encia escolhendo um m\'etodo particular chamado regulariza\c{c}\~ao da fun\c{c}\~ao $\zeta$, com $\z (s)=\sum^{\infty}_{n=1}n^{-s}$, que \'e anal\'{\i}tica para $s=-1$. Ent\~ao
\be
a=-\frac{D-2}{2}\z (-1)=\frac{D-2}{24}. \label{a}
\ee

        O estado fundamental, ou v\'acuo da corda, \'e definido por 
\be
\al^{i}_{n}|0;p\rangle _{\al} =0,     \ \ \ \ \ \ \ \ \ \ \ \ \ \ \ \ \ \ \ \ n>0,
\ee 
e
\be
p^{i}_{op}|0;p\rangle_{\al} =p^{i}|0;p\rangle_{\al} . 
\ee
Iremos frequentemente nos referir ao estado $|0;p\rangle _{\al} $ como $|0\rangle _{\al} $ em situa\c{c}\~oes onde os momentos da corda n\~ao t\^em import\^ancia. Estados excitados, vetores no espa\c{c}o de Fock, s\~ao criados por aplica\c{c}\~oes sucessivas dos operadores de cria\c{c}\~ao $(\al^{i}_{n})^\a= \al^{i}_{-n}$, $n>0$, no estado de v\'acuo $|0\rangle_{\al} $ .

        Os estados f\'{\i}sicos de uma part\'{\i}cula podem ser classificados de acordo com sua massa. As massas s\~ao autovalores de um operador massa relacionado com o momento de uma part\'{\i}cula pela condi\c{c}\~ao de concha de massa: $M^{2}=-p^{\mu}p_{\mu}$. O quadrado da massa da corda \'e o autovalor do operador $M^{2}$ definido como
\bea 
M^{2}&=&p^{\mu}p_{\mu} \non
&=&2p^{+}p^{-}-p^{i}p^{i} \non
&=&2\pi T H_{gcl}- \p T\al_{0}^{i}\al_{0}^{i} \non
&=& \pi T \le ( \sum^{\infty}_{m= -\infty}:\al^{i}_{m}\al^{i}_{-m}:-2a - \al_{0}^{i}\al_{0}^{i}\ri ) \non  
&=&\pi T \le ( \sum_{m \neq 0}:\al^{i}_{m}\al^{i}_{-m}:-2a \ri ) \non  
&=&2 \pi T(N-a),  \label{om}
\eea
onde
$$
N=\sum^{\infty}_{m=1}\al^{i}_{-m}\al^{i}_{m} 
$$
\'e o operador n\'umero usual que conta o n\'umero de excita\c{c}\~oes no estado. Esta \'e a mesma condi\c{c}\~ao de concha de massa encontrada no tratamento covariante, exceto que somente os modos transversais contribuem para $N$.

        O quadrado da massa da corda no estado fundamental, ($N=0$), \'e
\be
M^{2}|0\rangle =-2 \p Ta|0\rangle =M_{0}^{2}|0\rangle .
\ee
Como $a > 0$, isto nos conduz a uma teoria n\~ao-causal, uma vez que o v\'acuo possui o quadrado da massa negativa. A part\'{\i}cula representada por este estado \'e chamada de tachyon. 

       No calibre de cone de luz, quantizamos somente os graus de liberdade din\^amicos, excluindo ent\~ao a possibilidade da exist\^encia de estados de norma negativa (fantasmas). No entanto, para quantizarmos somente estes graus de liberdade din\^amicos quebramos a covari\^ancia  de Lorentz manifesta. 

       Agora, vamos examinar o primeiro estado excitado, ou seja, um estado contendo uma excita\c{c}\~ao transversal de um oscilador, $\al^{i}_{-1}|0\rangle_{\al}$ .
Este conjunto de estados forma um vetor com $D-2$ componentes f\'{\i}sicas transversalmente polarizadas. A falta de componentes longitudinais e a elimina\c{c}\~ao de duas componentes n\~ao f\'{\i}sicas via transforma\c{c}\~oes de calibre, como no caso do f\'oton no calibre de Coulomb, \'e uma caracter\'{\i}stica de campos de massa nula. Isto implica que este estado deve possuir massa nula se a invari\^ancia de Lorentz for mantida. A massa deste estado \'e obtida aplicando o operador de massa (\ref{om}) no v\'acuo $|0 \rangle _{\al}$ e resultando o seguinte autovalor
\be
M^{2}=2 \pi T (1-a).
\ee
O estado ter\'a massa nula somente se $a=1$. Portanto, da equa\c{c}\~ao~(\ref{a}) obtemos que a dimens\~ao do espa\c{c}o-tempo \'e $D=26$.              

       Os geradores de Lorentz s\~ao
\be
M^{\mu \nu}=\int^{\p}_{0} d\s M^{\mu \nu}_{\t}=\int^{\p}_{0} d\s \le ( X^{\mu}P^{\nu}_{\t}-X^{\nu}P^{\mu}_{\t} \ri) ,   
\ee
onde $M^{\mu \nu}_{\t}$ \'e a componente $\t$ da corrente conservada associada  \`a simetria de Lorentz da a\c{c}\~ao da corda. Classicamente, estes geradores satisfazem \`a \'algebra de Lorentz
\be
[M^{\mu \nu},M^{\s \r}]=i\h^{\nu \s}M^{\mu \r}-i\h^{\m \s}M^{\nu \r}-i\h^{\nu \r}M^{\mu \s}+i\h^{\m \r}M^{\nu \s}.
\ee
Esta simetria pode ser verificada se fizermos  uma transforma\c{c}\~ao infinitesimal de Lorentz, $\d X^{\mu}=\d \L^{\m \nu}X_{\nu}$, na a\c{c}\~ao da corda. Uma vez que esta \'e uma opera\c{c}\~ao de simetria, $\d S=0$, a corrente \'e conservada. $M^{\mu \nu}$ \'e a carga associada a esta corrente.

        No calibre covariante onde a componente $\t$ da corrente de momento \'e dada por $P^{\mu}_{\t}=T \dot{X}^{\mu}$, os geradores de Lorentz s\~ao 
\be
M^{\mu \nu}=T\int^{\p}_{0} d\s \le( X^{\mu}\dot{X}^{\nu}-X^{\nu}\dot{X}^{\mu} \ri).
\ee
Substituindo os modos de expans\~ao dos campos $X^{\mu}$, equa\c{c}\~ao~(\ref{sol}) nesta express\~ao e efetuando a integra\c{c}\~ao obtemos:
\be
M^{\mu \nu}=x^{\mu}p^{\nu}-x^{\nu}p^{\mu}-i\sum^{\infty}_{n=1}\frac{1}{n} \le (\al^{\mu}_{-n}\al^{\nu}_{n}-\al^{\nu}_{-n}\al^{\mu}_{n} \ri ).
\ee

        Classicamente, as componentes $M^{-i}$ destes geradores  satisfazem a seguinte rela\c{c}\~ao de comuta\c{c}\~ao 
\be
[M^{-i},M^{-j}]=0,
\ee
e o comutador qu\^antico correspondente \'e
\be
[M^{-i},M^{-j}]=\frac{2}{(p^{+})^{2}}\sum^{\infty}_{n=1} \le [m \le (1-\frac{1}{24}(D-2) \ri ) + \frac{1}{m}\le (\frac{1}{24}(D-2)-a \ri ) \ri ] \le (\al^{i}_{-m}\al^{j}_{m}-\al^{j}_{-m}\al^{i}_{m} \ri ).
\ee
Para a teoria permanecer invariante por transforma\c{c}\~oes de Lorentz, o lado direito deste comutador deve ser zero. Para isto, devemos ter a dimens\~ao cr\'{\i}tica do espa\c{c}o-tempo $D= 26$ e $a=1$.

        Agora, analisaremos a corda fechada no calibre de cone de luz. Esta an\'alise seguir\'a a mesma linha que a usada para a corda aberta. Os modos de expans\~ao da corda fechada s\~ao dados pela equa\c{c}\~ao~(\ref{scf}). No calibre de cone de luz, a componente $X^{+}$ do campo ainda \'e dado pela equa\c{c}\~ao~(\ref{X+}). Isto implica, se compararmos as equa\c{c}\~oes~(\ref{scf}) e (\ref{X+}), que 
\be
\al_{n}^{+}={\b}_{n}^{+}=0
\ee
para $n \neq 0$. Podemos subtrair a equa\c{c}\~ao~(\ref{X'}) da (\ref{X.})  para obter a seguinte combina\c{c}\~ao entre os v\'{\i}nculos de Virasoro:
\be
\dot{X}^{-}-X^{-'}=\frac{1}{4\al^{'}p^{+}}\le(\dot{X}^{i}-X^{'}_{i}\ri )^{2}. \label{vvf}
\ee
A componente $X^{-}$ da expans\~ao da solu\c{c}\~ao $X^{\mu}$ no calibre covariante \'e
\be
X^{-}(\t,\s)=X^{-}_{R}(\t,\s)+X^{-}_{L}(\t,\s),
\ee
onde
\be
X^{-}_{R}=\frac{1}{2}x^{-} +\al^{'} p^{-}(\t - \s) +\frac{i}{2}\sqrt{2\al^{'}}\sum _{n\neq 0}\frac{1}{n}\al^{-}_{n}e^{-2in(\t - \s)},
\ee
e
\be
X^{-}_{L}=\frac{1}{2}x^{-} +\al^{'} p^{-} (\t + \s) +\frac{i}{2}\sqrt{2\al^{'}}\sum _{n\neq 0}\frac{1}{n} {\b}^{-}_{n}e^{-2in(\t + \s)}.
\ee
Ent\~ao, substituindo esta solu\c{c}\~ao nas equa\c{c}\~oes de v\'{\i}nculo (\ref{vv}) e (\ref{vvf}), teremos, respectivamente 
\be
\dot{X}^{-}+X^{-'}=\sqrt{2\al^{'}}({\al}^{-}_{0}-{\b}^{-}_{0})+2\sqrt{2\al^{'}}\sum _{n=-\infty}^{\infty}{\b}^{-}_{n}e^{-2in(\t + \s)},
\ee
e
\be
\dot{X}^{-}-X^{-'}=\sqrt{2\al^{'}}({\b}^{-}_{0}-{\al}^{-}_{0})+2\sqrt{2\al^{'}}\sum _{n=-\infty}^{\infty}\al^{-}_{n}e^{-2in(\t - \s)},
\ee
onde definimos
\be
{\al}^{-}_{0}+{\b}^{-}_{0}=p^{-}\sqrt{2\al^{'}}.
\ee
Analogamente para o caso da corda aberta, podemos definir 
\be
{\al}^{i}_{0}={\b}^{i}_{0}=\frac{\sqrt{2\al^{'}}p^{i}}{2}
\ee
e, a partir dos v\'{\i}nculos, obter as componentes  ${\al}^{-}_{l}$ e ${\b}^{-}_{l}$ em termos das componentes ${\al}^{i}_{l}$ e ${\b}^{i}_{l}$, respectivamente, como segue:
\be
{\al}^{-}_{l}=\frac{1}{\sqrt{2\al^{'}}p^{+}}\sum _{n=-\infty}^{\infty}{\al}^{i}_{l}{\al}^{i}_{l-n} \ \ , \ \ {\b}^{-}_{\sqrt{2\al^{'}}}=\frac{1}{lp^{+}}\sum _{n=-\infty}^{\infty}{\b}^{i}_{l}{\b}^{i}_{l-n}.
\ee
Desta forma, a Hamiltoniana cl\'assica~(\ref{hlcg}) no calibre de cone de luz para a corda fechada \'e
\be
H_{lcg}=2\al^{'}p^{+}p^{-}=\sum _{n=-\infty}^{\infty}\le({\al}^{i}_{n}{\al}^{i}_{-n}+{\b}^{i}_{n}{\b}^{i}_{-n} \ri ) 
\ee 
e a express\~ao cl\'assica para o quadrado da massa da corda aberta \'e
\bea 
M^{2}&=&p^{\mu}p_{\mu} \non
&=&2p^{+}p^{-}-p^{i}p^{i} \non
&=&4 \pi T  \sum^{\infty}_{m=1} \le (\al^{i}_{m}\al^{i}_{-m}+{\b}^{i}_{m}{\b}^{i}_{-m} \ri ).
\eea

        Para quantizarmos a corda fechada, precisamos impor rela\c{c}\~oes de comuta\c{c}\~ao an\'alogas \`as  rela\c{c}\~oes (\ref{rclcg}). Ent\~ao, se substituirmos a componente $i$ da solu\c{c}\~ao para a corda fechada, equa\c{c}\~ao~(\ref{scf}), obtemos as seguintes rela\c{c}\~oes de comuta\c{c}\~ao entre os osciladores
$$ 
[\al^{i}_{n},\al^{j}_{m}]=[{\b}^{i}_{n},{\b}^{j}_{m}]=n\d^{ij}\d_{n+m} \ \ \ \ \mbox{e} \ \ \ \ [{\al}^{i}_{n},{\b}^{j}_{m}]=0,
$$
onde ${\al}^{i}_{-n}$ e ${\b}^{i}_{-n}$ para $n>0$ atuam como operadores de cria\c{c}\~ao e, ${\al}^{i}_{n}$ e ${\b}^{i}_{n}$  como operadores de destrui\c{c}\~ao. Como na quantiza\c{c}\~ao aparecem problemas de diverg\^encia, precisamos introduzir o ordenamento normal em todas as quantidades definidas anteriormente em termos dos  operadores de cria\c{c}\~ao e destrui\c{c}\~ao.
$$ 
{\al}^{-}_{l}=\frac{1}{\sqrt{2\al^{'}}p^{+}}\sum _{n=-\infty}^{\infty}:{\al}^{i}_{n}{\al}^{i}_{-n}: -a
$$
$$ 
{\b}^{-}_{l}=\frac{1}{\sqrt{2\al^{'}}p^{+}}\sum _{n=-\infty}^{\infty}:{\b}^{i}_{n}{\b}^{i}_{-n}: -a
$$  
$$ 
H= \sum _{n=-\infty}^{\infty}\le(:{\al}^{i}_{n}{\al}^{i}_{-n}:+:{\b}^{i}_{n}{\b}^{i}_{-n}: -2a \ri ) 
$$
$$
M^{2}=4 \pi T (N + \ti{N}-2a),
$$
onde 
$$
N=\sum _{n=1}^{\infty}{\al}^{i}_{-n}{\al}^{i}_{n} \ \ \ \ \ \ \ \ \mbox{e} \ \ \ \ \ \ \   \ti{N}=\sum _{n=1}^{\infty}{\b}^{i}_{-n}{\b}^{i}_{n}.
$$

          Como na teoria para a corda aberta, a exig\^encia de que no calibre de cone de luz a teoria seja invariante por Lorentz requer que $D=26$ e $a=1$. Para a corda fechada,  o estado de v\'acuo que satisfaz
\be
\al^{i}_{n}|0\rangle_{\al}|0\rangle_{\b} ={\b}^{i}_{n}|0\rangle_{\al}|0\rangle_{\b} =0, \ \ \ \ \ \ \ \ \ \ \  n > 0     
\ee
\'e um tachyon. O pr\'oximo estado permitido possui um modo de excita\c{c}\~ao para a direita balanceado por um modo de excita\c{c}\~ao para a esquerda,
\be
\al^{i}_{-1}{\b}^{i}_{-1}|0\rangle_{\al}|0\rangle_{\b},
\ee
e possui massa nula. Podemos tamb\'em tomar uma combina\c{c}\~ao de  estados que seja sim\'etrica e de tra\c{c}o nulo e que se transforme como uma part\'{\i}cula de spin 2 sem massa. Esta combina\c{c}\~ao de  estados \'e a seguinte
\be
\le ( \frac{1}{2}(\al^{i}_{-1}{\b}^{j}_{-1}+\al^{j}_{-1}{\b}^{i}_{-1})-\al^{i}_{-1}{\b}^{i}_{-1} \ri )|0\rangle_{\al}|0\rangle_{\b},
\ee
e a identificaremos com o gr\'aviton. O tra\c{c}o do estado escalar $\al^{i}_{-1}{\b}^{i}_{-1}|0\rangle_{\al}|0\rangle_{\b}$ \'e chamado de dilaton. Os estados restantes no espectro da corda fechada s\~ao massivos, $M^{2}> 0$.

        A corda aberta livre n\~ao cont\'em o gr\'aviton. Quando adicionamos intera\c{c}\~oes, uma corda aberta pode auto-interagir juntando  suas extremidades. Deste modo, uma teoria para a corda aberta com intera\c{c}\~ao cont\'em necessariamente um setor de corda fechada, e deste modo, o gr\'aviton. 



\section{Din\^amica de Campos T\'ermicos}
${}$
        Geralmente, n\~ao encontramos na natureza sistemas completamente isolados, mas sim em contato no m\'{\i}nimo com um reservat\'orio t\'ermico (estes sistemas podem estar tamb\'em em contato com um reservat\'orio de part\'{\i}culas). Deste fato, surge a necessidade de termos uma teoria de campos \`a temperatura finita para descrevermos tais sistemas.

        Existem diversos formalismos para se introduzir temperatura em teorias de campos. Aqui, em particular, apresentaremos o formalismo desenvolvido inicialmente por Umezawa e Takahashi~{\cite{ut}} conhecido como {\it Din\^amica de Campos T\'ermicos } (DCT).  Neste formalismo \'e definido um v\'acuo t\'ermico, $|0(\b) \ra $, tal que o valor esperado neste v\'acuo de um operador Hermitiano $A$ coincida com sua m\'edia estat\'{\i}stica
\be
\la A \ra=Z^{-1}(\b)tr\;[e^{-\b {\cal{H}}}A]=\la 0(\b)|A|0(\b) \ra \label{0},
\ee
onde ${\cal{H}}=H-\mu N$, $Z(\b)=tr\;[e^{-\b {\cal{H}}}]$ e $\b=\frac {1}{{k}_{B}T}$ sendo $H$ a Hamiltoniana total, $\mu$ o potencial qu\'{\i}mico, $N$ o n\'umero de part\'{\i}culas e ${k}_B$ a constante de Boltzmann. Esta id\'eia foi inspirada na proposta de Matsubara~{\cite{mt}} de que,  a m\'edia estat\'{\i}stica de um \\operador $\la A \ra$ tem propriedades semelhantes \`as do valor esperado deste operador no v\'acuo $\la 0|A|0 \ra$ em teoria qu\^antica de campos. Como veremos, estados de v\'acuo em diferentes temperaturas s\~ao conectados uns aos outros atrav\'es de uma transforma\c{c}\~ao de Bogoliubov. 
       
\subsection{Din\^amica de Campos T\'ermicos - Formalismo Can\^onico}
${}$
        A id\'eia central em DCT \'e expressar m\'edias estat\'{\i}sticas de uma vari\'avel din\^amica $A$ como o valor esperado deste operador em um v\'acuo dependente da temperatura {\cite{ut,hq,hu}}. Sendo assim, devemos construir tal v\'acuo de modo que  a equa\c{c}\~ao~({\ref{0}}) seja satisfeita para uma vari\'avel din\^amica arbitr\'aria $A$ 
\be
 \la 0(\b)|A|0(\b) \ra=Z^{-1}(\b) \sum_{n}\la n|A|n \ra e^{-\b {\o}_{n}},   \label{1111}
\ee
onde, por simplicidade, iremos supor que os autovalores de ${\cal {H}}$ s\~ao discretos e sua equa\c{c}\~ao de autovalor \'e dada por ${\cal{H}}|n \ra =\o_{n}|n\ra$. Os auto-estados $|n \ra $ s\~ao ortonormalizados de modo que $\la n|m \ra=\d_{nm}$.

        Expandindo o estado de v\'acuo $|0(\b) \ra $ em termos de uma base $|n \ra$ do espa\c{c}o de Hilbert na forma
\be
|0(\b) \ra=\sum _{n}|n \ra \la n |0(\b) \ra=\sum _{n}f_{n}(\b)|n \ra ,    \label{2a}
\ee
onde usamos a completeza do nosso espa\c{c}o de Hilbert, e substituindo no lado esquerdo da equa\c{c}\~ao~(\ref{1111}), obtemos
$$
\sum_{n}f^{\ast}_{n}(\b) \la n|A \sum_{m}f_{m}(\b)|m \ra = Z^{-1}(\b) \sum_{n} \langle n|A|n\rangle e^{-\b \o_{n}}, 
$$
$$
\sum_{n,m}f^{\ast}_{n}(\b)f_{m}(\b) \langle n|A|m \ra =Z^{-1}(\b) \sum_{n} \langle n|A|n \ra e^{-\b \o_{n}} , 
$$
$$
\sum_{n,m}f^{\ast}_{n}(\b)f_{m}(\b) \langle n|A|m \ra =Z^{-1}(\b) \sum_{n,m} \d_{nm} \langle n|A|m \ra e^{-\b \o_{n}} , 
$$
\be
f^{\ast}_{n}(\b)f_{m}(\b) = Z^{-1}(\b)e^{-\b \o_{n}} \d_{nm}. \label{3a}
\ee
Esta rela\c{c}\~ao n\~ao faz sentido, pois os coeficientes da expans\~ao (\ref{2a}) s\~ao escalares complexos e estes n\~ao possuem a propriedade de ortogonalidade entre si como expressa esta rela\c{c}\~ao. Portanto, se mantivermos o espa\c{c}o de Hilbert original, tal estado ($|0(\b) \ra$) n\~ao poder\'a ser constru\'{\i}do. No entanto, a rela\c{c}\~ao ({\ref{3a}}) mostra uma estrutura vetorial (ortogonalidade) e, portanto, os coeficientes da expans\~ao devem ser vetores. Em outras palavras, o estado $|0(\b) \ra $ dever\'a ser um vetor expandido por $|n \ra $ e $f_{n}(\b)$.

        Para constru\'{\i}rmos um estado que satisfa\c{c}a a equa\c{c}\~ao~({\ref{1111}}) devemos dobrar os graus de liberdade da teoria. Isto \'e feito introduzindo um sistema din\^amico n\~ao f\'{\i}sico representado por um estado $|\w{m} \ra $ (chamado de espa\c{c}o til) ortogonal e com as mesmas caracter\'{\i}sticas do sistema f\'{\i}sico original, por exemplo, energia, n\'umero de part\'{\i}culas. Denotaremos, ent\~ao, todas as quantidades associadas a este sistema n\~ao f\'{\i}sico com um til. A Hamiltoniana deste sistema $ \widetilde{\cal{H}}$ e  o espa\c{c}o dos vetores de estado $| \widetilde n \ra$  obedecem as seguintes rela\c{c}\~oes  
$$
 \widetilde {\cal{H}}| \widetilde n \rangle =\o_{n}| \widetilde n \ra, \ \ \ \ \ \ \ \ \ \ \langle  \widetilde n| \widetilde m \ra =\d_{mn},
$$ 
onde, como definimos anteriormente, $\o_{n}$ \'e o mesmo do sistema f\'{\i}sico. 

       O vetor de estado do sistema total \'e constru\'{\i}do a partir do produto direto dos vetores de estado de cada um dos sistemas, f\'{\i}sico e n\~ao f\'{\i}sico, $|n, \widetilde m \ra=|n\rangle \otimes | \widetilde m \ra$. Desta forma, os elementos de matriz dos operadores $A$ e $ \widetilde A$ s\~ao dados respectivamente por
\be
 \langle  \widetilde m,n|A|n^{'}, \widetilde m^{'} \ra = \langle n|A|n^{'} \ra \d_{mm{'}}, \ \ \langle  \widetilde m,n| \widetilde A|n^{'}, \widetilde m^{'} \ra = \langle  \widetilde m| \widetilde A| \widetilde m^{'} \ra \d_{nn{'}}.
\ee
Definindo o coeficiente vetorial da expans\~ao~({\ref{2a}}) como 
\be
f_{n}(\b)=e^{-\b \o_{n}/2}Z^{-1/2}(\b)| \widetilde n \ra,   \label{def}
\ee
podemos verificar que a rela\c{c}\~ao~{(\ref{3a})} \'e satisfeita:
\bea
f_{n}^{\ast}(\b)f_{m}(\b)&=&e^{-\b \o_{n}/2}Z^{-1/2}(\b) \langle  \widetilde n|e^{-\b \o_{m}/2}Z^{-1/2}(\b)| \widetilde m\rangle \non
&=&e^{-{\frac {\b}{2}} (\o_{n}+\o_{m})}Z^{-1}(\b) \langle  \widetilde n| \widetilde m\rangle \non
&=&Z^{-1}e^{-{\b}\o_{n}}(\b) \d_{nm}. \nonumber
\eea 
Ainda, usando a defini\c{c}\~ao~{(\ref{def})}, podemos construir o estado de v\'acuo t\'ermico a partir da equa\c{c}\~ao~{(\ref{2a})} como segue:
\be
|0(\b) \ra = \sum_{n}e^{-\b \o_{n}/2}Z^{-1/2}(\b) |\widetilde n \ra \otimes | n\rangle = \sum_{n}e^{-\b \o_{n}/2}Z^{-1/2}(\b)|n,\widetilde n \ra. \label{vab}
\ee

        A express\~ao da fun\c{c}\~ao de parti\c{c}\~ao $Z(\b)=\sum_{n}e^{-\b \o_{n}} \langle n|n\rangle=tr\;[e^{-\b {\cal{H}}}]$ pode ser obtida exigindo-se que o estado $ |0(\b) \ra$ satisfa\c{c}a a condi\c{c}\~ao  de normaliza\c{c}\~ao $\langle 0(\b)|0(\b) \ra=1$ como segue 
 \bea
 \langle 0(\b)|0(\b) \ra&=& \sum_{n}e^{-\b \o_{n}/2}Z^{-1/2}(\b) \langle  \widetilde n,n|\sum_{m}e^{-\b \o_{m}/2}Z^{-1/2}(\b)|m, \widetilde m\rangle \non
&=&\sum_{n,m}e^{-{\frac{\b}{2}}( \o_{n}+\o_{m})}Z^{-1}(\b) \langle  \widetilde n,n|m, \widetilde m\rangle \non
&=&\sum_{n,m}e^{-{\frac{\b}{2}}( \o_{n}+\o_{m})}Z^{-1}(\b) \langle n|m\rangle \d_{nm} \non
&=&\sum_{n}e^{-\b \o_{n}}Z^{-1}(\b) \langle n|n \ra =1 \nonumber.
\eea
Obtemos ent\~ao
\be
Z(\b)=\sum_{n}e^{-\b \o_{n}} \langle n|n\rangle=tr\;[e^{-\b {\cal{H}}}].
\ee

        Verificaremos, agora, que o valor m\'edio de um operador $A$ no estado de v\'acuo t\'ermico obtido, equa\c{c}\~ao~(\ref{vab}), est\'a de acordo com a hip\'otese de Umezawa, eq. (\ref{1111})
\bea
 \langle 0(\b)|A|0(\b) \ra &=&Z^{-1/2}(\b) \sum_{n}e^{-\b \o_{n}/2} \langle  \widetilde{n},n|AZ^{-1/2}(\b)\sum_{m} e^{-\b \o_{m}/2}|m, \widetilde{m}\rangle \non
&=&Z^{-1}(\b) \sum_{n,m}e^{-{\frac {\b}{2}} (\o_{n}+\o_{m})} \langle n|A|m\ra \d_{nm} \non
&=&Z^{-1}(\b) \sum_{n}e^{-\b \o_{n}} \langle n|A|n \ra . \nonumber
\eea

        Podemos notar que na constru\c{c}\~ao dos estados $|0(\b) \ra $ aparece o produto direto dos vetores $|n \ra$ e $| \widetilde n\ra$,  e como verificado acima, o vetor $| \widetilde n \ra$  seleciona o elemento diagonal do observ\'avel $A$.

        Portanto, mostramos que podemos introduzir um estado de v\'acuo dependente da temperatura tal que a m\'edia estat\'{\i}stica de qualquer operador pode ser identificado com o valor esperado do operador neste estado. Mas, isto implica obrigatoriamente em uma duplica\c{c}\~ao do espa\c{c}o de Hilbert. Por outro lado, a vantagem se encontra no fato de que podemos usar as t\'ecnicas de uma teoria de campos \`a temperatura nula nesta teoria \`a temperatura finita.   Usaremos agora um sistema de b\'osons livres como exemplo para ilustrar como construir um conjunto completo de vetores ortonormais ao qual pertence o vetor $|0(\b) \ra$. Veremos tamb\'em como construir o espa\c{c}o de Fock usando este exemplo.  

\subsection{Ensemble de B\'osons Livres com Freq\"u\^encia $\o$}
${}$

        Vamos agora detalhar o formalismo da DCT atrav\'es de um exemplo simples. Consideremos um oscilador bos\^onico de freq\"u\^encia $\o$ descrito por um par de vari\'aveis complementares $a$ e $a^{\a}$ n\~ao Hermitianas,  cuja  Hamiltoniana \'e
\be
{\bo{H}}=\o a^{\dag}a,
\ee
com $a$ e $a^{\a}$ satisfazendo as seguintes rela\c{c}\~oes de comuta\c{c}\~ao
\be
[a,a^{\a}]=aa^{\a}-a^{\a}a=1 \ \ \ \ \mbox{e} \ \ \ \ [a,a]=[a^{\a},a^{\a}]=0.  \label{com1}
\ee

       O espa\c{c}o de Hilbert \'e de dimens\~ao infinita e os auto-estados de $\bo{H}$ podem ser constru\'{\i}dos baseados nas rela\c{c}\~oes de comuta\c{c}\~ao (\ref{com1}). Os estados ser\~ao rotulados por um \'{\i}ndice inteiro $n$: 
\be
{\bo{H}}|n \ra =\o_{n}|n\rangle \ \ \ \ \ \ \ \ \ \ \ \ \ \ \ \ \  \ \ \ \ \ \ \ \ \ \ \ n=0,1,2, \dots, \infty.    
\ee 
 Partindo do estado $|n \ra$, encontramos que $a^{\a}|n \ra$ e  $a|n \ra$ s\~ao tamb\'em auto-estados da Hamiltoniana, mas com o autovalor da energia deslocado de um quantum, ou seja,
\be
{\bo{H}}a^{\a}|n \ra=\o a^{\a}aa^{\a}|n \ra=\o a^{\a}(1+a^{\a}a)|n \ra=(\o_{n}+\o )a^{\a}|n \ra
\ee
e
\be
{\bo{H}}a|n\ra =(\o_{n}-\o )a|n\ra .
\ee

        O estado fundamental $|0\ra $ \'e definido como sendo o estado de menor energia poss\'{\i}vel, que implica em 
$$
a|0\ra =0.
$$
Qualquer auto-estado do oscilador pode ser constru\'{\i}do atrav\'es de aplica\c{c}\~oes sucessivas do operador de cria\c{c}\~ao, $a^{\a}$, no estado fundamental, ou seja
$$
|n\ra =\frac{1}{\sqrt{n!}}(a^{\a})^{n}|0\ra .
$$
Portanto, o conjunto de estados ortonormais \'e
$$
|0 \ra , \ \ \ \frac{1}{\sqrt{1!}}a^{\a}|0\ra=|{1} \ra , \ \ \  \frac{1}{\sqrt{2!}}a^{\a}a^{\a}|0\ra =|{2} \ra , \ \dots, \ \frac{1}{\sqrt{n!}}(a^{\a})^{n}|0\ra=|{n} \ra \dots \ .
$$

        Mostraremos agora como construir o vetor de estado $|0(\b)\ra $ para este sistema. De acordo com a nossa discuss\~ao geral da DCT feita na se\c{c}\~ao anterior, devemos introduzir um sistema fict\'{\i}cio, id\^entico ao sistema original, descrito pela Hamiltoniana 
$$ 
 \widetilde{\bo {H}}=\o  \widetilde a^{\a}  \widetilde a,
$$
onde $ \widetilde a$ e ${ \widetilde a}^{\a}$ satisfazem as mesmas rela\c{c}\~oes de comuta\c{c}\~ao expressa pela equa\c{c}\~ao (\ref{com1})
$$
[ \widetilde a,  \widetilde a^{\a}]=1 \ \ \ \ \ \mbox{e} \ \ \ \ \  [ \widetilde a, \widetilde a]=[ \widetilde a^{\a}, \widetilde a^{\a}]=0. 
$$
Assumimos ainda que todas as vari\'aveis complementares do sistema real comutam com as vari\'aveis do sistema fict\'{\i}cio, ou seja, 
$$ 
[a, \widetilde a]=[a^{\a}, \widetilde a^{\a}]=[a, \widetilde a^{\a}]=[a^{\a}, \widetilde a]=0.
$$
Analogamente ao sistema f\'{\i}sico, teremos que os vetores de estado ortogonais s\~ao  
$$
| \widetilde 0\ra , \ \ \ \frac{1}{\sqrt{1!}} \widetilde a^{\a}| \widetilde 0\ra=| \widetilde{1} \ra , \ \ \ \frac{1}{\sqrt{2!}} \widetilde a^{\a} \widetilde a^{\a}| \widetilde 0\ra=| \widetilde{n} \ra , \ \dots ,\ \frac{1}{\sqrt{n!}}( \widetilde a^{\a})^{n}| \widetilde 0\rangle = | \widetilde{n} \ra \dots
$$
e, como o sistema fict\'{\i}cio deve possuir as mesmas caracter\'{\i}sticas do sistema f\'{\i}sico, os autovalores da Hamiltoniana $\widetilde{\bo{H}}$ s\~ao $n \o $, onde $n=0,1,2,\dots, \infty$ . Os vetores de estado ortogonais do sistema composto (produto direto dos estados $|n\ra $ e $| \widetilde n \ra $ ) ser\~ao
\be
|0\ra \! \ra, \ \ \ a^{\a}|0 \ra \! \ra, \ \ \   \widetilde a^{\a}|0 \ra \! \ra, \ \ \   a^{\a} \widetilde a^{\a}|0 \ra \! \ra,\ \ \  \frac{1}{n!}(a^{\a})^{n}( \widetilde a^{\a})^{n}|0 \ra \! \ra, \dots  \label{efd}
\ee
onde estamos denotando
$$
|0, \widetilde{0}\ra =|0 \ra \otimes \w {|0 \ra} = |0\ra \! \ra.
$$
Assim sendo, podemos construir o estado $|0(\b)\ra $, como indicado pela equa\c{c}\~ao (\ref{vab}), da seguinte forma
\bea
|0(\b)\ra &=& \sum_{n}e^{-\b \o_{n}/2}Z^{-1/2}(\b)|n, \widetilde n\rangle \non
&=&\sum_{n}e^{-\b \o_{n}/2}Z^{-1/2}(\b) \frac{1}{n!}(a^{\a})^{n}( \widetilde a^{\a})^{n}|0 \ra \! \ra. \label{vtb}
\eea

        Podemos ainda obter a fun\c{c}\~ao de parti\c{c}\~ao para este sistema, exigindo a ortonormaliza\c{c}\~ao do estado $|0(\b)\ra $ encontrado
\bea
 \langle 0(\b)|0(\b)\ra &=& Z^{-1}(\b) \sum_{m,n} \langle \! \langle 0| \frac{1}{n!}( \widetilde a)^{n}(a)^{n}e^{-\b (\o_{n}+\o_{m})/2} \frac{1}{m!}(a^{\a})^{m}( \widetilde a^{\a})^{m}|0 \ra \! \ra \non
&=& Z^{-1}(\b) \sum_{m,n}\langle \! \langle 0| \frac{1}{n!}( \widetilde a)^{n}(a)^{n}e^{-\b (\o_{n}+\o_{m})/2} \frac{1}{m!}(a^{\a})^{m}( \widetilde a^{\a})^{m}|0 \ra \! \ra \non
&=& Z^{-1}(\b) \sum_{m,n}e^{-\b (\o_{n}+\o_{m})/2} \langle   \widetilde{n},n |m, \widetilde{m} \rangle \non
&=& Z^{-1}(\b) \sum_{m,n} e^{-\b (\o_{n}+\o_{m})/2}  \langle  n |m \rangle \d_{m,n}\non
&=& Z^{-1}(\b) \sum_{n}e^{-\b \o n} \langle  n |n \rangle  \non
&=&Z^{-1}(\b) \sum_{n}e^{({-\b \o})^{n}} \langle  n |n \rangle  \non
&=&Z^{-1}(\b) \frac{1}{1-e^{-\b \o}}=1,  \nonumber
\eea
de forma que
$$
Z(\b)= \frac{1}{1-e^{-\b \o}} .
$$
Usando esta fun\c{c}\~ao de parti\c{c}\~ao, e rearranjando os termos, podemos reescrever o estado de v\'acuo t\'ermico, equa\c{c}\~ao (\ref{vtb}), na forma 
\bea
|0(\b)\ra &=& \sqrt{1-e^{-\b \o}}\sum_{n=0}^{\infty}e^{-\frac{\b}{2}n\o} \frac{1}{n!} \le(a^{\a} \widetilde a^{\a} \ri)^{n}|0 \ra \! \ra \non
&=& \sqrt{1-e^{-\b \o}}\sum_{n=0}^{\infty}e^{\le(-\frac{\b}{2}\o \ri )^{n}} \frac{1}{n!} \le(a^{\a} \widetilde a^{\a} \ri)^{n}|0 \ra \! \ra \non
&=& \sqrt{1-e^{-\b \o}}\sum_{n=0}^{\infty}\frac{1}{n!} \le(e^{-\frac{\b}{2}\o} a^{\a} \widetilde a^{\a} \ri)^{n}|0 \ra \! \ra \non
&=& \sqrt{1-e^{-\b \o}} \exp \le(e^{-\frac{\b}{2}\o} a^{\a} \widetilde a^{\a} \ri)|0 \ra \! \ra . \nonumber
\eea

        Mostraremos agora que o estado de v\'acuo t\'ermico $|0(\b)\ra $ pode ser mapeado no v\'acuo total $|0 \ra \! \ra $ em $T=0$ por uma dada transforma\c{c}\~ao de Bogoliubov. Para isto definimos
\be
u(\b)=(1-e^{-\b \o})^{-\frac{1}{2}}=\sqrt{1+f_{B}(\o)}=\cosh\th(\b)  \label{u11}
\ee
e
\be
\v(\b)=(e^{\b \o}-1)^{-\frac{1}{2}}=\sqrt{f_{B}(\o)}=\sinh\th(\b)  \label{v11}
\ee
onde $f_{B}$ \'e a distribui\c{c}\~ao de Bose {\cite{rk}}.
Podemos ent\~ao verificar que 
\be
u^{2}(\b)-\v^{2}(\b)=1.  \label{quadrados}
\ee
Usando estas defini\c{c}\~oes, podemos escrever o  vetor de estado $|0(\b)\ra $ como
\bea
|0(\b)\ra &=& u(\b)^{-1} \exp \le [e^{-\b \o /2} \frac{ (1-e^{-\b \o})^{1/2}}{(1-e^{-\b \o})^{1/2}} a^{\a} \widetilde a^{\a} \ri ]|0 \ra \! \ra  \non
&=& u(\b)^{-1} \exp \le[ \le (\frac{1}{(e^{\b \o}-1)} \ri)^{1/2}
(1-e^{-\b \o})^{1/2} a^{\a} \widetilde a^{\a}\ri ]|0 \ra \! \ra  \non
&=& u(\b)^{-1} \exp \le(\frac{\v (\b)}{u(\b)} a^{\a} \widetilde a^{\a}\ri )|0 \ra \! \ra , \nonumber
\eea
onde usamos as equa\c{c}\~oes para $u(\b)$ e $v(\b)$ definidas anteriormente. 

        Podemos ainda definir o seguinte operador Hermitiano 
\be
G(\th)=G(\th)^{\a}=-i\th(\b)( \widetilde{a}a- {a}^{\a}\widetilde{a}^{\a}),  \label{gtb}
\ee 
onde $\th(\b)$ \'e um par\^ametro real que depende da temperatura definido em (\ref{u11}) e (\ref{v11}), tal que o operador  $U(\th)=e^{-iG(\th)}$ seja unit\'ario. Este operador gera o v\'acuo t\'ermico a partir do espa\c{c}o dobrado como segue 
\be
|0(\b)\ra =e^{-iG(\th)}|0\ra \! \ra .  \label{tbv}
\ee
A transforma\c{c}\~ao que liga o estado de v\'acuo t\'ermico ao v\'acuo duplicado \`a tempe-\\ratura nula \'e chamada de transforma\c{c}\~ao de Bogoliubov, equa\c{c}\~ao~(\ref{tbv}), e o gerador desta transforma\c{c}\~ao, equa\c{c}\~ao~(\ref{gtb}), \'e chamado operador de Bogoliubov. Esta afirma\c{c}\~ao de que o estado $|0(\b)\ra$  pode ser colocado na forma (\ref{tbv}) est\'a demonstrado no ap\^endice A.

      Dadas as seguintes rela\c{c}\~oes de comuta\c{c}\~ao 
$$
[G,a]=-i\th (\b) \widetilde{a}^{\a}, \ \ \ \ [G, \widetilde{a}]=-i\th (\b){a}^{\a}, \ \ \ \ [G,a^{\a}]=-i\th (\b) \widetilde{a}, \ \ \ \ [G, \widetilde{a}^{\a}]=-i\th (\b)a
$$
podemos ver que os operadores de destrui\c{c}\~ao na temperatura finita s\~ao tamb\'em gerados por estas transforma\c{c}\~oes unit\'arias, transforma\c{c}\~oes de Bogoliubov, como segue
\bea
a(\b)&=&e^{-iG}ae^{iG} \non
&=&a-\fr{i}{1!}[G,a]+\fr{(i)^{2}}{2!}[G,[G,a]]-\fr{(i)^{3}}{3!}[G,[G,[G,a]]]+\fr{(i)^{4}}{4!}[G,[G,[G,[G,a]]]] + ....\non
&=& \le( \sum_{n=0}^{\infty}\frac{{\th}^{2n}(\b)}{2n!} \ri)a - \le ( \sum_{n=0}^{\infty}\frac{{\th}^{2n+1}(\b)}{(2n+1)!} \ri) \widetilde{a}^{\a} \non
&=& \cosh{\th(\b)}a-\sinh{\th(\b)} \widetilde{a}^{\a} \non
&=&u(\b)a-\v(\b) \widetilde{a}^{\a} \label{aa}
\eea
e
\bea
 \widetilde{a}(\b)&=&e^{-iG} \widetilde{a}e^{iG} \non
&=&a-\fr{i}{1!}[G, \widetilde{a}]+\fr{(i)^{2}}{2!}[G,[G, \widetilde{a}]]-\fr{(i)^{3}}{3!}[G,[G,[G, \widetilde{a}]]]+\fr{(i)^{4}}{4!}[G,[G,[G,[G, \widetilde{a}]]]] + ....\non
&=& \le( \sum_{n=0}^{\infty}\frac{{\th}^{2n}(\b)}{2n!} \ri) \widetilde{a} - \le ( \sum_{n=0}^{\infty}\frac{{\th}^{2n+1}(\b)}{(2n+1)!} \ri) \widetilde{a}^{\a} \non
&=& \cosh{\th(\b)}\widetilde{a}-\sinh{\th(\b)}{a}^{\a} \non
&=&u(\b) \widetilde{a}-\v(\b){a}^{\a} \label{at}
\eea
e seus Hermitianos conjugados (operadores de cria\c{c}\~ao) s\~ao respectivamente
\be
a^{\a}(\b)=e^{-iG}ae^{iG}=u(\b)a^{\a}-\v(\b) \widetilde{a} , \label{atd}
\ee
\be
 \widetilde{a}^{\a}(\b)=e^{-iG} \widetilde{a}^{\a}e^{iG}=u(\b) \widetilde{a}^{\a}-\v(\b){a}. \label{atd1}
\ee

        Abaixo est\~ao expressos  $a$ e $a^{\a}$ em termos dos operadores t\'ermicos, os quais foram obtidos simplesmente invertendo as equa\c{c}\~oes (\ref{aa}), (\ref{at}), (\ref{atd}) e (\ref{atd1}) 
$$
a=u(\b)a(\b)+\v(\b) \widetilde{a}^{\a}(\b)\ ,\ a^{\a}=u(\b)a^{\a}(\b)+\v(\b) \widetilde{a}(\b)
$$
$$
 \widetilde{a}=u(\b) \widetilde{a}(\b)+\v(\b){a}^{\a}(\b) \ , \
 \widetilde{a}^{\a}=u(\b) \widetilde{a}^{\a}(\b)+\v(\b){a}(\b).
$$

        Os operadores t\'ermicos podem tamb\'em ser obtidos de um modo geral se definirmos o seguinte dubleto como
\be 
A=\le(\begin{array}{c} a \\ \w{a}^{\a} \end{array}  \ri ),
\ee
ent\~ao podemos escrever
\bea
A(\b)&=&\le(\begin{array}{c} a(\b) \\ \w{a}^{\a}(\b) \end{array}  \ri )=U(\th)AU^{\a}(\th)=\overline{U}(\th)A \non
&=&\fourmat{\cosh {\th(\b)}}{-\sinh {\th(\b)}}{-\sinh {\th(\b)}}{\cosh {\th(\b)}} 
\le(\begin{array}{c} a \\ \w{a}^{\a} \end{array}  \ri ),
\eea
com $\th(\b)$ dado por (\ref{u11}) e (\ref{v11}) e 
\be
\overline{U}(\th)=\fourmat{\cosh {\th(\b)}}{-\sinh {\th(\b)}}{-\sinh {\th(\b)}}{\cosh {\th(\b)}}=\fourmat{u(\b)}{-\v(\b)}{-\v(\b)}{u(\b)}.
\ee
 
        A m\'edia estat\'{\i}stica do operador n\'umero $N=a^{\a}a$ pode ser calculada usando os operadores dependentes da temperatura, isto \'e,
\bea
 \langle 0(\b)|{a}^{\a}a|0(\b)\ra &=& \langle 0(\b)| \le( u(\b)a^{\a}(\b)+\v(\b) \widetilde{a}(\b) \ri) \le( u(\b)a(\b)+\v(\b) \widetilde{a}^{\a}(\b) \ri)|0(\b)\rangle \non
&=& \langle 0(\b)|u^{2}(\b){a}^{\a}(\b)a(\b)+\v^{2}(\b) \widetilde{a}(\b) \widetilde{a}^{\a}(\b) \non
&+&u(\b)\v(\b) \le ({a}^{\a}(\b) \widetilde{a}^{\a}(\b)+ a(\b) \widetilde{a}(\b) \ri)|0(\b)\rangle \non
&=&\v^{2}(\b) \langle 0(\b)|0(\b)\ra =\v^{2}(\b)=\frac{1}{e^{\b \o}-1}=f_{B}(\b) \label{n}
\eea
que \'e exatamente a distribui\c{c}\~ao de Bose que aparece nas equa\c{c}\~oes~({\ref{u11}}) e ({\ref{v11}}). Vemos que a constru\c{c}\~ao de Umezawa nos conduz a um resultado conhecido da mec\^anica estat\'{\i}stica. Mais adiante, obteremos outros resultados, tais como a entropia e a energia livre de Helmholtz.

        Definidos os operadores de cria\c{c}\~ao e destrui\c{c}\~ao t\'ermicos, podemos mostrar que o vetor de estado $|0(\b)\ra $ \'e realmente o v\'acuo t\'ermico, pois 
\be
a(\b)|0(\b)\ra =e^{-iG}{a}e^{iG}|0(\b)\ra =e^{-iG}{a}e^{iG}e^{-iG}|0\ra \! \ra=e^{-iG}{a}|0\ra \! \ra=0 \label{av}
\ee
e
\be
 \widetilde{a}(\b)|0(\b)\ra =e^{-iG} \widetilde{a}e^{iG}|0(\b)\ra =e^{-iG} \widetilde{a}e^{iG}e^{-iG}|0\ra \! \ra=e^{-iG} \widetilde{a}|0\ra \! \ra=0,\label{avt}
\ee
onde usamos as transforma\c{c}\~oes de Bogoliubov,  equa\c{c}\~oes~({\ref{tbv}}), ({\ref{aa}}) e ({\ref{at}}), para obtermos o v\'acuo e os operadores de destrui\c{c}\~ao t\'ermicos em termos dos respectivos operadores \`a temperatura nula. Verificamos portanto que o estado $|0(\b)\ra$ \'e aniquilado por operadores de destrui\c{c}\~ao t\'ermicos. Podemos ent\~ao pensar neste como sendo um estado de v\'acuo t\'ermico. Finalmente, podemos construir o espa\c{c}o t\'ermico de Hilbert atrav\'es de aplica\c{c}\~oes sucessivas dos operadores de cria\c{c}\~ao t\'ermicos no v\'acuo t\'ermico. Das equa\c{c}\~oes (\ref{u11}) e (\ref{v11}), podemos ainda obter
\be
\tanh \th (\b)=e^{-\b \o /2}
\ee
de modo que (\ref{av}) e (\ref{avt}) possam ser escritas, respectivamente, da seguinte forma:
\be
a|0(\b)\ra = e^{-\b \o /2}{\w{a}}^{\a}|0(\b)\ra=e^{-\b{\hat{H}} /2}{\w{a}}^{\a}|0(\b)\ra \label{vou}
\ee
e
\be
\w{a}|0(\b)\ra = e^{-\b \o /2}{a}^{\a}|0(\b)\ra=e^{-\b{\hat{H}} /2}{a}^{\a}|0(\b)\ra.\label{vou1}
\ee  
Juntas, elas conduzem as regras de conjuga\c{c}\~ao til que ser\~ao apresentadas adiante. 

No entanto, tais  vetores de estados n\~ao s\~ao auto-estados das Hamiltonianas dos sistema f\'{\i}sico e fict\'{\i}cio, entretanto podemos construir uma nova Hamiltoniana $(\hat{H})$ envolvendo $H$ e $ \widetilde{H}$ da qual estes s\~ao auto-estados. Usando as equa\c{c}\~oes que nos fornecem os operadores em $T=0$ em fun\c{c}\~ao dos operadores em $T \neq 0$ e a equa\c{c}\~ao~(\ref{quadrados}) podemos escrever esta nova Hamiltoniana como
\be
\hat{H}= H- \widetilde{H}=\o \le(a^{\a}a- \widetilde{a}^{\a} \widetilde{a} \ri)=\o \le( a^{\a}(\b)a(\b)- \widetilde{a}^{\a}(\b) \widetilde{a}(\b) \ri ).  \label{ht}
\ee
Isto mostra que a Hamiltoniana total \'e invariante por transforma\c{c}\~oes de Bogoliubov, logo, o gerador \'e conservado
\be 
[G,H]=0.         \label{gerador}
\ee
Podemos pensar nesta Hamiltoniana como sendo a que governa a din\^amica do sistema combinado (real e fict\'{\i}cio).    

\section{Formalismo Lagrangeano para os Campos Livres}
${}$

       Na se\c{c}\~ao anterior, estudamos um sistema quanto-mec\^anico simples \`a tempera-\\tura finita no formalismo da DCT. Agora, iremos analisar uma teoria de campos \`a temperatura finita neste formalismo. Como exemplo, vamos considerar um campo de  Schr\"{o}dinger em $3 + 1 $ dimens\~oes.  Para isto, faremos uma breve exposi\c{c}\~ao dos conceitos de campos cl\'assicos e quantiza\c{c}\~ao can\^onica dos campos usando como exemplo o campo de Schr\"{o}dinger. 

       Consideraremos um sistema descrito pela densidade de lagrangeana do campo de Schr\"{o}dinger {\cite{wgr}} livre
\be
{\cal{L}}(\ve{x},t)=i\psi^{\ast}\dot{\psi}-\frac{1}{2m}{\boldmath{\nabla}}{\psi}^{\ast}{\boldmath{\nabla}}{\psi}.
\ee
Neste caso, devemos considerar as fun\c{c}\~oes $\psi({\ve{x}},t)$ e $\psi^{\ast}({\ve{x}},t)$ como campos independentes. Fazendo a varia\c{c}\~ao da a\c{c}\~ao, obteremos duas equa\c{c}\~oes independentes de Euler-Lagrange

\be
\frac{\6 {\cal{L}}}{\6 {\psi}}-{\bo {\n}}\frac{\6{\cal{L}}}{\6({\bo {\n}\psi})} - \frac{\6}{\6 t}\frac{\6 {\cal{L}}}{\dot{\psi}}=0\ \ \ , \ \ \ \frac{\6 {\cal{L}}}{\6 {\psi}^{\ast}}-{\bo {\n}}\frac{\6{\cal{L}}}{\6({\bo {\n}\psi}^{\ast})} - \frac{\6}{\6 t}\frac{\6 {\cal{L}}}{\dot{\psi}^{\ast}}=0.
\ee
Podemos verificar que estas equa\c{c}\~oes de Euler-Lagrange nos conduzem \`as seguintes equa\c{c}\~oes de movimento
\be
i\frac{\6 \psi}{\6 t}=-\fr{1}{2m}{\bo {\n}}^{2} \psi \ \ \ \ \mbox{e} \ \ \ \  i\frac{\6 {\psi}^{\ast}}{\6 t}=\fr{1}{2m}{\bo {\n}}^{2} \psi^{\ast},
\ee
que s\~ao respectivamente a equa\c{c}\~ao de Schr\"{o}dinger livre e seu complexo conjugado. Para introduzirmos o formalismo Hamiltoniano, ser\'a necess\'ario introduzirmos os campos canonicamente conjugados associados aos campos $\psi ({\ve{x}},t)$ e $\psi^{\ast}({\ve{x}},t) $ que s\~ao respectivamente definidos por
\be
\p({\ve{x}},t)=\frac {\6{\cal{L}}}{\6 {\dot{\psi}}}=i\psi^{\ast}({\ve{x}},t) \ \ \ 
\mbox{e} \ \ \ \p^{\ast}({\ve{x}},t)=\fr{\6 {\cal{L}}}{\6 {\dot{\psi}}^{\ast}}=0,
\ee
portanto, existem somente dois campos independentes $\psi({\ve{x}},t)$ e $\p({\ve{x}},t)$. A densidade de Hamiltoniana \'e
\be
{\cal{H}}=\p \dot{\psi}-{\cal{L}}=\frac{1}{2m}{\nabla}{\psi}^{\ast}{\nabla}{\psi}
\ee
que depois de uma integra\c{c}\~ao por partes nos conduz \`a Hamiltoniana
\be
H= \int d^{3}x {\cal{H}}=\int d^{3}x \psi^{\ast}({\ve{x}},t) \le (- \frac{1}{2m}{\boldmath{\nabla}}^{2} \ri ) \psi({\ve{x}},t).
\ee

        Os colchetes de Poisson a tempos iguais s\~ao
\be
\{ \psi({\ve{x}},t),\p({\ve{x}}{'},t) \} = \d^{3}({\ve{x}} - {\ve{x}}{'}),
\ee
\be
\{ \psi({\ve{x}},t),\psi({\ve{x}}{'},t) \} = \{ \p({\ve{x}},t),\p({\ve{x}}{'},t) \}=0.
\ee

        No processo de quantiza\c{c}\~ao can\^onica, os campos cl\'assicos $\psi({\ve{x}},t)$ e $\p({\ve {x}},t)$ s\~ao trocados pelos operadores $\hat{\psi}({\ve {x}},t)$ e $\hat{\p}({\ve {x}},t)= i \hat{\psi}^{\a}({\ve{x}},t)$, onde o campo complexo $\psi^{\ast}({\ve {x}},t)$ foi trocado pelo operador de campo Hermitiano adjunto $\hat{\psi}^{\a}({\ve {x}},t)$. Trocamos tamb\'em os par\^enteses de Poisson por comutadores de acordo com a rela\c{c}\~ao
\be
\{ \ \ , \ \ \} \longrightarrow \frac{1}{i}[ \ \  ,\ \  ].  \label{pc}
\ee
Desta forma, as rela\c{c}\~oes de comuta\c{c}\~ao a tempos iguais s\~ao
\be
[\hat{\psi}({\ve {x}},t),\hat{\psi}^{\a}({\ve {x}}{'},t) ] = \d^{3}({\ve {x}} - {\ve {x}}{'})
\ee
\be
[ \hat{\psi}({\ve {x}},t),\hat{\psi}({\ve {x}}{'},t)  ] =[ \hat{\psi}^{\a}({\ve {x}},t),\hat{\psi}^{\a}({\ve {x}}{'},t) ] =0.
\ee
Este procedimento \'e chamado  Princ\'{\i}pio da Correspond\^encia {\cite{gsw}}. A evolu\c{c}\~ao din\^amica dos operadores de campo \'e dada pelas equa\c{c}\~oes de movimento de Heisenberg
\be
i\dot{\hat{\psi}}=[{\hat{\psi}}, \hat{H}] \ \ \ \ \ \ \ \ \ \mbox{e} \ \ \ \ \ \ \ \ \  i\dot{\hat{\p}}=[{\hat{\p}}, \hat{H}].   
\ee

          Usando as  rela\c{c}\~oes de comuta\c{c}\~ao para os operadores de campo, e estas equa\c{c}\~oes de movimento,  podemos obter a equa\c{c}\~ao de Schr\"{o}dinger. Para verificarmos isto, u-\\saremos ${\cal{D}}_{x}$ como abrevia\c{c}\~ao para o operador diferencial de Schr\"{o}dinger livre definido como
\be
 {\cal{D}}_{x}=- \frac{1}{2m}{\boldmath{\nabla}}^{2}
\ee
ent\~ao
\bea
i\dot{\hat{\psi}}&=&[{\hat{\psi}}, \hat{H}] \non
&=&\int d^{3}x^{'} [\hat{\psi}({\ve {x}},t),{\hat{\psi}}^{\a}({\ve {x}}{'},t) {\cal{D}}_{x{'}} {\hat{\psi}}({\ve {x}}{'},t)] \non
&=&\int d^{3}x^{'} \le ([\hat{\psi}({\ve {x}},t),\hat{\psi}^{\a}({\ve {x}}{'},t)] {\cal{D}}_{x^{'}}\hat{\psi}({\ve {x}},t) +\hat{\psi}^{\a}({\ve {x}}{'},t) {\cal{D}}_{x^{'}} [\hat{\psi}({\ve {x}},t), \hat{\psi}({\ve {x}}{'},t)] \ri ) \non
&=&{\cal{D}}_{x}\hat{\psi}({\ve {x}},t).
\eea
Portanto, o operador de campo tamb\'em satisfaz a equa\c{c}\~ao de Schr\"{o}dinger livre e dependente do tempo

\be
i\frac{\6 \hat{\psi}}{\6 t}=-\frac{1}{2m}{\boldmath{\nabla}}^{2}\hat{\psi}.
\ee

      Introduziremos o sistema fict\'{\i}cio (denotado com til), o qual \'e descrito pela densidade de Lagrangeana convenientemente definida como {\cite{ut}}
\be
{ \widetilde{\cal{L}}}(\ve{x},t)=-i {\widetilde{\psi}}^{\ast}{ \dot{\widetilde{\psi}}}-\frac{1}{2m}{\boldmath{\nabla}} {\widetilde{{\psi}}}^{\ast}{\boldmath{\nabla}} \widetilde{{\psi}}.
\ee
A raz\~ao para o sinal negativo no primeiro termo da densidade de Lagrangeana acima tornarse-a claro em breve. O momento canonicamente conjugado ao campo $ \widetilde{{\psi}}({\bo{x}},t)$ \'e dado por 
\be
 \widetilde{\p}({\ve {x}},t)=\frac {\6{ \widetilde{\cal{L}}}}{\6 { \dot{\widetilde{\psi}}}}=-i {\widetilde{\psi}}^{\ast}({\ve {x}},t).
\ee
Como no caso do sistema f\'{\i}sico real, o momento canonicamente conjugado ao campo $ \widetilde{\psi}^{\ast}({\ve{x}},t)$ \'e nulo,  existindo somente dois campos independentes $ \widetilde{\psi}({\ve {x}},t)$ e $ \widetilde{\p}({\ve {x}},t)$. A densidade de Hamiltoniana para este sistema fict\'{\i}cio \'e
\be
 \widetilde{\cal{H}}= \widetilde{\p}{ \dot{\widetilde{\psi}}}- \widetilde{\cal{L}}=\frac{1}{2m}{\boldmath{\nabla}} {\widetilde{\psi}}^{\ast}{\boldmath{\nabla}} \widetilde{\psi}
\ee
e os colchetes de Poisson a tempos iguais s\~ao
\be
\{  \widetilde{\psi}({\ve {x}},t), \widetilde{\p}(\ve{x}^{'},t) \} = \d^{3}({\ve {x}} - \ve{x}^{'}),
\ee
\be
\{  \widetilde{\psi}({\ve {x}},t), \widetilde{\psi}(\ve{x}^{'},t) \} = \{  \widetilde{\p}({\ve {x}},t), \widetilde{\p}(\ve {x}^{'},t) \}=0,
\ee 
al\'em disso
\be
\{{\psi}({\ve {x}},t), \widetilde{\psi}(\ve {x}^{'},t) \}=\{  \widetilde{\psi}({\ve {x}},t),{\psi}^{\a}(\ve{x}^{'},t) \}=0.
\ee

        Analogamente ao processo de quantiza\c{c}\~ao can\^onica feito para os campos cl\'assicos do sistema f\'{\i}sico,  ${ \widetilde{\psi}}({\ve {x}},t)$ e ${ \widetilde{\p}}({\ve {x}},t)$ s\~ao trocados pelos operadores ${\hat{ \widetilde{\psi}}}({\ve {x}},t)$ e ${\hat{ \widetilde{\p}}}({\ve {x}},t)= i { {\hat{\widetilde{\psi}}}^{\a}}({\ve {x}},t)$. Desta forma, as rela\c{c}\~oes de comuta\c{c}\~ao a tempos iguais s\~ao
\be
[ \hat{ \widetilde{\psi}}({\bo {x}},t),\hat{ \widetilde{\psi}}^{\a}({\ve {x}{'}},t)] = \d^{3}({\ve {x}} - {\ve {x}{'}})
\ee
\be
[\hat{ \widetilde {\psi}}({\ve {x}},t),\hat{ \widetilde {\psi}}({\ve {x}}{'} ,t) ] =[\hat{ \widetilde{\psi}}^{\a}({\ve {x}},t),\hat{ \widetilde{\psi}}^{\a}({\ve {x}}{'},t) ] =0
\ee
e tamb\'em 
\be
[{\hat{\psi}}({\ve {x}},t),\hat{ \widetilde{\psi}}^{\a}({\ve {x}}{'},t) ] =[ {\hat{\psi}}({\ve {x}},t),\hat{ \widetilde{\psi}}({\ve {x}}{'},t) ] =0.
\ee
A evolu\c{c}\~ao din\^amica dos operadores de campo \'e dada pelas equa\c{c}\~oes de movimento de Heisenberg e conduz \`as  equa\c{c}\~oes Schr\"{o}dinger livre.

       De agora em diante, trataremos somente do sistema quantizado, sendo assim, omitiremos o s\'{\i}mbolo $\hat{}$  para os operadores, e passaremos a us\'a-lo somente para notarmos quantidades relacionadas ao sistema total (f\'{\i}sico e fict\'{\i}cio), se houver alguma excess\~ao deixaremos claro no texto.

         O sistema total \'e caracterizado pela densidade de Lagrangeana ${\hat{\cal{L}}}(\ve{x},t)$
\be
{\hat{\cal{L}}}(\ve{x},t)={\cal{L}}(\ve{x},t)- \widetilde{\cal{L}}(\ve{x},t) ,      
\ee
e os momentos can\^onicos do sistema total s\~ao
\be
\frac {\6{\hat{\cal{L}}}}{\6 {\dot{\psi}}}=i{\psi}^{\a}({\ve {x}},t)\ \ \ \ \ ,\ \  \ \ \ \frac {\6 \bo{{\hat{\cal{L}}}}}{\6 {\dot{ \widetilde{\psi}}}}=i \widetilde{\psi}^{\a}({\bo {x}},t),
\ee
e a densidade de Hamiltoniana total \'e
\be
{\hat{\cal{H}}}={\cal{H}} -{ \widetilde{\cal{H}}}=\frac{1}{2m} \le ({\boldmath{\nabla}}{\psi}^{\a}{\boldmath{\nabla}}{\psi} -{\boldmath{\nabla}}{ {\widetilde{\psi}}}^{\a}{\boldmath{\nabla}} \widetilde{\psi} \ri ).
\ee
A Hamiltoniana total \'e dada pela integra\c{c}\~ao sobre todo o espa\c{c}o das coordenadas da densidade de Hamiltoniana total, a qual denotamos como
\be
{\bo{\hat{H}}}=\int d^{3}x ({\cal{H}} - { \widetilde{\cal{H}}})=H-{ \widetilde{H}}.
\ee
Esta Hamiltoniana total tamb\'em satisfaz as equa\c{c}\~oes de movimento de Heisenberg, ou seja,
\be
i\dot{\psi}=[{\psi},{\hat{H}}]=[{\psi},{H}]\ \ \ , \ \ \ i{ \dot{\widetilde{\psi}}}=[ \widetilde{\psi},{\hat{H}}]=-[ \widetilde{\psi},\widetilde{H}].
\ee

       Os operadores de campo de Schr\"odinger $\psi$ e $ \widetilde {\psi}$, podem ser representados em termos de uma decomposi\c{c}\~ao de Fourier em uma base de ondas planas como segue
\be 
\psi (\ve{x},t)=\frac{1}{\sqrt{V}} \sum_{\ve{k}}e^{i \ve{k} \cdot \ve{x}} \psi_{\ve{k}}(t)= \frac{1}{\sqrt{V}} \sum_{\ve{k}}e^{i \ve{k} \cdot \ve{x}}e^{-i\o_{\ve{k}}t}a_{\ve{k}}\ ,
\ee
\be
 \widetilde{\psi} (\ve{x},t)=\frac{1}{\sqrt{V}} \sum_{\ve{k}}e^{-i \ve{k} \cdot \ve{x}}  \widetilde{\psi}_{\ve{k}}(t)= \frac{1}{\sqrt{V}} \sum_{\ve{k}}e^{-i \ve{k} \cdot \ve{x}}e^{i\o_{\ve{k}}t} \widetilde{a}_{\ve{k}}\ ,
\ee
e seus respectivos Hermitianos conjugados s\~ao
\be
{\psi}^{\a} (\ve{x},t)=\frac{1}{\sqrt{V}} \sum_{\ve{k}}e^{-i \ve{k} \cdot \ve{x}} {\psi}^{\a}_{\ve{k}}(t)= \frac{1}{\sqrt{V}} \sum_{\ve{k}}e^{-i \ve{k} \cdot \ve{x}}e^{i\o_{\ve{k}}t}a^{\a}_{\ve{k}}\ ,
\ee 
e
\be
{\widetilde{\psi}}^{\a} (\ve{x},t)=\frac{1}{\sqrt{V}} \sum_{\ve{k}}e^{i \ve{k} \cdot \ve{x}}  {\widetilde{\psi}}^{\a}_{\ve{k}}(t)= \frac{1}{\sqrt{V}} \sum_{\ve{k}}e^{i \ve{k} \cdot \ve{x}}e^{-i\o_{\ve{k}}t} \widetilde{a}^{\a}_{\ve{k}}
\ee
onde a energia \'e discretizada e dada por ${\o_{\ve{k}}}=\frac{|\ve{k}|^{2}}{2m}$, devido ao fato da quantiza\c{c}\~ao ser realizada em uma caixa c\'ubica de volume finito $V$ (constante de normaliza\c{c}\~ao) do espa\c{c}o que nos conduz a valores discretos de $\ve{k}$. Usando estas expans\~oes em termos das componentes de Fourier (operadores de cria\c{c}\~ao e destrui\c{c}\~ao), a Hamiltoniana total $\hat{H}$ pode ser escrita como
\be
\hat{H}=\sum_{\ve{k}}\o_{\ve{k}}(a^{\a}_{\ve{k}}a_{\ve{k}}- \widetilde{a}^{\a}_{\ve{k}} \widetilde{a}_{\ve{k}}). \label{HT}
\ee

        O espa\c{c}o de Hilbert para o sistema combinado \'e constru\'{\i}do de maneira seme-\\lhante \`a da se\c{c}\~ao anterior, ou seja, podemos definir o v\'acuo t\'ermico como
\be 
|0(\b) \ra = \sum_{n_{\ve{k}}}\sum_{\ve{k}}e^{-\b \o_{n_{\ve{k}}}/2}Z^{-1/2}(\b)|n_{\ve{k}},\widetilde {n}_{\ve{k}} \ra,
\ee
onde a fun\c{c}\~ao de parti\c{c}\~ao do sistema \'e agora dada por
\be
Z(\b)= \sum_{\ve{k}}\frac{1}{1-e^{-\b \o_{\ve{k}}}}.
\ee
        O gerador das transforma\c{c}\~oes que nos permite mapear os operadores t\'ermicos nos operadores em $T=0$ \'e
\be
G=-i \sum_{\ve{k}} \th_{\ve{k}}(a^{\a}_{\ve{k}}  \widetilde{a}^{\a}_{\ve{\k}}-\widetilde{a}_{\ve{\k}}a_{\ve{k}}),  \label{TB}
\ee
o qual nos conduz formalmente ao operador unit\'ario
\be
U(\th)=e^{-iG(\th)}=e^{- \sum_{\ve{k}} \th_{\ve{k}}(a^{\a}_{\ve{k}}  \widetilde{a}^{\a}_{\ve{k}}-\widetilde{a}_{\ve{k}}a_{\ve{k}})}.
\ee
        Este operador unit\'ario ir\'a conectar o v\'acuo t\'ermico ao v\'acuo original de um modo padr\~ao (transforma\c{c}\~ao de Bogoliubov), ou seja, usaremos o gerador $G$ dado pela equa\c{c}\~ao~({\ref{TB}}) para  definir o espa\c{c}o de Fock transformado atrav\'es do novo v\'acuo
\bea
|0(\b)\ra &=&U(\th)|0(\b)\ra= e^{-iG}|0\ra \! \ra \non
&=&\exp{ \le( \sum_{\ve{k}} \th_{\ve{k}}({a}^{\a}_{\ve{k}}\widetilde a^{\a}_{\ve{k}}-\widetilde a_{\ve{k}}{a}_{\ve{k}}) \ri) }|0\ra \! \ra \non  
&=&\prod_{\ve{k}} \exp{\le( \th_{\ve{k}}(a^{\a}_{\ve{k}}  \widetilde{a}_{\ve{k}}^{\a}-a_{\ve{k}}  \widetilde{a}_{\ve{k}}) \ri )}|0\ra \! \ra \non
&=&\prod_{\ve{k}} \le [{\cosh}^{-1}{\th}_{\ve{k}}(\b) \exp{({\tanh}{\th}_{\ve{k}}(\b){a}^{\a}_{\ve{k}} \widetilde{a}^{\a}_{\ve{k}})}\ri ]|0\ra \! \ra . \label{vtc}
\eea 
onde 
\be
u_{\ve{k}}(\b)=(1-e^{-\b \o_{\ve{k}}})^{-\frac{1}{2}}=\sqrt{1+f_{B}(\o_{\ve{k}})}=\cosh\th_{\ve{k}}(\b) \label{uv}  
\ee
e
\be
\v_{\ve{k}}(\b)=(e^{\b \o_{\ve{k}}}-1)^{-\frac{1}{2}}=\sqrt{f_{B}(\o_{\ve{k}})}=\sinh\th_{\ve{k}}(\b)  \label{v}
\ee
Usando as equa\c{c}\~oes~({\ref{aa}}) e ({\ref{at}}), podemos verificar que estes operadores podem ser mapeados na forma
\bea
\le(\begin{array}{c} a_{\ve{k}}(\b) \\ \w{a}_{\ve{k}}^{\a}(\b) \end{array}  \ri )&=&\overline{U}(\th_{\ve{k}})\le(\begin{array}{c} a_{\ve{k}} \\ \w{a}_{\ve{k}}^{\a} \end{array}  \ri ) \non
&=&\fourmat{\cosh {\th_{\ve{k}}(\b)}}{-\sinh {\th_{\ve{k}}(\b)}}{-\sinh {\th_{\ve{k}}(\b)}}{\cosh {\th_{\ve{k}}(\b)}} 
\le(\begin{array}{c} a \\ \w{a}^{\a} \end{array}  \ri )
\eea
com $\th_{\ve{k}}(\b)$ dado por (\ref{uv})e (\ref{v}). Alternativamente, isto pode ser escrito em termos das vari\'aveis de campo como segue:
\be
\le(\begin{array}{c} \psi_{\ve{k}}(t,\b) \\ \w{\psi}_{\ve{k}}^{\a}(t,\b) \end{array}  \ri )=\overline{U}(\th_{\ve{k}})\le(\begin{array}{c} \psi_{\ve{k}}(t) \\ \w{\psi}_{\ve{k}}^{\a}(t) \end{array}  \ri ) 
=\fourmat{\cosh {\th_{\ve{k}}(\b)}}{-\sinh {\th_{\ve{k}}(\b)}}{-\sinh {\th_{\ve{k}}(\b)}}{\cosh {\th_{\ve{k}}(\b)}} 
\le(\begin{array}{c} \psi_{\ve{k}}(t) \\ \w{\psi}^{\a}(t) \end{array}  \ri ).
\ee
Podemos ainda escrever esta \'ultima  express\~ao na forma de duas equa\c{c}\~oes, a saber,
\be
{a}_{\ve{k}}(\b)=e^{-iG}a_{\ve{k}}e^{iG}={a}_{\ve{k}}\cosh{\th_{\ve{k}}}(\b)- \widetilde{a}^{\a}_{\ve{k}}\sinh{\th_{\ve{k}}}(\b), \label{oac}
\ee
\be
\widetilde{a}_{\ve{k}}(\b)=e^{-iG} \widetilde{a}_{\ve{k}}e^{iG}= \widetilde{a}_{\ve{k}}\cosh {\th_{\ve{k}}}(\b)-{a}^{\a}_{\ve{k}}\sinh {\th_{\ve{k}}}(\b)\ . \label{oatc}
\ee
e substitu\'{\i}-las na Hamiltoniana total, equa\c{c}\~ao~({\ref{HT}}), e usar a equa\c{c}\~ao~({\ref{quadrados}}) para mostrarmos que, como no caso da Hamiltoniana total para a part\'{\i}cula dada pela equa\c{c}\~ao~(\ref{ht}), a Hamiltoniana total \'e invariante por transforma\c{c}\~oes de Bogoliubov
\be
\hat{H}=\sum_{\ve{k}}\o_{\ve{k}} \le(a^{\a}_{\ve{k}}(\b)a_{\ve{k}}(\b)- \widetilde{a}^{\a}_{\ve{k}}(\b) \widetilde{a}_{\ve{k}}(\b) \ri)
\ee 
e portanto o gerador \'e conservado 
\be
i\dot{G}=[G,\hat{H}]=0.
\ee

       Os estados de uma part\'{\i}cula s\~ao $a^{\a}_{\ve{k}}(\b)|0(\b)\ra =e^{-iG}a^{\a}_{\ve{k}}e^{iG}e^{-iG}|0\ra \! \ra =e^{-iG}a^{\a}_{\ve{k}}|0\ra \! \ra, \\ \widetilde{a}^{\a}_{\ve{k}}(\b)|0(\b)\ra =e^{-iG} \widetilde{a}^{\a}_{\ve{k}}e^{iG}e^{-iG}|0\ra \! \ra =e^{-iG} \widetilde{a}^{\a}_{\ve{k}}|0\ra \! \ra,$ e assim por diante. Notemos ent\~ao que n\~ao somente o estado de v\'acuo pode ser obtido por uma transforma\c{c}\~ao de Bogoliubov, mas tamb\'em qualquer estado de part\'{\i}cula. Ainda, usando a propriedade $a|0\ra \! \ra =  \widetilde{a}|0\ra \! \ra =0$, podemos verificar que o estado $|0(\b)\ra $  \'e realmente o v\'acuo t\'ermico:
\be
a_{\ve{k}}(\b)|0(\b)\ra =e^{-iG}a_{\ve{k}}e^{iG}e^{-iG}|0\ra \! \ra =e^{-iG}a_{\ve{k}}|0\ra \! \ra =0, \label{va}
\ee
\be
 \widetilde{a}_{\ve{k}}(\b)|0(\b)\ra =e^{-iG} \widetilde{a}_{\ve{k}}e^{iG}e^{-iG}|0\ra \! \ra 
=e^{-iG} \widetilde{a}_{\ve{k}}|0\ra \! \ra 
=0.  \label{vat}
\ee
Se substituirmos a equa\c{c}\~ao~(\ref{oac}) em (\ref{va}) e no resultado obtido substituirmos a transforma\c{c}\~ao inversa da equa\c{c}\~ao~(\ref{oac}) obtemos um resultado importante relacionado ao novo estado de v\'acuo:
\be
a^{\a}_{\ve{k}}(\b)|0(\b)\ra =\frac{1}{\cosh{{\th}_{\ve{k}}}(\b)}a^{\a}_{\ve{k}}|0(\b)\ra =\frac{1}{\sinh{{\th}_{\ve{k}}}(\b)} \widetilde{a}_{\ve{k}}|0(\b)\ra \label{buraco1}.
\ee
Da mesma forma, podemos substituir a equa\c{c}\~ao~(\ref{oatc}) em (\ref{vat}) e em seguida a transforma\c{c}\~ao inversa de (\ref{oatc}) no resultado obtido por esta substitui\c{c}\~ao e obtermos
\be
 \widetilde{a}^{\a}_{\ve{k}}(\b)|0(\b)\ra =\frac{1}{\cosh{{\th}_{\ve{k}}}(\b)} \widetilde{a}^{\a}_{\ve{k}}|0(\b)\ra =\frac{1}{\sinh{{\th}_{\ve{k}}}(\b)}{a}_{\ve{k}}|0(\b)\ra. \label{buraco2}
\ee
Al\'em disso, da invari\^ancia da Hamiltoniana total temos
\be
\hat{H}|0(\b)\ra =\sum_{\ve{k}}\o_{\ve{k}} \le(a^{\a}_{\ve{k}}(\b)a_{\ve{k}}(\b)- \widetilde{a}^{\a}_{\ve{k}}(\b) \widetilde{a}_{\ve{k}}(\b) \ri)|0(\b)\ra =0. \label{htc}
\ee 

        Definindo  o operador n\'umero total de part\'{\i}culas $\hat{N}$  como
\bea
\hat{N}=N- \widetilde{N}&=&\sum_{\ve{k}}(a^{\a}_{\ve{k}}a_{\ve{k}}- \widetilde{a}^{\a}_{\ve{k}} \widetilde{a}_{\ve{k}}) \non
&=&\sum_{\ve{k}}\le[u^{2}(\b) \le(a^{\a}_{\ve{k}}(\b)a_{\ve{k}}(\b)- \widetilde{a}^{\a}_{\ve{k}}(\b) \widetilde{a}_{\ve{k}}(\b)\ri ) - \v^{2}(\b)\le(a^{\a}_{\ve{k}}(\b)a_{\ve{k}}(\b)- \widetilde{a}^{\a}_{\ve{k}}(\b) \widetilde{a}_{\ve{k}}(\b) \ri) \ri ] \non
&=&\sum_{\ve{k}}\le[ \le(u^{2}(\b)-v^{2}(\b)\ri)\le (a^{\a}_{\ve{k}}(\b)a_{\ve{k}}(\b)- \widetilde{a}^{\a}_{\ve{k}}(\b) \widetilde{a}_{\ve{k}}(\b)\ri )\ri ] \non
&=&\sum_{\ve{k}}\le (a^{\a}_{\ve{k}}(\b)a_{\ve{k}}(\b)- \widetilde{a}^{\a}_{\ve{k}}(\b) \widetilde{a}_{\ve{k}}(\b)\ri )   \label{on}
\eea
onde usamos a equa\c{c}\~ao~({\ref{quadrados}}). Desta forma, a  Hamiltoniana total pode ser escrita como
\be
\hat{H}=\sum_{\ve{k}}{\hat{N}}_{\ve{k}}\o_{\ve{k}},
\ee
onde ${\hat{N}}_{\ve{k}}=a^{\a}_{\ve{k}}(\b)a_{\ve{k}}(\b)- \widetilde{a}^{\a}_{\ve{k}}(\b) \widetilde{a}_{\ve{k}}(\b)$. Logo, da equa\c{c}\~ao~(\ref{htc}) temos um importante resultado:
\be
\la 0(\b)|\hat{H}|0(\b)\ra =\la 0(\b)|(H- \widetilde{H})|0(\b)\ra=\la 0(\b)|H|0(\b)\ra -\la 0(\b)| \widetilde{H}|0(\b)\ra =0  
\ee
logo
\be
\la 0(\b)|H|0(\b)\ra =\la 0(\b)| \widetilde{H}|0(\b)\ra ,
\ee
que demostra a nossa suposi\c{c}\~ao de que a energia do sistema fict\'{\i}cio \'e igual a do sistema f\'{\i}sico. Da equa\c{c}\~ao~(\ref{on}) obtemos ainda
\be
\la 0(\b)|\hat{N}|0(\b)\ra =\la 0(\b)|(N- \widetilde{N})|0(\b)\ra =\la 0(\b)|N|0(\b)\ra -\la 0(\b)| \widetilde{N}|0(\b)\ra =0  
\ee
que resulta em
\be
\la 0(\b)|N|0(\b)\ra =\la 0(\b)| \widetilde{N}|0(\b)\ra 
\ee
ou seja, o sistema possui o mesmo n\'umero de part\'{\i}culas f\'{\i}sicas  quanto de n\~ao f\'{\i}sicas, como tamb\'em j\'a hav\'{\i}amos suposto. Al\'em disso, as express\~oes ~(\ref{buraco1}) e ~(\ref{buraco2}) mostram que  acrescentar uma part\'{\i}cula f\'{\i}sica ao v\'acuo $|0(\b)\ra $ \'e equivalente a eliminar uma part\'{\i}cula n\~ao f\'{\i}sica, a menos de um fator constante,  e vice-versa. Portanto, a part\'{\i}cula n\~ao f\'{\i}sica pode ser interpretada como um buraco ou a falta de uma part\'{\i}cula f\'{\i}sica.
\section{Entropia}
${}$
       Nesta se\c{c}\~ao, obteremos duas grandezas de extrema import\^ancia em mec\^anica estat\'{\i}stica no formalismo da Din\^amica de Campos T\'ermicos, s\~ao elas a entropia e a energia livre de Helmholtz. Para isto, definiremos o operador
\be
K=-\sum_{\ve{k}} \le(a^{\a}_{\ve{k}}a_{\ve{k}}\ln{\sinh^{2}{\th_{\ve{k}}(\b)}}-a_{\ve{k}}a^{\a}_{\ve{k}}\ln{\cosh^{2}{\th_{\ve{k}}(\b)}}\ri ),  
\label{centropia}
\ee
onde o operador $ \widetilde{K}$ pode ser obtido deste simplesmente pela troca de $a^{\a}_{\ve{k}}$ e $a_{\ve{k}}$ por $ \widetilde{a}^{\a}_{\ve{k}}$ e $ \widetilde{a}_{\ve{k}}$ respectivamente. Usando as rela\c{c}\~oes de comuta\c{c}\~ao para vari\'aveis bos\^onicas, podemos escrever este operador na seguinte forma
\be 
K=-2\sum_{\ve{k}}\le (a^{\a}_{\ve{k}}a_{\ve{k}}\ln{\tanh{\th_{\ve{k}}}(\b)}-\ln{\cosh{\th_{\ve{k}}(\b)}}\ri).
\ee   

        Usando esta defini\c{c}\~ao para o operador $K$ definido em (\ref{centropia}), o estado de v\'acuo $|0(\b)\ra $ dado por (\ref{vtc}) pode ser expresso como
\be
|0(\b)\ra =e^{-K/2} \le \{\exp{\sum_{\ve{k}}a^{\a}_{\ve{k}} \widetilde{a}^{\a}_{\ve{k}}}\ri \}|0\ra \! \ra =e^{- \widetilde{K}/2}\le \{ \exp{\sum_{\ve{k}} \widetilde{a}^{\a}_{\ve{k}}{a}^{\a}_{\ve{k}}} \ri \}|0\ra \! \ra . \label{vtk}
\ee

       Para demostrarmos isto, usaremos os seguintes resultados
\be
e^{-K/2}a^{\a}_{\ve{k}}e^{K/2}=\tanh{\th_{\ve{k}}(\b)}a^{\a}_{\ve{k}} \ \ , \ \ e^{K/2}|0\ra \! \ra= \prod_{\ve{k}}\cosh{\th_{\ve{k}}(\b)}|0\ra \! \ra \ ,
\ee
e
\be
e^{-K/2}\le ( \widetilde{a}^{\a}_{\ve{k}}\ri )^{n}|0\ra \! \ra= \prod_{\ve{k}}\frac{1}{\cosh{\th_{\ve{k}}(\b)}}\le ( \widetilde{a}^{\a}_{\ve{k}}\ri )^{n}|0\ra \! \ra.
\ee
Logo
\bea
|0(\b)\ra &=&e^{-K/2}\exp{\sum_{\ve{k}}a^{\a}_{\ve{k}} \widetilde{a}^{\a}_{\ve{k}}}|0\ra \! \ra \non
&=&e^{-K/2}\prod_{\ve{k}}\exp \le({a^{\a}_{\ve{k}} \widetilde{a}^{\a}_{\ve{k}}}\ri )|0\ra \! \ra \non
&=&\prod_{\ve{k}} \le \{ \frac{1}{\cosh{\th_{\ve{k}}(\b)}} + \frac{1}{1!}[\tanh{\th_{\ve{k}}(\b)}]{a}^{\a}_{\ve{k}}e^{-K/2} \widetilde{a}^{\a}_{\ve{k}} +  \frac{1}{2!}[\tanh{\th_{\ve{k}}(\b)}]^{2}{a}^{\a}_{\ve{k}}a^{\a}_{\ve{k}}e^{-K/2} \le ( \widetilde{a}^{\a}_{\ve{k}} \ri )^{2} \ri . \non
 &+& \le . \frac{1}{3!}[\tanh{\th_{\ve{k}}(\b)}]^{3}{a}^{\a}_{\ve{k}}a^{\a}_{\ve{k}}a^{\a}_{\ve{k}}e^{-K/2} \le ( \widetilde{a}^{\a}_{\ve{k}} \ri )^{3} +\dots  \ri \}|0\ra \! \ra \non
&=&\prod_{\ve{k}} \le \{ \frac{1}{\cosh{\th_{\ve{k}}(\b)}} + \frac{1}{1!}[\tanh{\th_{\ve{k}}(\b)}]{a}^{\a}_{\ve{k}}\frac{1}{\cosh{\th_{\ve{k}}(\b)}} \widetilde{a}^{\a}_{\ve{k}} +  \frac{1}{2!}[\tanh{\th_{\ve{k}}(\b)}]^{2}{a}^{\a}_{\ve{k}}a^{\a}_{\ve{k}}\frac{1}{\cosh{\th_{\ve{k}}(\b)}} \le ( \widetilde{a}^{\a}_{\ve{k}} \ri )^{2} \ri . \non
 &+& \le . \frac{1}{3!}[\tanh{\th_{\ve{k}}(\b)}]^{3}{a}^{\a}_{\ve{k}}a^{\a}_{\ve{k}}a^{\a}_{\ve{k}}\frac{1}{\cosh{\th_{\ve{k}}(\b)}} \le ( \widetilde{a}^{\a}_{\ve{k}} \ri )^{3} +\dots \ri \}|0\ra \! \ra \non
&=&\prod_{\ve{k}}  \frac{1}{\cosh{\th_{\ve{k}}(\b)}}\le \{1 + \frac{1}{1!}[{a}^{\a}_{\ve{k}} \widetilde{a}^{\a}_{\ve{k}}\tanh{\th_{\ve{k}}(\b)}] +  \frac{1}{2!}[{a}^{\a}_{\ve{k}} \widetilde{a}^{\a}_{\ve{k}}\tanh{\th_{\ve{k}}(\b)}]^{2} \ri . \non
 &+& \le .  \frac{1}{3!}[{a}^{\a}_{\ve{k}} \widetilde{a}^{\a}_{\ve{k}}\tanh{\th_{\ve{k}}(\b)}]^{3} + \dots \ri \}|0\ra \! \ra \nonumber
\eea
ou seja
$$
|0(\b)\ra =\prod_{\ve{k}}  \frac{1}{\cosh{\th_{\ve{k}}(\b)}}\exp \le \{{a}^{\a}_{\ve{k}} \widetilde{a}^{\a}_{\ve{k}}\tanh{\th_{\ve{k}}(\b)} \ri \}|0\ra \! \ra
$$ 
que \'e a mesma express\~ao obtida atrav\'es do mapeamento com o operador de Bogoliubov. Procedimento completamente an\'alogo pode ser usado para mostrarmos a validade da express\~ao com til, uma vez que ${a}^{\a}_{\ve{k}}$ e $ \widetilde{a}^{\a}_{\ve{k}}$ comutam.

        Tamb\'em \'e conveniente definirmos a quantidade
$$
|\hat{I}\ra =\exp \le \{\sum_{\ve{k}}a^{\a}_{\ve{k}} \widetilde{a}^{\a}_{\ve{k}}\ri \}|0\ra \! \ra,
$$
que em termos da nota\c{c}\~ao usada na se\c{c}\~ao anterior \'e
$$
|\hat{I}\ra =\sum_{n}|n, \widetilde{n}\ra .
$$
Usando esta quantidade, a equa\c{c}\~ao~(\ref{vtk}) pode ser escrita como
$$
|0(\b)\ra =e^{-K/2}|\hat{I}\ra =e^{- \widetilde{K}/2}|\hat{I}\ra .
$$

    Pode-se mostrar que o operador $\hat{K}= K- \widetilde{K}$ satisfaz a rela\c{c}\~ao
$$ 
[K- \widetilde{K},G]=0,
$$
com $G$ dado pela equa\c{c}\~ao~(\ref{TB}). Uma outra propriedade envolvendo os operadores $K$ e $ \widetilde{K}$ aplicados no estado $|0(\b)\ra $ \'e a seguinte: 
\bea 
(K- \widetilde{K})|0(\b)\ra &=&(K- \widetilde{K})e^{-iG}|0\ra \! \ra \non
&=&e^{-iG}(K- \widetilde{K})|0\ra \! \ra \non
&=&-2e^{-iG} \sum_{\ve{k}}(\ln{\tanh{\th_{\ve{k}}(\b)}})(a^{\a}_{\ve{k}}a_{\ve{k}}- \widetilde{a}^{\a}_{\ve{k}} \widetilde{a}_{\ve{k}})|0\ra \! \ra \non
&=&0. \nonumber
\eea

        At\'e o momento, nada foi dito sobre os valores assumidos pelo `` \^angulo'' $\th_{\ve{k}}(\b)$, que est\'a sendo considerado completamente arbitr\'ario. Agora, iremos exigir que o valor esperado no v\'acuo do operador $\hat{H}=\sum_{\ve{k}}\o_{\ve{k}}{\hat{N}}_{k}$ seja constante e analisar quais as condi\c{c}\~oes que  $\th_{\ve{k}}(\b)$ deve satisfazer para que isto ocorra. Nestas condi\c{c}\~oes calcularemos o valor esperado no v\'acuo do operador $K$. Usando a equa\c{c}\~ao (\ref{n}) podemos fixar o valor para  $\th_{\ve{k}}(\b)$ da seguinte forma
\be 
 \langle 0(\b)|a^{\a}_{\ve{k}}a_{\ve{k}}|0(\b)\ra =\v_{\ve{k}}^{2}(\b)=\sinh^{2}{\th_{\ve{k}}(\b)}=n_{\ve{k}}, \label{nk}
\ee
ent\~ao calculando o valor m\'edio de $K$ dado pela equa\c{c}\~ao~({\ref{centropia}) obtemos
$$
 \langle 0(\b)|K|0(\b)\ra =\sum_{\ve{k}}\le \{ (1+n_{\ve{k}})\ln(1+n_{\ve{k}})-n_{\ve{k}}\ln n_{\ve{k}} \ri \}= \langle K\ra .
$$

        Este resultado \'e proporcional \`a entropia calculada no sistema Gr\~a-Can\^onico{\cite{rk}}, ou seja,
$$
S=k_{B}\sum_{{k}}\le \{ (1+ \langle n_{{k}}\ra )\ln(1+ \langle n_{{k}}\ra )- \langle n_{{k}}\ra \ln  \langle n_{{k}}\rangle \ri \}
$$
onde $n_{k}$ representa o n\'umero m\'edio de ocupa\c{c}\~ao do estado $k$. Portanto, o valor esperado no v\'acuo do operador $K$ multiplicado pela constante de Boltzmann ($k_{B}$) \'e a entropia do sistema f\'{\i}sico. Desta forma, iremos nos referir ao operador $K$ como sendo o operador Entropia
$$ 
S=k_{B} \langle K\ra =k_{B} \langle 0(\b)|K|0(\b)\ra . 
$$

        Seja a energia livre de Helmholtz
\bea
\O &=& -TS+ \langle H\ra -\mu \langle N\rangle \non
&=&-\fr{1}{\b} \langle 0(\b)|K|0(\b)\ra + \langle 0(\b)|H|0(\b)\ra -\mu \langle 0(\b)|N|0(\b)\rangle \non
&=&-\fr{1}{\b} \langle 0(\b)|K|0(\b)\ra + \langle 0(\b)|\sum_{\ve{k}}\o_{\ve{k}}a^{\a}_{\ve{k}}a_{\ve{k}}|0(\b)\ra -\mu \langle 0(\b)|\sum_{\ve{k}}a^{\a}_{\ve{k}}a_{\ve{k}}|0(\b)\rangle \non
&=&\sum_{{k}}\le \{-\fr{1}{\b}[(1+n_{\ve{k}})\ln(1+n_{\ve{k}})-n_{\ve{k}}\ln n_{\ve{k}}] + (\e_{\ve{k}}-\mu)n_{\ve{k}} \ri \}, \label{elh} 
\eea
onde usamos a equa\c{c}\~ao~(\ref{nk}). Supondo que o sistema esteja em equil\'{\i}brio, ou seja, $ \langle H\ra =constante$, ent\~ao
$$
\d \O=\sum_{\ve{k}}\fr{\6 \O}{\6 n_{\ve{k}}} \d n_{\ve{k}}=0.
$$
Como a varia\c{c}\~ao $\d n_{\ve{k}}$ \'e arbitr\'aria, temos que
$$
\fr{\6 \O}{\6 n_{\ve{k}}}=0.
$$
Para calcularmos esta derivada, devemos usar a regra da cadeia
\bea
\fr{\6 \O}{\6 n_{\ve{k}}}&=& \sum_{\ve{k}^{'}}\fr{dn_{\ve{k}^{'}}}{d n_{\ve{k}}}\fr{d \O}{d n_{\ve{k}^{'}}} \non
&=& \sum_{\ve{k}^{'}}\d_{\ve{k}\ve{k}^{'}} \le \{ \fr{1}{\b}\ln{ \fr{n_{\ve{k}}}{1+n_{\ve{k}}}+ (\e_{\ve{k}}-\mu)} \ri \}=0. \label{dp}
\eea
Da equa\c{c}\~ao~({\ref{dp}}) obtemos
$$
\b \e_{\ve{k}}=-\ln{ \fr{n_{\ve{k}}}{1+n_{\ve{k}}}},
$$
onde $\e_{\ve{k}}=\o_{\ve{k}}-\mu$ e $n_{\ve{k}}=\sinh^{2}{\th_{\ve{k}}}(\b)$. Esta express\~ao especifica o \^angulo ${\th_{\ve{k}}}(\b)$ e nos fornece tamb\'em a distribui\c{c}\~ao estat\'{\i}stica de Bose. Observemos que estes resultados foram obtidos somente porque consideramos que o sistema est\'a no estado de equil\'{\i}brio t\'ermico. 

         Se escrevermos 
$$
|0(\b)\ra =\sum_{\ve{k}} \sqrt{\e_{\ve{k}}}|n, \widetilde{n}\ra 
$$
podemos verificar que
$$
 \langle 0(\b)|K|0(\b)\ra =-\sum_{\ve{k}}{\e_{\ve{k}}}\ln n_{\ve{k}}
$$
e, da normaliza\c{c}\~ao do estado $|0(\b)\ra $, obtemos
$$
\sum_{\ve{k}}{\e_{\ve{k}}}=1.
$$

         As express\~oes (\ref{buraco1}) e (\ref{buraco2}) podem ser generalizadas. Para isto, consideremos os funcionais 
\be
F=\sum_{m,n}\sum_{\ve{k},\ve{p}}C(\ve{k}_{1},\ve{k}_{2},\dots \ve{k}_{m};\ve{p}_{1},\ve{p}_{2},\dots,\ve{p}_{n})a^{\a}_{\ve{k}_{1}}a^{\a}_{\ve{k}_{2}}\dots a^{\a}_{\ve{k}_{m}}a_{\ve{p}_{1}}a_{\ve{p}_{2}}\dots a_{\ve{p}_{n}} \label{F1}
\ee
e
\be
 \widetilde{F}=\sum_{m,n}\sum_{\ve{k},\ve{p}}C^{\ast}(\ve{k}_{1},\ve{k}_{2}, \dots \ve{k}_{m};\ve{p}_{1},\ve{p}_{2}, \dots ,\ve{p}_{n}) \widetilde{a}^{\a}_{\ve{k}_{1}} \widetilde{a}^{\a}_{\ve{k}_{2}} \dots  \widetilde{a}^{\a}_{\ve{k}_{m}} \widetilde{a}_{\ve{p}_{1}} \widetilde{a}_{\ve{p}_{2}} \dots  \widetilde{a}_{\ve{p}_{n}} \label{F2}
\ee
onde $C$ e $C^{\ast}$ s\~ao coeficientes complexos. Ent\~ao as rela\c{c}\~oes (\ref{buraco1}) e (\ref{buraco2}) podem ser escritas genericamente na forma
$$
e^{ \widetilde{K}/2}F|0(\b)\ra =e^{{K}/2} {\widetilde{F}}^{\a}|0(\b)\ra 
$$
e seu dual
$$
 \langle 0(\b)|Fe^{ \widetilde{K}/2}= \langle 0(\b)| {\widetilde{F}}^{\a}e^{{K}/2}.
$$
Para provarmos isto, inicialmente iremos definir
\be
A=a^{\a}_{\ve{k}_{1}}a^{\a}_{\ve{k}_{2}} \dots a^{\a}_{\ve{k}_{m}}a_{\ve{p}_{1}}a_{\ve{p}_{2}} \dots a_{\ve{p}_{n}}|0(\b)\ra .\label{A1}
\ee
Usando as express\~oes (\ref{buraco1}) e (\ref{buraco2}) e o fato de que as vari\'aveis bos\^onicas do sistema f\'{\i}sico comutam com as do sistema fict\'{\i}cio, podemos escrever
\be 
A=(\tanh{\th_{\ve{k}}} \widetilde{a}^{\a}_{\ve{p}_{m}}) \dots (\tanh{\th_{\ve{k}}} \widetilde{a}^{\a}_{\ve{p}_{1}})(\coth{\th_{\ve{k}}} \widetilde{a}_{\ve{k}_{m}}) \dots (\coth{\th_{\ve{k}}} \widetilde{a}_{\ve{k}_{1}})|0(\b)\ra \label{A2}
\ee
e, substituindo as equa\c{c}\~oes 
$$
e^{-K/2}a^{\a}_{\ve{k}}e^{K/2}=\tanh{\th_{\ve{k}}(\b)}a^{\a}_{\ve{k}}
$$e 
$$
e^{- \widetilde{K}/2} \widetilde{a}_{\ve{k}}e^{ \widetilde{K}/2}=\coth{\th_{\ve{k}}(\b)} \widetilde{a}_{\ve{k}}
$$
na equa\c{c}\~ao~(\ref{A2}) obtemos
$$
A=e^{- \widetilde{K}/2} \widetilde{a}^{\a}_{\ve{p}_{m}} \dots  \widetilde{a}^{\a}_{\ve{p}_{1}} \widetilde{a}_{\ve{k}_{m}} \dots  \widetilde{a}_{\ve{k}_{1}}e^{\widetilde{K}/2}|0(\b)\ra ,
$$
mas
$$
e^{ \widetilde{K}/2}|0(\b)\ra =e^{ \widetilde{K}/2}[e^{-\widetilde {K}/2}| \hat{I}\ra ]=| \hat{I}\ra =e^{{K}/2}|0(\b)\ra. 
$$
Logo, obtemos a seguinte equa\c{c}\~ao para $A$
\be
A=e^{- \widetilde{K}/2}e^{{K}/2} \widetilde{a}^{\a}_{\ve{p_{m}}} \dots  \widetilde{a}^{\a}_{\ve{p_{1}}} \widetilde{a}_{\ve{k_{m}}} \dots  \widetilde{a}_{\ve{k_{1}}}|0(\b)\ra . \label{A3}
\ee
Portanto, igualando as equa\c{c}\~oes (\ref{A1}) e (\ref{A3}) vem
$$
a^{\a}_{\ve{k}_{1}}a^{\a}_{\ve{k}_{2}} \dots a^{\a}_{\ve{k}_{m}}a_{\ve{p}_{1}}a_{\ve{p}_{2}} \dots a_{\ve{p}_{n}}|0(\b)\ra =e^{- \widetilde{K}/2}e^{{K}/2} \widetilde{a}^{\a}_{\ve{p}_{m}} \dots  \widetilde{a}^{\a}_{\ve{p}_{1}} \widetilde{a}_{\ve{k}_{m}} \dots  \widetilde{a}_{\ve{k}_{1}}|0(\b)\ra .
$$
Agora usando a defini\c{c}\~ao do funcional generalizado, eqs. (\ref{F1}) e (\ref{F2}), obtemos o resultado desejado, ou seja,
$$
e^{ \widetilde{K}/2}F|0(\b)\ra =e^{{K}/2} \widetilde{F}^{\a}|0(\b)\ra .
$$
Observemos que esta equa\c{c}\~ao \'e dita ser a generaliza\c{c}\~ao das equa\c{c}\~oes~(\ref{buraco1}) e (\ref{buraco2}) devido ao fato que, para obtermos esta, n\'os partimos da  equa\c{c}\~ao~(\ref{A1}) que, como podemos observar, \'e a forma mais geral das equa\c{c}\~oes~(\ref{buraco1}) e (\ref{buraco2}).

       Usando o formalismo desenvolvido neste cap\'{\i}tulo, pretendemos obter os estados de v\'acuo e a Hamiltoniana \`a temperatura finita para a corda bos\^onica, assim como o operador entropia e a energia livre de Helmholtz. Visando isto, discutiremos no pr\'oximo cap\'{\i}tulo a quantiza\c{c}\~ao da corda bos\^onica.
        
\subsection{Condi\c{c}\~oes de Kubo-Martin-Schwinger (KMS)} 
${}$

       As condi\c{c}\~oes de KMS possuem uma grande import\^ancia para a mec\^anica estat\'{\i}stica no equil\'{\i}brio t\'ermico. Elas surgem da necessidade de calcularmos um tra\c{c}o quando desejamos calcular uma m\'edia em um ensemble. Na DCT, as condi\c{c}\~oes de KMS seguem das condi\c{c}\~oes t\'ermicas (\ref{vou}) e (\ref{vou1}). Usaremos tais condi\c{c}\~oes para calcularmos a m\'edia em um ensemble da correla\c{c}\~ao de dois operadores arbitr\'arios.
\bea
\la 0(\b)|A(t)B(t^{'})|0(\b) \ra  &=& \la 0(\b)|\w{A}^{\a}(t)e^{\b \hat{H}/2}B(t^{'})|0(\b) \ra \non
&=& \la 0(\b)|\w{A}^{\a}(t+i\b/2)B(t^{'})|0(\b) \ra \non
&=& \la 0(\b)|B(t^{'})e^{\b \hat{H}/2}\w{A}^{\a}(t+i\b/2)|0(\b) \ra \non
&=& \la 0(\b)|B(t^{'})A(t+i\b)|0(\b) \ra . \label{kms}
\eea
Por simplicidade, omitimos aqui as coordenada espaciais. A import\^ancia desta rela\c{c}\~ao \'e que em certas situa\c{c}\~oes, \'e interessante termos uma regra para trocarmos as posi\c{c}\~oes de dois operadores quando tomamos o valor esperado do produto destes no v\'acuo.

\section{V\'acuo T\'ermico e Axiomas B\'asicos}
${}$      
        Como vimos anteriormente, para constru\'{\i}rmos o estado de v\'acuo t\'ermico $ |0(\b) \ra$ precisamos dobrar os graus de liberdade dos operadores, ou seja, para cada operador $A$ deve existir um operador correspondente $\w{A}$. Quaisquer operadores $A$ e $\w{A}$ s\~ao independentes, isto \'e, comutam entre si, e al\'em disso, existe um mapeamento entre os conjuntos de operadores $\{A\}$ e $\{\w{A}\}$ chamado {\it regras de conjuga\c{c}\~ao til}. A temperatura entra na teoria atrav\'es de condi\c{c}\~oes que relacionam a forma na qual $A$ e $\w{A}^{\a}$ atuam no estado de v\'acuo t\'ermico $|0(\b) \ra $. Esta condi\c{c}\~ao \'e chamada de {\it condi\c{c}\~ao de estado t\'ermico} ou {\it regra de substitui\c{c}\~ao til}. Uma teoria din\^amica de campos t\'ermicos para teoria qu\^antica de campos (TQC) pode ser melhor constru\'{\i}da a partir dos seguintes axiomas {\cite{m}}:

        Sejam dois conjuntos de operadores  $\Im=\{A\}$  e $\w{\Im}=\{\w{A}\}$, ent\~ao: \\
{\it Axioma 1 }. A tempos iguais, vari\'aveis din\^amicas pertencentes a diferentes subespa\c{c}os ($A \in \Im \ \mbox{e} \ \w{B} \in \w{\Im} $) s\~ao independentes, ou seja, comutam:
\be
[A,\w{B}]=0.
\ee
{\it Axioma 2 }. Existe um mapeamento um a um entre dois subespa\c{c}os ortogonais, $\Im$ e $\w{\Im}$, chamado conjuga\c{c}\~ao til, que para quaisquer $A$ e $B$ $\in \Im $, $\w{A}$ e $\w{B}$ $\in \w{\Im}$ e $c_{1}, c_{2}$ n\'umeros complexos, valem as seguintes regras:
\bea
& \mbox{a)}& \ \w{(AB)}=\w{A}\w{B},  \\ 
& \mbox{b)} & \ \w{{(c_{1}A+c_{2}B)}}=c^{\ast}_{1}\w{A}+c^{\ast}_{2}\w{B}, \\
& \mbox{c)}& \ \w{A^{\a}}=\w{A}^{\a}.
\eea
{\it Axioma 3 }. O v\'acuo t\'ermico \'e invariante sob as regras de conjuga\c{c}\~ao til:
\be
\w{|0(\b) \ra}=|0(\b) \ra.
\ee
{\it Axioma 4 }. Transla\c{c}\~oes espa\c{c}o-temporais de um operador $A \in \Im $ \'e induzida pelo operador energia-momento $P_{\mu} \in \Im $ da seguinte maneira
\be
A(x)=e^{iP_{\mu}x^{\mu}}Ae^{-iP_{\mu}x^{\mu}}.
\ee
{\it Axioma 5 }. O v\'acuo t\'ermico \'e definido pelas seguintes rela\c{c}\~oes operatoriais chamadas condi\c{c}\~oes de estado t\'ermico:
\be
A(t,\ve{x})|0(\b) \ra=\s \w{A}^{\a}(t-i\b/2, \ve{x})|0(\b) \ra, \label{cet1}
\ee 
\be
\la O(\b)|A(t,\ve{x})=\la O(\b)| \w{A}^{\a}(t+i\b/2), \ve{x}) \s^{\ast}, \label{cet2}
\ee 
onde no caso em que $A$ \'e uma vari\'avel bos\^onica, escolhemos $\s = 1$. \\
{\it Axioma 6 }. Definiremos a dupla conjuga\c{c}\~ao til como segue:
\be
\w{\w{A}}= \s A.
\ee

        Existe uma certa liberdade na escolha da defini\c{c}\~ao das condi\c{c}\~oes de estado t\'ermico, axioma 5, e dupla conjuga\c{c}\~ao til, axioma 6,  devido a presen\c{c}a do fator de fase $\s$ que deve satisfazer somente $|\s|=1$. No trabalho de H.Matsumoto, Y.Nakano e H.Umezawa~{\cite{mnu}}, s\~ao apresentadas regras gerais para a escolha deste fator de fase para operadores de Heisenberg arbitr\'arios. 

        A import\^ancia das regras de conjuga\c{c}\~ao til, axioma 2, est\'a no fato de que todas as rela\c{c}\~oes usuais em TQC, tais como rela\c{c}\~oes de comuta\c{c}\~ao e equa\c{c}\~oes de Heisenberg, podem ser generalizadas para DCT atrav\'es destas regras (como veremos adiante).

        Podemos observar tamb\'em que a condi\c{c}\~ao de estado t\'ermico define o v\'acuo t\'ermico, sendo esta uma das mais importantes rela\c{c}\~oes em DCT. Esta rela\c{c}\~ao mostra tamb\'em que sempre existe uma certa combina\c{c}\~ao dos operadores $A(x)$ e ${\w{A}}^{\a}(x)$ que aniquila o v\'acuo t\'ermico. Esta \'e uma importante caracter\'{\i}stica da DCT que n\~ao existe na TQC usual.

        Podemos agora generalizar o axioma 1 usando os axiomas 2 e 4 do seguinte modo. Se  $A(x) \in \Im \ \mbox{e} \ \w{B}(y) \in \w{\Im} $, ent\~ao eles comutam em todo o espa\c{c}o-tempo:
\be 
[A(x),\w{B}(y)]=0.
\ee

       Se realizarmos uma opera\c{c}\~ao til-dagger ou dagger-til, podemos verificar a partir do axioma 2 que os coeficientes dos operadores permanecem inalterados. Desta forma, introduziremos a nota\c{c}\~ao de dubleto t\'ermico como:
\be
A^{\al}= \cases {A, \ \mbox{se} \ \al=1 \cr
\w{A}^{\a}, \ \mbox{se} \ \al=2.  \cr }   \label{dt}
\ee

        Deste modo, qualquer fun\c{c}\~ao de um operador $A$, digamos $F(A)$, pode ser colocada na forma de dubletos t\'ermicos como
\be
[F(A)]^{\al}=P_{\al}F(A^{\al}), \label{oter}
\ee
onde $P_{\al}$ \'e o operador de ordenamento t\'ermico definido como 
\be 
P_{\al}[A^{\al}B^{\al}\dots C^{\al}]=\cases {A^{1}B^{1} \dots C^{1}, \ \mbox{se} \ \al=1  \cr
 C^{2} \dots B^{2}A^{2}, \ \mbox{se} \ \al=2. \cr} \label{operador}
\ee
Como exemplo, estenderemos a equa\c{c}\~ao de Heisenberg quadridimensional 
\be
i\6_{\mu}\ps(x)=[\ps(x),P_{\mu}]
\ee 
para a nota\c{c}\~ao de dubleto t\'ermico como
\be 
i\6_{\mu}\ps^{\al}(x)=\e^{\al}[\ps^{\al}(x),P^{\al}_{\mu}], \label{eht}
\ee                             
onde $\e^{\al}=1(\al=1) \ \mbox{e} \ -1(\al=2)$ \'e introduzido para preservar as rela\c{c}\~oes de comuta\c{c}\~ao que s\~ao alteradas devido ao ordenamento t\'ermico definido em (\ref{operador}). 

        O gerador total de transla\c{c}\~oes espa\c{c}o-temporais em DCT \'e dado por
\be
\hat{P}_{\mu}=\sum_{\al}\e^{\al}P^{\al}_{\mu}=P_{\mu}-{\w{P}}_{\mu},
\ee
e desta forma, a eq.~(\ref{eht}) pode ser escrita como
\be 
i\6_{\mu}\ps^{\al}(x)=[\ps^{\al}(x),\hat{P}_{\mu}]. \label{ehts}
\ee 
Podemos verificar que esta constru\c{c}\~ao est\'a de acordo com o axioma 2. Por exemplo, se fizermos $\al=1$ na eq.~(\ref{ehts}), obtemos 
\be 
i\6_{\mu}\ps(x)=[\ps(x),P_{\mu}], 
\ee  
e seu associado til, de acordo com o axioma 2, \'e 
\be 
i\6_{\mu}{\w{\ps}}(x)=-[{\w{\ps}}(x),\w{P}_{\mu}], \label{al=1}
\ee
que \'e justamente a express\~ao para $\al=2$ na  eq.~(\ref{ehts}) supondo que os campos sejam Hermitianos. Caso os campos n\~ao sejam Hermitianos, esta constru\c{c}\~ao tamb\'em \'e v\'alida, basta tomarmos o Hermitiano conjugado da eq.~(\ref{al=1}).

        Dada a Lagrangeana ou a Hamiltoniana de um sistema f\'{\i}sico, podemos obter a Lagrangeana ou a Hamiltoniana t\'ermicas, respectivamente, da seguinte maneira:
\be
\hat{H}=\sum_{\al}\e^{\al}H^{\al}=H-\w{H}, \ \  \ \hat{L}=\sum_{\al}\e^{\al}L^{\al}=L-\w{L}.
\ee

        O comutador $[A(x),B(y)]=C(x,y)$ pode ser estendido como:
\be
[A^{\al}(x),B^{\al}(y)]=\t^{\al \g}C^{\g}(x,y),
\ee
onde
\be
\t=\fourmat{1}{0}{0}{-1}  \label{tau}.
\ee

        A condi\c{c}\~ao de estado t\'ermico nos conduz \`a exist\^encia de um operador que aniquila o v\'acuo t\'ermico
\be
a(\b,t)|0(\b)\ra= \w{a}(\b,t)|0(\b)\ra=\la 0(\b)|a^{\a}(\b,t)=\la 0(\b)|\w{a}^{\a}(\b,t)=0. \label{oavt}
\ee 
Estes operadores de aniquila\c{c}\~ao, $a(\b,t)$ e $\w{a}(\b,t)$, podem ser constru\'{\i}dos de qualquer outro operador, o que expressa a  condi\c{c}\~ao de estado t\'ermico. Assumindo que $A(t)$ tenha um Hermitiano conjugado $A^{\a}(t)$ tal que 
\be
A(t)|0(\b)\ra= {\w{A}}^{\a}(t-i\b/2)|0(\b)\ra,
\ee
\be
A^{\a}(t)|0(\b)\ra={\w{A}}(t-i\b/2)|0(\b)\ra,
\ee
onde estamos omitindo as vari\'aveis espaciais, pois neste caso somente as vari\'aveis temporais t\^em import\^ancia, estes operadores podem ser constru\'{\i}dos na forma
\be
a(\b,t)=f^{1/2}(-i\6_{t}) \left (A(t+i\b/2)-{\w{A}}^{\a}(t) \ri ),
\ee 
\be
\w{a}(\b,t)=f^{1/2}(-i\6_{t})^{\ast} \left (\w{A}(t-i\b/2)-A^{\a}(t) \ri),
\ee
onde $f^{1/2}$ \'e um operador diferencial que ser\'a determinado adiante. Usando a nota\c{c}\~ao de dubleto t\'ermico, temos que os operadores t\'ermicos est\~ao relacionados com os operadores \`a temperatura nula atrav\'es de uma transforma\c{c}\~ao de Bogoliubov na forma
\be
a^{\al}(\b,t)=U^{-1}(-i\6_{t})^{\al \g}A^{\g}(t), \label{tb}
\ee 
onde 
\be
U^{-1}(\o)=f^{1/2}(\o) \fourmat{\e ^{\b \o /2}}{-1}{-1}{\e ^{\b \o /2}}.
\ee
Ela est\'a normalizada como
\be
U^{-1}(\o)\t U^{-1}(\o)^{\a}=\t,
\ee
onde $\t$ est\'a definido em~(\ref{tau}). Desta condi\c{c}\~ao de normaliza\c{c}\~ao, obtemos a seguinte express\~ao para $f(\o)$
\be
f(\o)=\frac{1}{e^{\b \o} - 1}. \label{f}
\ee
Podemos tamb\'em definir a transforma\c{c}\~ao inversa de~(\ref{tb}) atav\'es de uma superposi\c{c}\~ao n\~ao local no tempo dos operadores que aniquilam o v\'acuo t\'ermico $|0(\b) \ra $ e seu dual $\la 0(\b)|$
\be
A^{\al}(t)=U(-i\6_{t})^{\al \g}a^{\g}(\b,t). \label{tbi}
\ee 

        Podemos mostrar que os operadores \`a temperatura nula est\~ao conectados aos operadores \`a temperatura finita na forma can\^onica como segue
\be
A^{(~)}=e^{iG(\b)}a(\b,t)^{(~)}e^{-iG(\b)},
\ee
e o v\'acuo t\'ermico pode ser mapeado no v\'acuo duplicado \`a temperatura nula na forma
\be
|O(\b)\ra=e^{-iG(\b)}|0,\w{0}\ra, \label{mv}
\ee
com o gerador $G$ satisfazendo
\be
G(\b)=G^{\a}(\b)=-\w{G}(\b).
\ee
Este gerador \'e chamado de {\it operador de Bogoliubov }. Podemos ainda obter a eq.~(\ref{vab}) partindo da eq.~(\ref{mv}) com um $G(\b)$ apropriado. 



\section{Teoria de Corda \`a Temperatura Finita}

${}$ 

       Neste cap\'{\i}tulo, construiremos uma teoria de cordas bos\^onicas \`a temperatura finita usando o formalismo de H. Umezawa e Y. Takahashi desenvolvido no cap\'{\i}tulo 3 para descrevermos uma Din\^amica de Campos T\'ermicos. Este formalismo pode ser convenientemente aplicado em sistemas descritos por osciladores (um exemplo desta aplica\c{c}\~ao foi feito na se\c{c}\~ao 3.1.1). Deste modo, tal constru\c{c}\~ao ser\'a poss\'{\i}vel, uma vez que o estado de uma corda, que pode ser completamente representado pelo campo $X^{\mu}$, pode ser expandido em termos de operadores de cria\c{c}\~ao $\al^{\m \dagger}_n$  e destrui\c{c}\~ao $ \al^{\m}_n $. De acordo com a Din\^amica de Campos T\'ermicos, a termodin\^amica deste sistema pode ser descrita em um espa\c{c}o de Fock, composto do espa\c{c}o de Fock original da corda e de uma c\'opia id\^entica a este (que ser\'a denotada por $~\tilde{}~$), ou seja, estamos duplicando o sistema. Novos graus de liberdade surgir\~ao devido a esta duplica\c{c}\~ao. As duas c\'opias s\~ao independentes e o espa\c{c}o de Fock total \'e dado pelo produto direto dos espa\c{c}os de Fock das c\'opias.

        Como j\'a exposto, para implementarmos esta constru\c{c}\~ao para o caso de uma corda bos\^onica livre, iremos escrever os operadores de cria\c{c}\~ao e destrui\c{c}\~ao da corda f\'{\i}sica, cujos quais denotaremos por 
\be
A^{\m}_n = \frac{1}{\sqrt{n}}\al^{\m}_n ~~~;~~~A^{\m \dagger}_n = 
\frac{1}{\sqrt{n}}\al^{\m}_{-n}~~,
\ee
\be
B^{\m}_n = \frac{1}{\sqrt{n}}\b^{\m}_n ~~~;~~~B^{\m \dagger}_n = 
\frac{1}{\sqrt{n}}\b^{\m}_{-n}~~,
\ee
onde $n>0$. C\'opias id\^enticas existem para ambos os conjuntos de osciladores do sistema,
\be
\w{A}^{\m}_n = \frac{1}{\sqrt{n}}\w{\al}^{\m}_n ~~~;~~~\w{A}^{\m \dagger}_n = 
\frac{1}{\sqrt{n}}\w{\al}^{\m}_{-n}~~,
\ee
\be
\w{B}^{\m}_n = \frac{1}{\sqrt{n}}\w{\b}^{\m}_n ~~~;~~~\w{B}^{\m \dagger}_n = 
\frac{1}{\sqrt{n}}\w{\b}^{\m}_{-n}.
\ee
Estes operadores satisfazem duas \'algebras independentes como segue
\be
[A^{\m}_n,A^{\nu \a}_m]=[\w{A}^{\m}_n,\w{A}^{\nu \a}_m]=\d_{n+m} \h^{\mu \nu} \ \ , \ \ [B^{\m}_n,B^{\nu \a}_m]=[\w{B}^{\m}_n,\w{B}^{\nu \a}_m]=\d_{n+m} \h^{\mu \nu}, \label{comut1}
\ee
\be
[A^{\m}_{n},{\w{A}}^{\nu}_{m}]=[A^{\m}_n,{\w{A}}^{\nu \a}_m]=[A^{\m}_{n},{\w{B}}^{\nu}_{m}]= \cdots =0.\label{comut2}
\ee
O espa\c{c}o de Fock extendido do sistema total \'e dado pelo produto direto de dois espa\c{c}os de Fock das cordas
\be
\hat{\cal H}={\cal H} \bigotimes \w{\cal H} .
\ee
Denotaremos um estado de $\hat{\cal H}$ por $ | ~~ \ra \ra $. Os estados de v\'acuo da corda para cada um dos setores podem ser escritos, como: 
\be 
{|0 \ra \ra} _{\al}={|0 \ra}_{\al} \bigotimes \w{|0 \ra}_{\al}=|0,0 \ra_{\al} \ \ \mbox{e}\ \ \ {|0 \ra \ra} _{\b}={|0 \ra}_{\b} \bigotimes \w{|0 \ra}_{\b}=|0,0 \ra_{\b}. \label{etca}
\ee
Portanto, o v\'acuo total para a corda fechada pode ser escrito como:
\bea
|0 \ra \ra={|0 \ra \ra} _{\al}{|0 \ra \ra} _{\b}= \le ({|0 \ra}_{\al} \bigotimes \w{|0 \ra}_{\al} \ri )\le ({|0 \ra}_{\b} \bigotimes \w{|0 \ra}_{\b} \ri ) \non =\le ({|0 \ra}_{\al} \bigotimes {|0 \ra}_{\b} \ri )\le (\w{|0 \ra}_{\al} \bigotimes \w{|0 \ra}_{\b} \ri ), \label{etcf}
\eea
onde a \'ultima equa\c{c}\~ao \'e consequ\^encia do fato de que as cordas original e a til s\~ao independentes, e tamb\'em  mostra explicitamente a duplica\c{c}\~ao do estado de cada oscilador e a estrutura do estado de v\'acuo da corda. Ainda, para obtermos o estado fundamental, devemos multiplicar (\ref{etca}) e (\ref{etcf}) respectivamente por $ |p \ra $ e $ |p \ra \bigotimes \w{|p \ra }$. 

        Uma vez duplicado o sistema, sua descri\c{c}\~ao t\'ermica pode ser obtida atuando os operadores unit\'arios de Bogoliubov $G^{\al}_{n}$ e $G^{\b}_{n}$   em cada um dos setores dos espa\c{c}os f\'{\i}sico e extendido de Hilbert e nos operadores de cria\c{c}\~ao e destrui\c{c}\~ao de acordo com as regras estabelecidas para a DCT. Os operadores $G$ s\~ao definidos como segue
\bea 
G^{\al}_{n} &=& -i\th(\b_{T})\le (A_{n} \cdot {\w{A}}_{n}-A^{\a}_{n} \cdot {\w{A}}^{\a}_{n} \ri ) \non
 G^{\b}_{n}&=& -i\th(\b_{T})\le (B_{n} \cdot {\w{B}}_{n}-B^{\a}_{n} \cdot {\w{B}}^{\a}_{n} \ri ), \label{obtf}
\eea
onde $\b_{T}=\frac{1}{k_{B}T}$ e $\th_{n}(\b_{T})$ \'e um par\^ametro real que depende da estat\'{\i}stica do $nth$ modo do oscilador (como mostrado no cap.3)
\be
\cosh {\th_{n}}(\b _{T})=(1-e^{\b_{T}n})^{-1}.
\ee
No entanto, $\th_{n}(\b _{T})$ \'e o mesmo tanto para o movimento para a esquerda quanto para a direita para um dado modo $n$ dos osciladores. O ponto em (\ref{obtf}) representa o produto escalar no espa\c{c}o de Minkowski $A_{n} \cdot {\w{A}}_{n}=A^{\mu}_{n} {\w{A}}_{n\mu}$. Sendo os modos para a direita e esquerda independentes, os operadores de Bogoliubov comutam como segue:
\be
[G_{n}^{\al},G_{m}^{\al}]=[G_{n}^{\b},G_{m}^{\b}]=[G_{n}^{\al},G_{m}^{\b}]=0.\label{c}
\ee
Das defini\c{c}\~oes expressas em (\ref{obtf}) para os operadores $G$, podemos observar que estes s\~ao Hermitianos, ou seja,
\be
(G_{n}^{\al})^{\a}=G_{n}^{\al} \ \ \ \ \mbox{e} \ \ \ \  (G_{n}^{\b})^{\a}=G_{n}^{\b}. \label{herm}
\ee
E para $n$'s negativos temos
\be
G_{|n|}^{\al}=-G_{-n}^{\al} \label{neg}.
\ee

Uma \'algebra simples nos fornece as rela\c{c}\~oes de comuta\c{c}\~ao entre os operadores $G$ e os osciladores
$$
[G_{n}^{\al},A^{\mu}_{n}]=-i \th_{n}({\b}_{T}){\w{A}}_{n}^{\mu \a}, \ \ \ \  [G_{n}^{\al},B^{\mu}_{n}]=-i \th_{n}({\b}_{T}){\w{B}}_{n}^{\mu \a}, 
$$
$$
[G_{n}^{\al},A^{\mu \a}_{n}]=-i \th_{n}({\b}_{T}){\w{A}}_{n}^{\mu}, \ \ \ \  [G_{n}^{\al},B^{\mu \a}_{n}]=-i \th_{n}({\b}_{T}){\w{B}}_{n}^{\mu}, 
$$
$$
[G_{n}^{\al},{\w{A}}^{\mu}_{n}]=-i \th_{n}({\b}_{T}){A}_{n}^{\mu \a}, \ \ \ \  [G_{n}^{\al},{\w{B}}^{\mu}_{n}]=-i \th_{n}({\b}_{T}){B}_{n}^{\mu \a}, 
$$
\be
[G_{n}^{\al},{\w{A}}^{\mu \a}_{n}]=-i \th_{n}({\b}_{T}){A}_{n}^{\mu}, \ \ \ \  [G_{n}^{\al},{\w{B}}^{\mu \a}_{n}]=-i \th_{n}({\b}_{T}){B}_{n}^{\mu}. \label{rcobo}
\ee

        Vamos agora construir o estado de v\'acuo e operadores de cria\c{c}\~ao e destrui\c{c}\~ao em  $T \neq 0 $. Atuando no v\'acuo \`a temperatura nula com o operador de Bogoliubov adequado, obtemos estados dependentes explicitamente da temperatura como segue
\be
\left. \left| 0(\b_T ) \right\rangle \! \right\rangle ~= ~\prod_{ m > 0} 
e^{iG_{m}}
\left. \left| 0 \right\rangle \! \right\rangle=\prod_{ m > 0}|0(\b_{T})_{m} \ra \ra  ,
\label{vacT}
\ee
para a corda aberta e 

\be
\left. \left| 0(\b_T ) \right\rangle \! \right\rangle = \prod_{ m > 0} 
e^{iG_m^{\al}}
\left. \left| 0 \right\rangle \! \right\rangle_{\al} ~\prod_{ n > 0} 
e^{iG_n^{\b}} 
\left. \left| 0 \right\rangle \! \right\rangle_{\b}=\prod_{ m > 0}|0(\b_{T})_{m} \ra \ra _{\al} \prod_{ n > 0}|0(\b_{T})_{n} \ra \ra _{\b}  
\label{vacfT}
\ee
para a fechada. Uma vez que os operadores $G$ n\~ao misturam os estados dos modos para a direita e para a esquerda, podemos construir o estado de v\'acuo para a corda fechada \`a temperatura finita na forma
\be
|0(\b_{T})_{n} \ra \ra = |0(\b_{T})_{n} \ra \ra _{\al}\bigotimes|0(\b_{T})_{n} \ra \ra _{\b}.
\ee
Os operadores de cria\c{c}\~ao e destrui\c{c}\~ao em $T \neq 0$, que aniquilam estes estados de v\'acuo, s\~ao obtidos a partir dos conjuntos de operadores $\{ A, A^{\a},\w{A}, \w{A}^{\a} \}$ e $\{ B, B^{\a},\w{B}, \w{B}^{\a} \}$ por transforma\c{c}\~ao de Bogoliubov como segue:

\bea
A^{\m}_{n}(\b_T) &= & e^{iG_{n}^{\al}}A ^{\m}_{n}e^{-iG^{\al}_n}=u_{n}(\b_{T}){A}^{\m}_{n}-v_{n}(\b_{T})\w{A}^{\m \a}_{n}, \non
\tilde{A}^{\m}_{n}(\b_T) &=& e^{iG^{\al}_n}\tilde{A} ^{\m}_{n}e^{-iG_{n}^{\al}}=u_{n}(\b_{T})\w{A}^{\m}_{n}-v_{n}(\b_{T}){A}^{\m \a}_{n},
\label{annihT}
\eea  
para a corda aberta e para o setor de movimento para a direita da corda fechada. Para o setor de movimento para a esquerda da corda fechada temos   
\bea
B^{\m}_{n}(\b_T) &= & e^{iG_{n}^{\b}}B ^{\m}_{n}e^{-iG^{\b}_n}=u_{n}(\b_{T}){B}^{\m}_{n}-v_{n}(\b_{T})\w{B}^{\m \a}_{n}, \non
\tilde{B}^{\m}_{n}(\b_T) &=& e^{iG^{\b}_n}\tilde{B} ^{\m}_{n}e^{-iG_{n}^{\b}}=u_{n}(\b_{T})\w{B}^{\m}_{n}-v_{n}(\b_{T}){A}^{\m \a}_{n}.
\label{annihT1}
\eea 
Estes resultados foram obtidos usando (\ref{herm}), (\ref{neg}), (\ref{c}) e (\ref{rcobo}). Aqui,
\be
u_{n}(\b_{T})=\cosh {\th_{n}}(\b_{T}) \ \ \ \ , \ \ \ \ v_{n}(\b_{T})=\sinh \th_{n}(\b_{T}).
\ee
Desde que os operadores $\hat{p} , \ \hat{X}, \ \hat{\w{p}}, \ \hat{\w{X}}$ comutam com todos os operadores dos osciladores, eles n\~ao s\~ao afetados pelas transforma\c{c}\~oes de Bogoliubov, ou seja, as coordenadas de momento e centro de massa s\~ao invariantes por transforma\c{c}\~oes de Bogoliubov.

        Os operadores \`a temperatura finita satisfazem a \'algebra dos osciladores (\ref{comut1}) e (\ref{comut2}) para cada modo, em cada setor e para ambas as c\'opias da corda original, e todas estas \'algebras s\~ao independentes. Sendo assim, os estados de um sistema \`a temperatura finita s\~ao obtidos atuando no v\'acuo t\'ermico os operadores de cria\c{c}\~ao e destrui\c{c}\~ao \`a temperatura finita, definidos em (\ref{vacT}) e (\ref{vacfT}). Os estados obtidos desta forma pertencem a um espa\c{c}o de Fock t\'ermico.

       Uma importante quest\~ao \'e se a constru\c{c}\~ao desta teoria de cordas \`a temperatura finita possui a mesma estrutura da teoria em $T=0$, ou se existe algum mapeamento entre elas. Se substituirmos os operadores \`a temperatura finita na solu\c{c}\~ao da equa\c{c}\~ao de movimento da corda bos\^onica, obtemos uma solu\c{c}\~ao que depende de $T$ e mistura as cordas original e a til. Este resultado \'e uma consequ\^encia da defini\c{c}\~ao dos operadores $G$ que n\~ao afeta a estrutura da folha mundo, mas somente os coeficientes de Fourier. Podemos mostrar que todas as propriedades da corda bos\^onica \`a temperatura zero s\~ao satisfeitas. Em particular, podemos construir das solu\c{c}\~oes da equa\c{c}\~ao de movimento da corda \`a temperatura finita, o tensor energia momento que possui a mesma forma do tensor para $T=0$. Ent\~ao, os seguintes operadores 
\be
L_{n}^{\al}(\b_{T})= \frac{1}{2} \sum_{k} \al_{-k}(\b_{T}) \cdot \al_{k+m}(\b_{T}), \ \ \ \ {\w{L}}_{n}^{\b}(\b_{T})= \frac{1}{2} \sum_{k} \b_{-k}(\b_{T}) \cdot \b_{k+m}(\b_{T}),
\ee
podem ser constru\'{\i}dos e a \'algebra de Virasoro pode ser mostrada usando as propriedades dos operadores de Bogoliubov (\ref{herm}), (\ref{neg}, (\ref{c}) e (\ref{rcobo}).  Uma vez que estamos trabalhando com duas c\'opias do mesmo sistema, tudo que foi dito acima \'e v\'alido tamb\'em para a corda til. No entanto, nesta teoria de temperatura finita, a no\c{c}\~ao de corda e corda til s\~ao um pouco diferentes desde que ambas misturam os operadores da corda e da corda til como definidos inicialmente em $T=0$. Deste modo, o s\'{\i}mbolo til nos lembra somente que os operadores foram obtidos de uma c\'opia do sistema original.

\subsection{ A Entropia dos Estados de uma Corda Bos\^onica Aberta com Dependencia das Condi\c{c}\~oes de Contorno}
${}$

        Consideremos uma corda aberta no espa\c{c}o de Minkowisk. Para evitarmos a introdu\c{c}\~ao de fantasmas na teoria, iremos trabalhar no calibre de cone de luz $X^{0} \pm X^{25}$ apresentado no cap\' {\i}tulo 2. Neste calibre, o conjunto dos \'{\i}ndices correm de $ \mu, \nu = 1,2, \cdots 24$. Consideraremos aqui somente \`as solu\c{c}\~oes sujeitas as condi\c{c}\~oes de contorno de Neumann em ambas as extremidades. Os outros casos podem ser tratados da mesma maneira. Neste caso, escreveremos o operador de Bogoliubov definido em  (\ref{obtf}) como 
\be
G_n = \sum_{\m=1}^{24}G^{\m}_{n}.
\label{opbogoltot}
\ee 
e denotaremos o v\'acuo t\'ermico total do sistema da seguinte forma  
\be
\left. \left| \O (\b_T) \right\rangle \right\rangle ~=~ \left. \left| 0(\b_T) \right\rangle \right\rangle
\left| p \right\rangle \left| \tilde{p} \right\rangle. 
\label{totvacT}
\ee

        Uma vez que estamos tratando a corda como um conjunto de osciladores bos\^onicos, o operador entropia para a corda bos\^onica pode ser escrito, da defini\c{c}\~ao (\ref{centropia}), como segue: 
\bea
K ~&=& ~\sum_{\m = 1}^{24}\sum_{n=1}^{\infty}( A^{\m \dagger}_n A^{\m}_n 
\log \sinh^2 \th_n -A^{\m}_n A^{\m \dagger}_n \log \cosh^2 \th_n )
\label{entropy}\\
\tilde{K} ~&=&  \sum_{\m = 1}^{24}\sum_{n=1}^{\infty}(\tilde{A}^{\m \dagger}_n
\tilde{A}^{\m}_n \log \sinh^2 \th_n -
\tilde{A}^{\m}_n \tilde{A}^{\m \dagger}_n \log \cosh^2 \th_n ).
\label{entropytilde}
\eea
Como j\'a foi dito no cap.3, o v\'acuo da teoria \'e invariante sob a opera\c{c}\~ao til e toda a informa\c{c}\~ao f\'{\i}sica est\'a contida no sistema sem til, deste modo, como a entropia de um estado \'e dada pelo valor m\'edio do operador entropia neste estado, iremos somente calcul\'a-lo para o operador (\ref{entropy}). Para realizarmos os c\'alculos, \'e conveniente escrevermos (\ref{entropy}) na forma
\be
K ~ =~ \sum_{\m = 1}^{24} K^{\m}.
\label{decomentrop}
\ee

A id\'eia b\'asica destes c\'alculos \'e primeiramente calcularmos a contribui\c{c}\~ao dos elementos de matriz do operador entropia para cada dire\c{c}\~ao do espa\c{c}o-tempo. Como resultado obtido da eq. (\ref{0}), a entropia da corda representa a soma das entropias de todos os osciladores. 

        Nosso objetivo \'e encontrarmos a entropia da corda associada \`a solu\c{c}\~ao mais geral da equa\c{c}\~ao de movimento. Neste caso, a entropia \'e uma fun\c{c}\~ao do campo que descreve a folha mundo e a depend\^encia com os par\^ametros da folha mundo s\~ao ditados pelas condi\c{c}\~oes de contorno impostas nas equa\c{c}\~oes de movimento. Um vetor de estado geral para a corda \`a temperatura finita $\left.\left| X^{\m}(\b_T) \right\rangle\right\rangle$ pode ser obtido operando com a solu\c{c}\~ao (\ref{sol}) no estado de v\'acuo t\'ermico (\ref{totvacT}), uma vez que quantizando o sistema, a solu\c{c}\~ao (\ref{sol}) torna-se um operador que atua no espa\c{c}o de Fock.

        O elemento de matriz  
$\left\langle\left\langle X^{\m}(\b_T)\left| K^{\rho} \right| X^{\m}(\b_T )
\right\rangle\right\rangle$
pode ser ser dividido em duas partes, uma contendo somente contribui\c{c}\~oes dos osciladores e outra contendo as contribui\c{c}\~oes das coordenadas e momentos do centro de massa $(CM)$
\be
\left\langle\left\langle X^{\m}(\b_T)\left| K^{\rho} \right| X^{\m}(\b_T )
\right\rangle\right\rangle
~=~{\mbox{ CM}} 
- 2\al ' \sum_{n,k,l >0}\frac{e^{i(l-n)\t}}{\sqrt{ln}}\cos n\s \cos l\s
\left[ (T_1)^{\m\rho\nu}_{nkl} + (T_2)^{\m\rho\nu}_{nkl}\right], \ \ 
\label{matrentr}
\ee
onde definimos
\bea
(T_1)^{\m\rho\nu}_{nkl}=
\left\langle \tilde{0} \left| \left\langle 1^{\m}_n \left|
\prod_{m>0}e^{-iG_m}A^{\rho \dagger}_{k}A^{\rho}_{k}\log \sinh^2 \theta_k 
\prod_{s>0}e^{iG_s}
\right| 1^{\nu}_l \right\rangle \right| \tilde{0} \right\rangle
\left\langle \tilde{p} \left|  \tilde{q} \right\rangle \right.
\left\langle p \left| q \right \rangle \right.\nonumber\\
(T_2)^{\m\rho\nu}_{nkl}=
- \left\langle \tilde{0} \left| \left\langle 1^{\m}_n \left|
\prod_{m>0}e^{-iG_m}A^{\rho }_{k}A^{\rho \dagger}_{k}\log \cosh^2 \theta_k 
\prod_{s>0}e^{iG_s}
\right| 1^{\nu}_l \right\rangle \right| \tilde{0} \right\rangle
\left\langle \tilde{p} \left|  \tilde{q} \right\rangle \right.
\left\langle p \left| q \right \rangle \right.
\label{Ts}
\eea
e
\be
\left| 1^{\m}_{l} \right\rangle ~= ~A^{\m \dagger}_{l} \left| 0 \right\rangle.
\label{field}
\ee
Aqui, consideramos a normaliza\c{c}\~ao usual dos estados de momento em um volume $V_{24}$ no espa\c{c}o transverso 
\bea
\left.\left\langle p \right| q \right\rangle &~=~& 2 \pi 
\delta^{(24)} (p-q) \label{normstate1}\\
(2\pi )^{24}\delta^{(24)}(0) &~=~& V_{24}.
\label{normstate2}
\eea
Manipulando algebricamente (\ref{matrentr}), a contribui\c{c}\~ao dos osciladores para o elemento de matriz do operador entropia toma a forma
\bea
\left\langle\left\langle X^{\m}(\b_T)\left| K^{\rho} \right| X^{\m}(\b_T )
\right\rangle\right\rangle
&=&{\mbox{CM}} 
-2 \al ' (2 \pi)^{(48)} \delta^{\m \nu} \delta^{(24)}(p-q)\delta^{(24)}
(\tilde{p}-\tilde{q}) \times 
\nonumber\\
& &\sum_{n>0}\frac{1}{n}\cos^2 n\s [\log (\tanh \th_n )^2\delta^{\rho \nu } -
\delta^{\rho \rho }\sum_{k>0} \delta_{k k}].
\label{matrelemCM}
\eea

 O termo contendo  as CM cont\'em tanto as contribui\c{c}\~oes das coordenadas e momentos quanto dos osciladores. Para escrevermos explicitamente a forma desta contribui\c{c}\~ao para o elemento de matriz do operador entropia, devemos dividir ainda este termo em duas partes, uma contendo somente contribui\c{c}\~oes das coordenadas e centro de massa e outra dos osciladores. Para o c\'alculo das contribui\c{c}\~oes devido \`as coordenadas e momentos, usaremos as rela\c{c}\~oes de completeza dos auto-estados dos operadores de momento junto com o elemento de matriz   
\be
\left. \left\langle x \right| p \right\rangle ~=~(2\pi \hbar ) ^{-12}
e^{i x \cdot p / \hbar}.
\label{matrixxp}
\ee
Para calcularmos a contribui\c{c}\~ao do termo devido aos osciladores, podemos expressar os osciladores em $T=0$ em termos dos osciladores em $T \neq 0$ ou escrever o v\'acuo \`a temperatura finita em termos do v\'acuo \`a temperatura nula. Os dois modos conduzem ao mesmo resultado, no entanto, o primeiro nos fornece somente rela\c{c}\~oes polinomiais entre os operadores de cria\c{c}\~ao e destrui\c{c}\~ao \`a temperatura finita. Usando as propriedades dos operadores de Bogoliubov mostradas na se\c{c}\~ao anterior, podemos mostrar que as contribui\c{c}\~oes vindas dos termos que misturam os operadores do centro de massa com a parte dos osciladores s\~ao canceladas. Os termos diferentes de zero s\~ao todos proporcionais a $\left\langle \left \langle 0(\b_T) \left| K^{\rho} \right| 0(\b_T)\right\rangle \right\rangle$, que representa a entropia de um n\'umero infinito de osciladores bos\^onicos na $\rho$'\'esima dire\c{c}\~ao do espa\c{c}o-tempo. A rela\c{c}\~ao final obtida para a entropia levando em conta a contribui\c{c}\~ao de todos os termos \'e  

\bea
& &\left\langle\left\langle X^{\mu}(\b_T)\left| K^{\rho} \right| 
X^{\m}(\b_T )\right\rangle\right\rangle
~=~ \nonumber\\
&- &(2\pi \hbar)^{-24}\left[ (2\pi\hbar)^{24}(2\al ' \t)^2 p^{\m} p'^{\nu}
\delta^{(24)}(p-p') 
\right.
+ \left.2\al ' \t (I^{\m}_2p'^{\nu} + I'^{\nu}_2 p^{\m}) + 
I^{\m}_2I^{\nu}_2\prod_{j \neq \m , \nu}I^j_1 \right]\nonumber\\ 
&\times &\delta^{(24)}(\tilde{p} - \tilde{p'})\sum_{m=1} 
\left[{\mbox n}^{\rho}_m 
\log {\mbox  n}^{\rho}_m + (1- {\mbox  n}^{\rho}_m) 
\log( 1- {\mbox  n}^{\rho}_m ) 
\right]
- 2 \al ' (2 \pi)^{(48)} \delta^{\m \nu} \delta^{(24)}(p-p')\nonumber\\
&\times &\delta^{(24)}(\tilde{p}-\tilde{p'})  
\sum_{n>0}\frac{1}{n}\cos^2 n\s \left[ \log (\tanh \th_n )^2\delta^{\rho \nu } -
\delta^{\rho \rho }\sum_{k>0} \delta_{k k} \right]
\label{entropyRho},
\eea
onde as integrais unidimencionais no dominio finito
$ x \in \left[ x_0, x_1\right] $ 
s\~ao dadas por
\bea
I_1 &~=~& -i \hbar (p'-p)^{-1}\left[ e^{\frac{i}{\hbar}x_1(p'-p)} - 
e^{\frac{i}{\hbar}x_0(p'-p)}\right]
\label{int1}\\
I_2 &~=~& -i \hbar (p'-p)^{-1}\left[  -i\hbar I_1 + 
x_1e^{\frac{i}{\hbar}x_1(p'-p)} 
- x_0 e^{\frac{i}{\hbar}x_0(p'-p)}\right],
\label{int2}
\eea 
e os estados de momento final e inicial s\~ao denotados, respectivamente, por
$\left| p \right\rangle$ ,
$\left| p' \right\rangle$, e o termo
\be
{\mbox n}^{\rho}_{m} ~=~ \left\langle \left\langle 0(\b_T) \left| 
A^{\rho \dagger}_m A^{\rho}_m \right| 0(\b_T) \right\rangle\right\rangle 
= \sinh^2 \th_m \label{Nnumber} 
\ee 
representa o n\'umero de excita\c{c}\~oes da corda no v\'acuo t\'ermico.

        A equa\c{c}\~ao (\ref{entropyRho}) representa o elemento de matriz do operador entropia $K^{\rho}$ entre dois estados gerais, descritos pela equa\c{c}\~ao de movimento para a corda bos\^onica  com condi\c{c}\~oes de contorno de Neunmam em ambas as extremidades. Como expressa a equa\c{c}\~ao (\ref{decomentrop}), a entropia total \'e a soma das entropias em todas as dire\c{c}\~oes do espa\c{c}o transverso. Uma vez que as condi\c{c}\~oes de contorno s\~ao impostas nas coordenadas da folha mundo, podemos obter de modo similar express\~oes para as entropias das cordas sujeitas as outras condi\c{c}\~oes de contorno DD, DN e ND, cujas solu\c{c}\~oes est\~ao expressas no cap. 2. \'E importante observarmos que nestes casos n\~ao existem operadores associados com as coordenadas e momentos do centro de massa da corda, mas sim, vetores de posi\c{c}\~ao constantes associados com suas extremidades, n\~ao havendo, ent\~ao, contribui\c{c}\~ao destes termos para a entropia. Tais termos que misturam as coordenadas das extremidades da corda tornam-se zero pelas mesmas raz\~oes do caso NN j\'a estudado em detalhes. Os termos dos elementos de matriz diferentes de zero obtidos para cada um dos novos casos s\~ao       

\bea 
{ \mbox {DD}} &: &\left\langle\left\langle X^{\m}(\b_T)\left| K^{\rho} 
\right| X^{\m}(\b_T ) \right\rangle\right\rangle 
~=~ 
2 \al ' (2 \pi)^{(48)} \delta^{\m \nu} \delta^{(24)}(p-p')\delta^{(24)}
(\tilde{p}-\tilde{p'})  
\nonumber\\
&&\sum_{n>0}\frac{1}{n}\sin^2 n\s \left[ \log (\tanh \th_n)^2\delta^{\rho \nu } -
\delta^{\rho \rho }\sum_{k>0} \delta_{k k} \right]
\label{entrDD}\\
{\mbox {DN}}&: &\left\langle\left\langle X^{\m}(\b_T)\left| K^{\rho} \right| 
X^{\m}(\b_T )
\right\rangle\right\rangle
~=~ 
2 \al ' (2 \pi)^{(48)} \delta^{\m \nu} \delta^{(24)}(p-p')\delta^{(24)}
(\tilde{p}-\tilde{p'})  
\nonumber\\
&&\sum_{r={\cal{Z}} + 1/2 }\frac{1}{r}\sin^2 r\s \left[ \log (\tanh \th_r)^2
\delta^{\rho \nu } -
\delta^{\rho \rho }\sum_{k>0} \delta_{k k} \right]
\label{entrDN}\\
{\mbox {ND}}&: &\left\langle\left\langle X^{\m}(\b_T)\left| K^{\rho} \right| 
X^{\m}(\b_T )
\right\rangle\right\rangle
~=~ 
2 \al ' (2 \pi)^{(48)} \delta^{\m \nu} \delta^{(24)}(p-p')
\delta^{(24)}(\tilde{p}-\tilde{p'})  
\nonumber\\
&& \sum_{r={\cal{Z}} + 1/2}\frac{1}{r}\cos^2 r\s \left[ \log (\tanh \th_r)^2
\delta^{\rho \nu } -
\delta^{\rho \rho }\sum_{k>0} \delta_{k k} \right].
\label{entrND}
\eea
Aqui, ${\cal{Z}} + 1/2$ s\~ao n\'umeros inteiros.
As rela\c{c}\~oes (\ref{entropyRho}), (\ref{entrDD}), (\ref{entrDN}) e 
(\ref{entrND}) representam a entropia dos estados associados \`as solu\c{c}\~oes gerais das equa\c{c}\~oes de movimento. Elas d\~ao a entropia como fun\c{c}\~ao da folha mundo, no entanto, esta entropia n\~ao pode ser pensada como a entropia do v\'acuo da corda bos\^onica que \'e dada como a soma em todas as dire\c{c}\~oes espa\c{c}otemporais da entropia dos b\'osons escalares sem massa e n\~ao dependem da condi\c{c}\~oes de contorno.

\subsection{ Corda Bos\^onica Fechada no Grupo SU(1,1) T\'ermico}
${}$

       Nesta se\c{c}\~ao, ser\'a obtida uma express\~ao para a entropia da corda fechada de um modo um pouco diferente do usado para obtermos a express\~ao para a entropia da corda aberta na se\c{c}\~ao anterior. Aqui, consideraremos a constru\c{c}\~ao desenvolvida em {\cite{chu,hu}}, na qual podemos obter um gerador geral para as transforma\c{c}\~oes de Bogo-\\liubov, partindo de uma combina\c{c}\~ao linear de tr\^es geradores individuais que sa- \\ tisfazem independentemente as condi\c{c}\~oes necess\'arias para gerar uma transforma\c{c}\~ao de Bogoliubov. 

         Para termos uma teoria de cordas \`a temperatura finita, temos que inicialmente gerar um v\'acuo t\'ermico que, como sabemos, pode ser obtido do v\'acuo dobrado (\ref{etcf}) partindo de qualquer transforma\c{c}\~ao que misture os operadores   $A_{k}^{\mu }$, $\tilde{A}_{k}^{\mu \dagger }$ para os modos direitos, e $B_{k}^{\mu }$, $\tilde{B}_{k}^{\mu \dagger }$ para os modos esquerdos, e cujos geradores comutam com a Hamiltoniana total
\be
\hat{H} ~=~H-\tilde{H}  
~=~\sum_{n >0}^{\infty }
n\left( A_{n}^{\dagger }\cdot A_{n}+ B_{n}^{\dagger }\cdot B_{n}
 -  \tilde{A}_{n}^{\dagger }\cdot \tilde{A}_{n}- 
\tilde{B}_{n}^{\dagger }\cdot \tilde{B}_{n}\right) .
\label{exthamilt}
\ee
Al\'em de satisfazer estas condi\c{c}\~oes, tais transforma\c{c}\~oes devem ter a forma geral das transforma\c{c}\~oes de Bogoliubov, que fixam a forma dos geradores pelas seguintes rela\c{c}\~oes
\begin{equation}
\left( 
\begin{array}{c}
A^{\prime } \\ 
\tilde{A}^{\dagger \prime }
\end{array}
\right) =e^{-iG}\left( 
\begin{array}{c}
A \\ 
\tilde{A}^{\dagger }
\end{array}
\right) e^{iG}={\cal B}\left( 
\begin{array}{c}
A \\ 
\tilde{A}^{\dagger }
\end{array}
\right) ,\quad \left( 
\begin{array}{cc}
A^{\dagger ^{\prime }} & -\tilde{A}^{\prime }
\end{array}
\right) =\left( 
\begin{array}{cc}
A^{\dagger } & -\tilde{A}
\end{array}
\right) {\cal B}^{-1},
\label{Bogoliubov}
\end{equation}
onde $\cal B $ \'e uma matriz de transforma\c{c}\~ao $2 \times 2$ complexa e unit\'aria 
 \begin{equation}
{\cal B}=\left( 
\begin{array}{cc}
u & v \\ 
v^{*} & u^{*}
\end{array}
\right) ,\qquad \left| u\right| ^{2}-\left| v\right| ^{2}=1,
\label{Bmatrix}
\end{equation}
e $G$ \'e o gerador da transforma\c{c}\~ao conhecido como operador de Bogoliubov (j\'a citado no cap.1). Segundo {\cite{chu}} os operadores que satisfazem as condi\c{c}\~oes (\ref{Bogoliubov}) e (\ref{Bmatrix}) possuem a seguinte forma 
\begin{eqnarray}
g_{1_{k}}^{\alpha } &=&\theta _{1_{k}}\left( A_{k}\cdot \tilde{A}_{k}+\tilde{%
A}_{k}^{\dagger }\cdot A_{k}^{\dagger }\right) ,~~~~\qquad g_{1_{k}}^{\beta
}=\theta _{1_{k}}\left( B_{k}\cdot \tilde{B}_{k}+\tilde{B}_{k}^{\dagger
}\cdot B_{k}^{\dagger }\right) ,  \nonumber \\g_{2_{k}}^{\alpha } &=&i\theta _{2_{n}}\left( A_{k}\cdot \tilde{A}_{k}-%
\tilde{A}_{k}^{\dagger }\cdot A_{k}^{\dagger }\right) ,~~~\qquad
g_{2_{k}}^{\beta }=i\theta _{2_{k}}\left( B_{k}\cdot \tilde{B}_{k}-\tilde{B}%
_{k}^{\dagger }\cdot B_{k}^{\dagger }\right) ,  \nonumber \\
g_{3_{k}}^{\alpha } &=&\theta _{3_{n}}\left( A_{k}^{\dagger }\cdot A_{k}+%
\tilde{A}_{k}^{\dagger }\cdot \tilde{A}_{k}+1\right) ,\qquad
g_{3_{k}}^{\beta }=\theta _{3_{k}}\left( B_{k}^{\dagger }\cdot B_{k}+\tilde{B%
}_{k}^{\dagger }\cdot \tilde{B}_{k}+1\right),
\label{generators}
\end{eqnarray} 
onde os supra-\'{\i}ndices  $\al$ e $\b$ referem-se, respectivamente, aos modos dos movimentos para a direita e para a esquerda e $\theta$ \'e um par\^ametro real que depende da temperatura e foi convenientemente inclu\'{\i}do nos operadores. Estes operadores satisfazem as seguintes rela\c{c}\~oes de  comuta\c{c}\~ao: 
\be
\left[ g_{1_{k}}^{\alpha ,\beta },g_{2_{k}}^{\alpha ,\beta }\right]
=-i\Theta _{123}g_{3_{k}}^{\alpha ,\beta },\quad \left[ g_{2_{k}}^{\alpha
,\beta },g_{3_{k}}^{\alpha ,\beta }\right] =i\Theta _{231}g_{1_{k}}^{\alpha
,\beta },\quad \left[ g_{3_{k}}^{\alpha ,\beta },g_{1_{k}}^{\alpha ,\beta
}\right] =i\Theta _{312}g_{2_{k}}^{\alpha ,\beta }.
\label{su11}
\ee
Podemos observar destas rela\c{c}\~oes de comuta\c{c}\~ao que os geradores (\ref{generators}) satisfazem a \'algebra $SU(1,1)$ para a qual definimos 
\begin{equation}
\Theta _{ijk}\equiv 2\frac{\theta _{i_{k}}\theta _{j_{k}}}{\theta _{k_{k}}}.
\label{thetas}
\end{equation}
Como j\'a foi dito no in\'{\i}cio desta se\c{c}\~ao, podemos escrever um gerador geral como uma combina\c{c}\~ao linear dos geradores independentes escritos em (\ref{generators}) em uma forma compacta como segue
\begin{equation}
G=\sum_{k}\left( G_{k}^{\alpha }+G_{k}^{\beta }\right) ,
\label{gentransf}
\end{equation}
com os geradores das transforma\c{c}\~oes de Bogoliubov para os modos dos osciladores direitos e esquerdos da corda dados respectivamente por
\begin{eqnarray}
G_{k}^{\alpha } &=&\lambda _{1_{k}}\tilde{A}_{k}^{\dagger }\cdot
A_{k}^{\dagger }-\lambda _{2_{k}}A_{k}\cdot \tilde{A}_{k}+\lambda
_{3_{k}}\left( A_{k}^{\dagger }\cdot A_{k}+\tilde{A}_{k}^{\dagger }\cdot 
\tilde{A}_{k}+1\right) , \\
G_{k}^{\beta } &=&\lambda _{1_{k}}\tilde{B}_{k}^{\dagger }\cdot
B_{k}^{\dagger }-\lambda _{2_{k}}B_{k}\cdot \tilde{B}_{k}+\lambda
_{3_{k}}\left( B_{k}^{\dagger }\cdot B_{k}+\tilde{B}_{k}^{\dagger }\cdot 
\tilde{B}_{k}+1\right)
\label{rlgen}
\end{eqnarray}
onde os coeficientes $\lambda$ representam uma combina\c{c}\~ao linear complexa dos par\^ametros  $\theta$ que est\~ao relacionados \`a distribui\c{c}\~ao de Bose-Einstein, como segue
\begin{equation}
\lambda _{1_{k}}=\theta _{1_{k}}-i\theta _{2_{k}},\quad \lambda
_{2_{k}}=-\lambda _{1_{k}}^{*},\quad \lambda _{3_{k}}=\theta _{3_{k}}.
\label{lambdas}
\end{equation}
Desta forma, o operador (\ref{gentransf}) gera as transforma\c{c}\~oes t\'ermicas e a depend\^encia com a temperatura est\'a contida em $\lambda$. 

        Existe uma certa liberdade na escolha do par\^ametro  $\theta$, e podemos usar esta liberdade para fixarmos o tipo de transforma\c{c}\~ao. Existem duas condi\c{c}\~oes as quais uma transforma\c{c}\~ao t\'ermica deve satisfazer: i) unitariedade e  ii) e invari\^ancia sob conjuga\c{c}\~ao til de operadores arbitr\'arios como segue

\be
\tilde{(AB)} = \tilde{A}\tilde{B}~~~,~~~
\tilde{\alpha A} = \alpha^{*}\tilde{A},
\label{tildeop}
\ee
onde $\alpha$ \'e um n\'umero complexo e ${*}$ representa uma conjuga\c{c}\~ao complexa. A invari\^ancia sob conjuga\c{c}\~ao til garante a invari\^ancia do v\'acuo sob a mesma opera\c{c}\~ao. No entanto, o uso do grupo  $SU(1,1)$ t\'ermico implica na escolha de somente um tipo de transforma\c{c}\~ao. A unitariedade e a invari\^ancia por conjuga\c{c}\~ao til nem sempre s\~ao simultaneamente compat\'{\i}veis \cite{chu,hu}, sendo que, em geral, esta compatibilidade reduz esta teoria a DCT de um \'unico gerador (cap.3), pois somente o gerador $g_2$ \'e selecionado. Aqui, consideraremos que a condi\c{c}\~ao de unitariedade \'e a mais natural para o nosso sistema, sendo a outra escolha comentada na conclus\~ao.

\subsection{V\'acuo e Operadores T\'ermicos para a Corda Fechada}
${}$

          O v\'acuo t\'ermico \`a temperatura finita pode ser obtido atrav\'es da transforma\c{c}\~ao \cite{ut}

\begin{equation}
\left. \left| 0\left( \theta \right) \right\rangle \!\right\rangle
=e^{-iG}\left. \left| 0\right\rangle \!\right\rangle ,
\label{thermvac}
\end{equation}
onde, neste caso, $G$ \'e o operador geral de Bogoliubov escrito em (\ref{gentransf}) e o v\'acuo t\'ermico $\left. \left| 0\right\rangle \!\right\rangle$ \'e dado por (\ref{etcf}). Como os termos dos modos para a direita e para a esquerda comutam entre si, existem duas contribui\c{c}\~oes distintas dos setores direito $(\al)$ e  esquerdo $(\b)$. Uma vez que os termos que comp\~oem os geradores $G^{\al}_{k}$ e $G^{\al}_{k}$ satisfazem a \'algebra $SU \left( 1,1 \right)$, fazendo uso do Teorema do ``Desentrela\c{c}amento''  \cite{cha,eber}, podemos escrever o v\'acuo t\'ermico da seguinte forma
\begin{equation}
\left. \left| 0\left( \theta \right) \right\rangle \!\right\rangle _{\alpha
}=\prod_{k}e^{\Gamma _{1_{k}}\left( \tilde{A}_{k}^{\dagger }\cdot
A_{k}^{\dagger }\right) }e^{\log \left( \Gamma _{3k}\right) \left(
A_{k}^{\dagger }\cdot A_{k}+\tilde{A}_{k}^{\dagger }\cdot \tilde{A}%
_{k}+1\right) }e^{\Gamma _{2_{k}}\left( A_{k}\cdot \tilde{A}_{k}\right)
}\left. \left| 0\right\rangle \!\right\rangle _{\alpha },  \label{vt}
\label{vacalpha}
\end{equation}
para o qual os coeficientes dos v\'arios geradores s\~ao dados pelas seguintes rela\c{c}\~oes
\be
\Gamma _{1_{k}}=\frac{-\lambda _{1_{k}}\sinh \left( i\Lambda _{k}\right) }{%
\Lambda _{k}\cosh \left( i\Lambda _{k}\right) +\lambda _{3_{k}}\sinh \left(
i\Lambda _{k}\right) },\quad \Gamma _{2_{k}}=\frac{\lambda _{2_{k}}\sinh
\left( i\Lambda _{k}\right) }{\Lambda _{k}\cosh \left( i\Lambda _{k}\right)
+\lambda _{3_{k}}\sinh \left( i\Lambda _{k}\right) },
\label{lambda12}
\ee
\begin{equation}
\Gamma _{3_{k}}=\frac{\Lambda _{k}}{\Lambda _{k}\cosh \left( i\Lambda
_{k}\right) +\lambda _{3_{k}}\sinh \left( i\Lambda _{k}\right) },
\label{lambda3}
\end{equation}
e
\begin{equation}
\Lambda _{k}^{2}\equiv \left( \lambda _{3_{k}}^{2}+\lambda _{1_{k}}\lambda
_{2_{k}}\right) .
\label{biglambda}
\end{equation}
 Uma vez que os operadores $A_{k}^{\mu }$ e $\tilde{A}_{k}^{\mu }$ aniquilam o v\'acuo do modo direito \`a tempe-\\ratura nula, expandindo as expon\^enciais dos operadores na express\~ao (\ref{vacalpha}), encontramos que somente um termo contribui para o v\'acuo t\'ermico, ou seja,

\begin{equation}
\left. \left| 0(\theta )\right\rangle \!\right\rangle _{\alpha
}=\prod_{k}\Gamma _{3_{k}}e^{\Gamma _{1_{k}}\left( \tilde{A}_{k}^{\dagger
}\cdot A_{k}^{\dagger }\right) }\left. \left| 0\right\rangle \!\right\rangle
_{\alpha }.
\label{thermvacfin}
\end{equation}
O v\'acuo t\'ermico dos modos do movimento para a esquerda ($\left. \left| 0(\theta )\right\rangle \!\right\rangle _{\b}$) podem ser obtidos da mesma forma. O v\'acuo total \`a temperatura finita \'e dado pelo produto direto dos v\'acuos t\'ermicos $\al$ e $\beta$. A express\~ao obtida desta opera\c{c}\~ao \'e
\begin{equation}
\left. \left| 0(\theta )\right\rangle \!\right\rangle _{\alpha
}=\prod_{k}(\Gamma _{3_{k}})^{2tr \eta_{\mu \nu}} e^{\Gamma _{1_{k}}\left( \tilde{A}_{k}^{\dagger
}\cdot A_{k}^{\dagger }\right) }e^{\Gamma _{1_{k}}\left( \tilde{B}_{k}^{\dagger
}\cdot B_{k}^{\dagger }\right) }\left. \left| 0\right\rangle \!\right\rangle .
\end{equation}   
Os operadores s\~ao mapeados na temperatura finita pelos correspondentes geradores de Bogoliubov como segue
\begin{eqnarray}
A_{k}^{\mu }\left( \theta \right)  &=&e^{-iG_{k}^{\alpha }}A_{k}^{\mu
}e^{iG_{k}^{\alpha }},\qquad \tilde{A}_{k}^{\mu }\left( \theta \right)
=e^{-iG_{k}^{\alpha }}\tilde{A}_{k}^{\mu }e^{iG_{k}^{\alpha }},  \nonumber \\
B_{k}^{\mu }\left( \theta \right)  &=&e^{-iG_{k}^{\beta }}B_{k}^{\mu
}e^{iG_{k}^{\beta }}~~\qquad \tilde{B}_{k}^{\mu }\left( \theta \right)
=e^{-iG_{k}^{\beta }}\tilde{B}_{k}^{\mu }e^{iG_{k}^{\beta }}.
\label{thermop}
\end{eqnarray}
Usando as propriedades dos operadores $G$ descritas no cap.3, podemos obter das express\~oes acima rela\c{c}\~oes semelhantes para os operadores de cria\c{c}\~ao. Pode ser mostrado que estes operadores t\'ermicos satisfazem as mesmas rela\c{c}\~oes de comuta\c{c}\~ao dos operadores \`a temperatura nula.  Podemos ainda escrever este mapeamento em termos dos dubletos t\'ermicos \cite{chu,hu}, como segue 
\begin{equation}
\left( 
\begin{array}{c}
A_{k}^{\mu }\left( \theta \right)  \\ 
\tilde{A}_{k}^{\mu \dagger }\left( \theta \right) 
\end{array}
\right) ={\cal B}_{k}\left( 
\begin{array}{c}
A_{k}^{\mu } \\ 
\tilde{A}_{k}^{\mu \dagger }
\end{array}
\right) ,
\label{doublet}
\end{equation}
onde a matriz de transforma\c{c}\~ao que atua nos dubletos \`a temperatura nula  ${\cal B}_k$ \'e escrita na forma 
\begin{equation}
{\cal B}_{k}=\cosh \left( i\Lambda _{k}\right) {\cal I} +\frac{\sinh \left(
i\Lambda _{k}\right) }{\left( i\Lambda _{k}\right) }\left( 
\begin{array}{cc}
i\lambda _{3_{k}} & i\lambda _{1_{k}} \\ 
i\lambda _{2_{k}} & -i\lambda _{3_{k}},
\end{array}
\right) 
\label{explB}
\end{equation}
onde $\cal I$ \'e uma matriz identidade. Devemos novamente salientar que o sistema \`a temperatura finita satisfaz todas as propriedades do sistema \`a temperatura nula, pois os operadores t\'ermicos satisfazem as mesmas rela\c{c}\~oes de comuta\c{c}\~ao dos operadores em $T=0$. Deste modo, podemos construir solu\c{c}\~oes com condi\c{c}\~oes de contorno peri\'odicas para uma equa\c{c}\~ao de movimento \`a temperatura finita somente trocando os osciladores em $T=0$ pelos seus correspondentes em $T \neq 0$, uma vez que as coordenadas de posi\c{c}\~ao e momento do centro de massa s\~ao invariantes por transforma\c{c}\~oes de Bogoliubov. Desta equa\c{c}\~ao de movimento podemos tamb\'em construir um tensor de energia-momento que possui a mesma forma que o tensor \`a tempe- \\ ratura nula e que nos conduzem aos seguintes geradores da \'algebra de Virasoro  
\begin{equation}
L_{m}^{\alpha }\left( \theta \right) =\frac{1}{2}\sum_{k\in {\cal Z}}\alpha
_{-k}\left( \theta \right) \alpha _{k+m}\left( \theta \right) ,\quad
L_{m}^{\beta }\left( \theta \right) =\frac{1}{2}\sum_{k\in {\cal Z}}\beta
_{-k}\left( \theta \right) \beta _{k+m}\left( \theta \right) .
\label{thermalVir}
\end{equation}
Isto nos garante que realmente estamos trabalhando com cordas fechadas \`a tempe-\\ratura finita \cite{ivv,agv}. Portanto, temos agora todos os componentes necess\'arios para uma descri\c{c}\~ao t\'ermica da corda fechada.  

\subsection{Operador Entropia para a Corda Fechada}

No cap.3, descrevemos um operador cujo o valor esperado no v\'acuo t\'ermico, multiplicado pela constante de Boltzmann, recebe o nome de operador de entropia para um campo bos\^onico na aproxima\c{c}\~ao de Stirling, quando o sistema esta em equil\'{\i}brio, a saber:  

\begin{equation}
\frac{1}{k_{B}}\left\langle \left\langle 0\left( \theta \right) \right|
\right. K\left. \left| 0\left( \theta \right) \right\rangle \!\right\rangle
=\left\{ \sum_{k}\left[ \left( 1+n_{k}\right) \log \left( 1+n_{k}\right)
-n_{k}\log \left( n_{k}\right) \right] \right\},
\label{entrboson}
\end{equation}
onde $n_k$ \'e a densidade de n\'umero de part\'{\i}culas e $k_{B}$ a constante de Boltzmann. Definiremos para a nossa constru\c{c}\~ao o operador entropia para a corda bos\^onica fechada como sendo
\begin{equation}
K=K^{\alpha }+K^{\beta },
\label{entrstring}
\end{equation}
onde as entropias dos modos de movimento para a direita e para a esquerda s\~ao dadas, respectivamente, pelas seguintes rela\c{c}\~oes 
\bea
K^{\alpha } &=& \sum_{k}\left[ A_{k}^{\dagger }\cdot A_{k}\log \left( g\frac{%
\lambda _{1_{k}}\lambda _{2_{k}}}{\Lambda _{k}^{2}}\sinh ^{2}\left( i\Lambda
_{k}\right) \right) -A_{k}\cdot A_{k}^{\dagger }\log \left( 1+g\frac{\lambda
_{1_{k}}\lambda _{2_{k}}}{\Lambda _{k}^{2}}\sinh ^{2}\left( i\Lambda
_{k}\right) \right) \right] \label{rightentr}\non
K^{\beta }&=&-\sum_{k}\left[ B_{k}^{\dagger }\cdot B_{k}\log \left( g\frac{%
\lambda _{1_{k}}\lambda _{2_{k}}}{\Lambda _{k}^{2}}\sinh ^{2}\left( i\Lambda
_{k}\right) \right) -B_{k}\cdot B_{k}^{\dagger }\log \left( 1+g\frac{\lambda
_{1_{k}}\lambda _{2_{k}}}{\Lambda _{k}^{2}}\sinh ^{2}\left( i\Lambda_{k}
\right) \right) \right] . \nonumber
\label{leftentr}
\eea
 A escolha desta forma para o operador entropia se encontra no fato de podermos reproduzir a entropia apresentada em \cite{agv} para o caso de existir uma \'unica transforma\c{c}\~ao gerada por $g_{2k}$, isto \'e, quando   
$\theta _{1_{k}}=\theta _{3_{\kappa }}=0$. Tal escolha tamb\'em nos fornece o operador definido em  \cite{hu}
quando  $g = 1$. Portanto, realizando a m\'edia do operador (\ref{entrstring}) no v\'acuo t\'ermico e multiplicando o resultado pela constante de Boltzmann, obtemos a entropia para a corda bos\^onica fechada, ou seja,
\begin{eqnarray}
S &=&k_{B}\left\langle \left\langle 0\left( \beta \right) \right| \right.
K\left. \left| 0\left( \beta \right) \right\rangle \!\right\rangle  
\nonumber \\
&=&\sum_{k}\left[ \left( 1+n_{k}\right) \log \left( 1+n_{k}\right)
-n_{k}\log \left( n_{k}\right) -\frac{1}{2}\log \left( 1+n_{k}\right)
\right], 
\label{closedstringentr}
\end{eqnarray}
onde
\begin{equation}
n_{k}=g\left[ \frac{\lambda _{1_{k}}\lambda _{2_{k}}}{\Lambda _{k}^{2}}
\sinh^{2}\left( i\Lambda _{k}\right) \right] ,
\label{nnumb}
\end{equation}
e
\begin{equation}
g=\left\langle \left\langle 0\right| \right. \tilde{A}_{k}\cdot \tilde{A}%
_{k}^{\dagger }+\tilde{B}_{k}\cdot \tilde{B}_{k}^{\dagger }\left. \left|
0\right\rangle \!\right\rangle .
\label{gaver}
\end{equation}
O operador entropia para a corda bos\^onica fechada foi constru\'{\i}do como a soma das entropias dos modos dos movimentos para a esquerda e para a direita, tratados como dois subsistemas independentes do sistema total (corda bos\^onica fechada), ou seja, usamos a propriedade de extensividade do operador entropia. Uma observa\c{c}\~ao importante \'e que a entropia do sistema, eq.(\ref{closedstringentr}), vai a zero quando o sistema est\'a em equil\'{\i}brio, isto \'e, tomando a express\~ao para $n_{k}$, eq.(\ref{nk}), 
\begin{equation}
n_{k}=\frac{e^{-\left( k_{B}T\right) ^{-1}\omega }}{1-e^{-\left(
k_{B}T\right) ^{-1}\omega }},
\label{nequi}
\end{equation}
e tomando o limite  $T\rightarrow 0$. Isto garante que a terceira lei da termodin\^amica \'e satisfeita \cite{rk}. 


\section{Conclus\~ao}
${}$
       Inicialmente, descrevemos a a\c{c}\~ao de uma part\'{\i}cula cl\'assica puntual relativ\'{\i}stica, a qual, vimos ter uma interpreta\c{c}\~ao geom\'etrica: \'e proporcional ao comprimento da trajet\'oria (linha mundo) desta part\'{\i}cula no espa\c{c}o-tempo no qual est\'a imersa. Generalizamos ent\~ao este conceito para obtermos uma express\~ao para a a\c{c}\~ao de um objeto matem\'atico extenso e unidimensional (corda bos\^onica cl\'assica), ou seja, neste caso, a a\c{c}\~ao \'e proporcional \`a \'area delimitada pela trajet\'oria da corda no espa\c{c}o-tempo $D$-dimensional. Esta superf\'{\i}cie \'e chamada de folha mundo e \'e parametrizada por dois par\^ametros: $\t$ (tipo tempo) e $\s$ (tipo espa\c{c}o). Devido \`a presen\c{c}a de uma raiz quadrada nesta a\c{c}\~ao, foi necess\'ario escrevermos uma outra classicamente equi-\\valente, e que possui claramente invari\^ancia por reparametriza\c{c}\~ao local e reescalo-\\namento conforme da m\'etrica (Weyl), e uma outra  invari\^ancia que reflete a simetria global do espa\c{c}o-tempo (Lorentz ou Poincar\'e), no qual a corda est\'a se propagando. Do fato da a\c{c}\~ao ser uma invariante de Weyl, o tensor energia-momento possui tra\c{c}o nulo. Uma outra propriedade deste tensor, que nos conduz \`a v\'{\i}nculos na teoria que podem ser implementados somente no n\'{\i}vel cl\'assico, \'e $T_{ab}=0$. Nas coordenadas de cone de luz, estes v\'{\i}nculos tomam a forma $T_{++}=T_{--}=0$, e suas componentes de Fourier nos conduzem \`a uma \'algebra de Virasoro que n\~ao \'e v\'alida quanticamente, devido a problemas de ordenamento normal dos operadores.

       Classicamente, a teoria de cordas livre pode ser  formulada consistentemente em qualquer dimens\~ao espa\c{c}o-temporal, mas quando quantizamos, o espectro \'e livre de fantasmas somente para $D \leq 26$ e para $a=1$, onde $a$ \'e uma constante que surge devido ao ordenamento normal do produto dos operadores de cria\c{c}\~ao e aniquila\c{c}\~ao. A quantiza\c{c}\~ao can\^onica foi realizada em termos dos campos $X^{\mu}(\t,\s)$, somente com restri\c{c}\~oes f\'{\i}sicas no espa\c{c}o de Fock, originadas dos v\'{\i}nculos sobre o tensor energia-momento que d\~ao origem a graus de liberdade n\~ao f\'{\i}sicos (fantasmas). Vimos tamb\'em que surgem anomalias na \'algebra de Virasoro devido ao produto  normalmente ordenado dos operadores de cria\c{c}\~ao e aniquila\c{c}\~ao. Esta anomalia desaparece para determinados valores cr\'{\i}ticos de $D$  e $a$ como j\'a foi dito. Vimos tamb\'em que a quantiza\c{c}\~ao no calibre do cone de luz apresenta a vantagem de ser livre de fantasmas, embora n\~ao seja manifestamente covariante. A escolha deste calibre n\~ao manifestamente covariante nos permite resolver as equa\c{c}\~oes de v\'inculo de Virasoro e descrever a teoria em um espa\c{c}o de Fock que descreve somente graus de liberdade f\'{\i}sicos. Neste calibre vimos tamb\'em a necessidade de escolhermos $D=26$ e $a=1$.

      A corda fechada \'e descrita dobrando os graus de liberdade da corda aberta. Os modos para a direita ou para a esquerda s\~ao descritos matematicamente da mesma forma que a corda aberta, e seu estado \'e dado pelo produto direto do estado que representa o modo para a direita e do que representa o modo para a esquerda. Os dois modos s\~ao independentes, exceto para uma \'unica rela\c{c}\~ao, $L_{0}={\overline L}_{0}$. A corda fechada possui como import\^ancia particular, o fato do seu espectro conter gr\'avitons sem massa. 

       No cap\'{\i}tulo 3, mostramos como a m\'edia em um ensemble estat\'{\i}stico, que \'e dado por uma opera\c{c}\~ao de tra\c{c}o, pode ser trocado por uma m\'edia em mec\^anica qu\^antica. Este \'e o princ\'{\i}pio fundamental na contru\c{c}\~ao de uma Din\^amica de Campos T\'ermicos, que consiste basicamente em determinarmos um v\'acuo \`a temperatura finita que possa ser mapeado em um outro \`a temperatura nula atrav\'es de uma transforma\c{c}\~ao de Bogoliubov. Nesta constru\c{c}\~ao houve necessidade de dobrarmos os graus de liberdade do sistema, atrav\'es da introdu\c{c}\~ao de um sistema fict\'{\i}cio, que como vimos, possui todas as propriedades do sistema f\'{\i}sico real. Um espa\c{c}o de Fock t\'ermico p\^ode ser constru\'{\i}do a partir de aplica\c{c}\~oes dos operadores de cria\c{c}\~ao e destrui\c{c}\~ao t\'ermicos, cujos quais s\~ao obtidos por transforma\c{c}\~oes de Bogoliubov dos operadores em $T=0$, no v\'acuo t\'ermico. Verificamos a validade desta constru\c{c}\~ao por aplica\c{c}\~ao em um campo de Schr\"odinger, da qual obtivemos grandezas termodin\^amicas tais como entropia, energia livre de Helmholtz e tamb\'em a distribui\c{c}\~ao estat\'{\i}stica de Bose-Einstein. Em seguida, apresentamos regras gerais para a generaliza\c{c}\~ao de tal formalismo para uma teoria qu\^antica de campos arbitr\'aria. 

       Aplicamos os conceitos apresentados no cap\'{\i}tulos 2 e 3 para construirmos a corda bos\^onica t\'ermica no cap\'{\i}tulo 4. De acordo com a din\^amica de campos t\'ermicos, primeiramente duplicamos o sistema, e a c\'opia id\^entica que deve ser denotada por til, deve ser idependente do sistema original e n\~ao pode representar um sistema f\'{\i}sico. O espa\c{c}o de Hilbert do sistema total \'e dado pelo produto direto dos dois espa\c{c}os de Hilbert $(\cal{H} \ \mbox{e} \ \w{\cal{H}})$, e os operadores das duas cordas comutam entre si. Duplicando o sistema, surgem novos graus de liberdade, nos quais s\~ao atribu\'{\i}das as propriedades t\'ermicas da corda, que podem ser implementados no espa\c{c}o de Hilbert atrav\'es das transforma\c{c}\~oes t\'ermicas de Bogoliubov que leva o v\'acuo da corda em um v\'acuo t\'ermico definido por (\ref{0}). Como vimos, existem diversas formas de fazer tal mapeamento \cite{chu,hu} usando operadores que geram um grupo t\'ermico $SU(1,1)$, mas se exigirmos que esta transforma\c{c}\~ao seja unit\'aria e invariante por transforma\c{c}\~ao til, iremos selecionar somente um operador neste caso.  Os operadores de cria\c{c}\~ao e destrui\c{c}\~ao, equa\c{c}\~oes (\ref{annihT}) e (\ref{annihT1}), formam um conjunto de osciladores t\'ermicos, uma vez que estes satisfazem as mesmas rela\c{c}\~oes de comuta\c{c}\~ao dos operadores em $T=0$. Uma vez que as coordenadas e momentos do centro de massa s\~ao invariantes por transforma\c{c}\~oes de Bogoliubov, podemos construir a solu\c{c}\~ao $X^{\mu}(\b_{T})$ em $T \neq 0$ trocando os operadores em (\ref{sol}) pelos correspondentes em $T \neq 0 $. Do mesmo modo, constru\'{\i}mos os geradores da \'algebra de Virasoro em termos dos osciladores t\'ermicos e mostramos que a simetria conforme da solu\c{c}\~ao \'e preservada. Consequentemente, as transfoma\c{c}\~oes de Bogoliubov mapeiam as duas c\'opias da corda em duas c\'opias da corda t\'ermica. No entanto, a interpreta\c{c}\~ao de corda e corda til desaparece em $T\neq 0$, uma vez que os operadores de Bogoliubov misturam as duas c\'opias da corda. O v\'acuo t\'ermico tamb\'em \'e invariante sob opera\c{c}\~ao til. Podemos dizer que uma excita\c{c}\~ao da corda t\'ermica, sendo uma mistura de excita\c{c}\~oes em $T=0$ da corda e da corda til, carregam graus de liberdade t\'ermicos junto com os graus de liberdade din\^amicos.  Ap\'os introduzida a temperatura na corda bos\^onica, calculamos a entropia da corda em estados que dependem explicitamente das condi\c{c}\~oes de contorno. Para isto, partimos das solu\c{c}\~oes mais gerais das equa\c{c}\~oes de movimento, para cada uma das possibilidades de condi\c{c}\~oes de contorno e calculamos o valor esperado do operador entropia nos estados correspondentes \`a cada uma delas. As rela\c{c}\~oes importantes s\~ao (\ref{entropyRho}), (\ref{entrDD}), (\ref{entrDN}) e (\ref{entrND}). Estes elementos de matriz podem ser usados para calcularmos a entropia de v\'arios estados de corda aberta com diferentes condi\c{c}\~oes de contorno. Devemos observar que somente a estropia do estado correspondente \`as condi\c{c}\~oes de contorno de Neumann em ambas as extremidades depende de $\hbar$. No limite semicl\'assico, $\hbar \rightarrow 0$, a contribui\c{c}\~ao dos momentos tornam-se irrelevantes. O termo que domina \'e o mesmo que domina no limite da tens\~ao indo para o infinito, $\al^{'} \rightarrow 0$, e neste caso, a entropia dos estados correspondentes \`as outras condi\c{c}\~oes de contorno, (\ref{entrDD}), (\ref{entrDN}) e (\ref{entrND}) s\~ao nulas. Para finalizar, analisamos o grupo t\'ermico $SU(1,1)$ formado por todos os poss\'{\i}veis geradores unit\'arios de Bogoliubov no caso da corda bos\^onica fechada. A raz\~ao desta an\'alise \'e que constru\'{\i}mos a corda fechada t\'ermica no contexto da DCT onde o grupo $SU(1,1)$ representa a estrura mais geral das transforma\c{c}\~oes de Bogoliubov. Escolhendo um determinado tipo de parametro $\t$ as transforma\c{c}\~oes de Bogoliubov podem ser fixadas como sendo unit\'arias ou n\~ao unit\'arias. Escolhemos as transforma\c{c}\~oes unit\'arias para preservarmos  a estrutura do espa\c{c}o de Hilbert \`a temperatura zero e a interpreta\c{c}\~ao usual da mec\^anica qu\^antica. No entanto, com esta escolha, a invari\^ancia til do v\'acuo t\'ermico n\~ao \'e preservada e isto n\~ao \'e uma caracter\'{\i}stica desej\'avel na DCT. A solu\c{c}\~ao \'e construir o v\'acuo t\'ermico como um produto direto entre o v\'acuo t\'ermico original, obtido atrav\'es das transforma\c{c}\~oes de Bogoliubov e o conjugado sob opera\c{c}\~ao til. Se exigirmos que a transforma\c{c}\~ao geral de Bogoliubov seja unit\'aria e preserve a invari\^ancia til do v\'acuo, dois geradores do grupo $SU(1,1)$ ser\~ao eliminados. Deste modo, podemos concluir que a menos que o v\'acuo t\'ermico seja um produto do v\'acuo t\'ermico original e do conjugado til, n\~ao existe um grupo t\'ermico $SU(1,1)$ compat\'{\i}vel com a unitariedade da mec\^anica qu\^antica e a invari\^ancia til da DCT. 
        


\begin{appendix}

\section{Mapeamento do V\'acuo}
${}$
 
        Desejamos mostrar que o mapeamento do v\'acuo, em $T=0$, no v\'acuo t\'ermico, em $T \neq 0$, definido por
$$
|0(\b)\ra =e^{-iG}|0\ra \! \ra,
$$
onde $G=G^{\a}=-i\th(\b)(\widetilde{a}a-{a}^{\a}\widetilde{a}^{\a})$  o gerador das transforma\c{c}\~oes de Bogoliubov, nos conduz \`a equa\c{c}\~ao  
\bea
|0(\b)\ra &=&u(\b)^{-1} \exp(\frac{\v (\b)}{u(\b)} a^{\a}\ti a^{\a})|0 \ra \! \ra \non &=& \frac{1}{\cosh\th(\b)} \exp(\tanh\th(\b) a^{\a} \widetilde a^{\a})|0 \ra \! \ra, \nonumber
\eea
com $u(\b)$ e $\v(\b)$ dados pela equa\c{c}\~oes~(\ref{u11}) e (\ref{v11}) respectivamente.
 
        Das duas equa\c{c}\~oes anteriores, podemos escrever
\bea
|0(\b)\ra &=&e^{-\th(\b)( \widetilde{a}a- {\widetilde{a}}^{\a}a^{\a})}|O\ra \! \ra   \non
&=&e^{\t(A+B)}|0\ra \! \ra,   \nonumber
\eea
onde definimos $\t = \th(\b)$, $A=-  \widetilde{a}a$, $B=  \widetilde{a}^{\a}a^{\a}$. Para escrevermos a exponencial da soma como um produto de exponenciais, definiremos 
$$
E(\t)=e^{\t(A+B)}=e^{\al(\t)B}\k e^{\g(\t)A},
$$
onde $\k$ \'e um operador. Derivando $E(\t)$ em rela\c{c}\~ao a $\t$ obtemos
\be
\frac{d\k}{d\t}=(e^{-\al(\t)B}Be^{\al(\t)B})(1-\frac{d\al}{d\t})\k +(e^{-\al(\t)B}Ae^{\al(\t)B})\k-\frac{d\g(\t)}{d\t}\k A .   \label{ed1}
\ee  
Seja
\bea
f_{\th}(\t)&=&e^{-\al(\t)B}{\th}e^{\al(\t)B} \non
&=& \th + \frac{\al}{1!}[B,\th ]+ \frac{{\al}^{2}}{2!}[B,[B,\th ]] - \frac{{\al}^{3}}{3!}[B,[B,[B,\th ]]]+ \dots .  \label{f1}
\eea
ent\~ao
\be
f_{B}(\t)=e^{-\al(\t)B}Be^{\al(\t)B}=B,
\ee
portanto, a express\~ao (\ref{ed1}) fica
\be
\frac{d\k}{d\t}=B(1-\frac{d\al (\t)}{d\t})\k +(e^{-\al(\t)B}Ae^{\al(\t)B})\k-\frac{d\g (\t)}{d\t}\k A. \label{2}
\ee  

        Supondo que $\k$ seja uma exponencial, ou seja,
\be
\k=e^{\b(\t)C},   \label{3}
\ee
onde $C=[A,B]$, teremos
\be
\frac{d\k}{d\t}=\frac{d\b(\t)}{d\t}Ce^{\b(\t)C}.  \label{4}
\ee
As equa\c{c}\~oes (\ref{2}),(\ref{3}) e (\ref{4}) levam a
$$
\frac{d\b(\t)}{d\t}Ce^{\b(\t)C}= B \le (1-\frac{d\al(\t)}{d\t} \ri )e^{\b(\t)C} +(e^{-\al(\t)B}Ae^{\al(\t)B})e^{\b(\t)C}-\frac{d\g(\t)}{d\t}e^{\b(\t)C} A 
$$
\be
\frac{d\b(\t)}{d\t}C= \le(B(1-\frac{d\al(\t)}{d\t} )+e^{-\al(\t) B}Ae^{\al(\t) B} \ri )-\frac{d\g(\t)}{d\t}e^{\b(\t)C} A  e^{-\b(\t)C}  \label{5}
\ee 
e pela express\~ao (\ref{f}) teremos que
\be
e^{-\al(\t)B}Ae^{\al(\t)B} =A - \frac{\al}{1!}[B,A]+ \frac{{\al}^{2}}{2!}[B,[B,A ]] - \frac{{\al}^{3}}{3!}[B,[B,[B,A ]]]+ \dots .  \label{6}
\ee
e
\be
e^{\b(\t)C}Ae^{-\b(\t)C} =A + \frac{\b}{1!}[C,A]+ \frac{{\b}^{2}}{2!}[C,[C,A ]] + \frac{{\b}^{3}}{3!}[C,[C,[C,A ]]]+ \dots .  \label{7}
\ee
        Agora, verificaremos que os operadores $A$, $B$, e $C=[A,B]$ formam uma \'algebra fechada
\be 
[A,B]=[- \widetilde{a}a, \widetilde{a}^{\a}{a}^{\a}]=- \widetilde{a} \widetilde{a}^{\a} -{a}^{\a}a=C,   \label{8}
\ee
\be
[B,C]=[ \widetilde{a} \widetilde{a}^{\a} +{a}^{\a}a, \widetilde{a}^{\a}{a}^{\a}]=2 \widetilde{a}^{\a}{a}^{\a}=2 B, \label{9}
\ee
\be
[A,C]=[ \widetilde{a}a, \widetilde{a} \widetilde{a}^{\a} +{a}^{\a}a]=2 \widetilde{a}a=-2 A. \label{10}
\ee
Podemos ainda ver, destes tr\^es \'ultimos comutadores, que os operadores $A$, $B$, e $C$ satisfazem a identidade de Jacobi, ou seja
$$
[[A,B],C]+[[B,C],A]+[[C,A],B]=0.
$$

        Calculando explicitamente os comutadores que aparecem na express\~ao (\ref{6}) 
$$
[B,A]=-C,
$$
$$
[B,[B,A]]=-2 B,
$$
$$
[B,[B,[B,A]]]=0,
$$
e os demais termos tamb\'em s\~ao nulos, logo a equa\c{c}\~ao~(\ref{6}) fica
\be
e^{-\al(\t)B}Ae^{\al(\t)B} =A +\al C-{{\al}^{2}} B.  \label{11}
\ee
Para a  equa\c{c}\~ao~(\ref{7}) obtemos
$$
[C,A]=\l A,
$$
$$
[C,[C,A]]=\l [C, A]=2^{2} A,
$$
$$
[C,[C,[C,A]]]=[C,\l^{2} A]=2^{3} A,
$$
e desta forma, o e-n\'ezimo termo ser\'a
$$
[C,[C,\dots,[C,A]\dots]]=2^{n}A
$$
ent\~ao 
\bea
e^{\b(\t)C}Ae^{-\b(\t)C}&=& A + \frac{2\b}{1!}A+ \frac{(2\b)^{2}}{2!}A + \frac{(2\b)^{3}}{3!}A+ \dots . \non
&=&e^{2\b(\t)}A.   \label{12}
\eea

         Substituindo os resultados (\ref{11}) e (\ref{12}) na express\~ao (\ref{5}), teremos 
\be
\frac{d\b (\t)}{d\t}C= (1-\frac{d\al (\t)}{d\t})B+(A +\al(\t) C-\al^{2}(\t) B)-\frac{d\g (\t)}{d \t}e^{2\b(\t)}A
\ee  
ou ainda
\be
(1-\frac{d\g(\t)}{d \t}e^{2\b(\t)})A+(1-\frac{d\al(\t)}{d\t}-{{\al^{2}(\t)}}) B+(\al(\t) -\frac{d\b(\t)}{d\t})C=0. \label{13}
\ee

        Como cada operador n\~ao pode ser escrito como uma combina\c{c}\~ao linear dos dois outros, para que (\ref{13}) se verifique deveremos ter que 
\be
1-\frac{d\g (\t)}{d \t}e^{2\b(\t) }=0,   \label{14}
\ee
\be
1-\frac{d\al(\t)}{d\t}-\al^{2}(\t)=0,   \label{15}
\ee
\be
\al(\t) -\frac{d\b(\t)}{d\t}=0.     \label{16}
\ee
Integrando a equa\c{c}\~ao (\ref{15}) obtemos 
\be
\t = \int ^{\al(\t)}_{0} \frac{{d\al^{'}(\t)}}{1- {\al^{'2}(\t)}} \ .
\ee
Resolvendo esta integral resulta
\[\t = \int ^{\al(\t)}_{0} \frac{{d\al^{'}(\t)}}{1- {\al^{'2}(\t)}}=
 \left \{
\begin{array}{rl}
\frac{1}{2}\ln \left | \frac{\al^{'}(\t) +1}{\al^{'}(\t) -1} \right | \left| ^{\al(\t)}_{0} \right. &, \mbox{ para $|\al(\t)| >1$}; \\
 \mbox{arctanh} \ {\al^{'}(\t)} \left| ^{\al(\t)}_{0} \right. &,  \mbox{ para $|\al(\t)|  < 1$}.
\end{array}
\right.
\]
Agora, vamos analisar cada um dos dois casos. Primeiramente analisaremos o caso para o qual $|\al(\t)| >1$
\be
\t = \int ^{\al_{1}(\t)}_{0} \frac{{d\al^{'}(\t)}}{1- {\al^{'2}(\t)}}=\ln \le ( \frac{\al_{1}(\t)+1}{\al_{1}(\t)-1} \ri )^{\frac{1}{2}}  \label{17}
\ee
desta \'ultima equa\c{c}\~ao resulta 
\be
 \le ( \frac{\al_{1}(\t)+1}{\al_{1}(\t)-1} \ri )^{\frac{1}{2}}=e^{\t}
\ee
que, resolvendo para $\al_{1}(\t)$ obtemos
\be
\al_{1}(\t)=-\frac{1+e^{2\t}}{1-e^{2\t}}=\frac{e^{\t}+e^{-\t}}{e^{\t}-e^{-\t}}=\frac{\cosh{\t}}{\sinh{\t}}=\coth{\t}
\ee
da equa\c{c}\~ao (\ref{16}) e deste \'ultimo resultado temos
\be
\b_{1}(\t)= \int ^{\t}_{0}\al_{1}(\t^{'}) d\t^{'}=\int ^{\t}_{0} \coth{\t^{'}} d\t^{'}
\ee
cujo resultado diverge para $\b_{1}  \rightarrow -\infty$ quando $\t \rightarrow \infty$. Portando, este primeiro caso, n\~ao nos leva a uma solu\c{c}\~ao satisfat\'oria para
\be
e^{\t (A+B)}=e^{\al(\t) B} e^{\b(\t) C} e^{\g (\t) A}.  
\ee
Assim, assumiremos a outra hip\'otese, ou seja,  $|\al_{2}(\t)|  < 1$ . Desta forma, a equa\c{c}\~ao (\ref{17}) fica
\be
\t = \int ^{\al_{2}(\t)}_{0} \frac{{d\al^{'}(\t)}}{1- {\al^{'2}(\t)}}=\mbox{arctgh}\al_{2}(\t)
\ee
que resolvendo para $\al_{2}(\t)$ obtemos
\be
\b_{2}(\t) = \int^{\t}_{0} {\tanh {\t^{'}}} d \t
\ee
\be
\b_{2}(\t)=\ln{\cosh{\t}}.
\ee
Substituindo este resultado na equa\c{c}\~ao (\ref{14}) obtemos
\bea
\frac{d\g_{2}(\t)}{d \t}&=&e^{-2\b_{2}(\t) }  \non
\frac{d\g_{2}(\t)}{d \t}&=&e^{-2 \ln{\cosh{\t}} }  \non
d{\g_{2}}(\t)&=&\cosh^{-2}{\t} d\t \non
{\g_{2}}(\t)&=&\tanh{\t}.
\eea
Portanto
\bea
|O(\b(\t))\ra &=&e^{\t (A+B)}|0\ra \! \ra \non
&=&e^{\al(\t) B} e^{\b (\t) C} e^{\g (\t) A} |0\ra \! \ra \non
&=&e^{\tanh{\t}a^{\a}  {\widetilde{a}}^{\a}}e^{-\ln{\cosh{\t}( \widetilde{a}  \widetilde{a}^{\a}+a^{\a}a)}}e^{-\tanh{\t}  \widetilde{a}a} |0\ra \! \ra \non
&=&e^{\tanh{\th (\b)}a^{\a}  \widetilde{a}^{\a}}e^{-\ln{\cosh{\th (\b)}( \widetilde{a}  \widetilde{a}^{\a}+a^{\a}a)}}(1-\tanh{\t}  \widetilde{a}a +\frac{1}{2!}(\tanh{\t}  \widetilde{a}a)^{2}-.... |0\ra \! \ra \non
&=&e^{\tanh{\th (\b)}a^{\a}  \widetilde{a}^{\a}}e^{\ln{\cosh{\th (\b)}(1+  \widetilde{a}^{\a} \widetilde{a}+a^{\a}a)}}|0\ra \! \ra \non
&=&e^{\tanh{\th (\b)}a^{\a}  \widetilde{a}^{\a}}e^{-\ln{\cosh{\th (\b)}}}e^{-\ln{\cosh{\th (\b)}} \widetilde{a}^{\a} \widetilde{a}}e^{-\ln{\cosh{\th (\b)}}a^{\a}a}|0\ra \! \ra \non
&=&e^{\tanh{\th (\b)}a^{\a}  \widetilde{a}^{\a}}(\cosh{\th (\b)})^{-1}|0\ra \! \ra \non
&=&\frac{1}{\cosh{\th (\b)}}\exp{\tanh{\th (\b)}a^{\a}  \widetilde{a}^{\a}}|0\ra \! \ra
\eea
como quer\'{\i}amos demostrar. 

\end{appendix}



%
%
%
%
%

\end{document}